\documentclass[twocolumn]{aastex62}

\usepackage{mathptmx}
\usepackage[T1]{fontenc}
\usepackage{ae,aecompl}
\usepackage{graphicx}	
\graphicspath{{./Figures/}} 
\usepackage{amsmath}	
\usepackage{amssymb}	
\usepackage{subfigmat}
\usepackage{enumerate}
\usepackage{multirow}
\usepackage{natbib}

\received{-}
\revised{-}
\accepted{-} 

\shorttitle{The JCMT Transient Survey: A Four Year Summary}
\shortauthors{Y.-H. Lee et al.}

\begin{document}

\title{The JCMT Transient Survey: Four Year Summary of Monitoring the Submillimeter Variability of Protostars}

\author[0000-0001-6047-701X]{Yong-Hee Lee}
\affil{School of Space Research and Institute of Natural Sciences, Kyung Hee University, 1732 Deogyeong-daero, Giheung-gu, Yongin-si, Gyeonggi-do 446-701, Korea}

\author[0000-0002-6773-459X]{Doug Johnstone}
\affiliation{NRC Herzberg Astronomy and Astrophysics, 5071 West Saanich Rd, Victoria, BC, V9E 2E7, Canada}
\affiliation{Department of Physics and Astronomy, University of Victoria, 3800 Finnerty Road, Elliot Building, Victoria, BC, V8P 5C2, Canada}

\author[0000-0003-3119-2087]{Jeong-Eun Lee}
\affiliation{School of Space Research and Institute of Natural Sciences, Kyung Hee University, 1732 Deogyeong-daero, Giheung-gu, Yongin-si, Gyeonggi-do 446-701, Korea}

\author{Gregory Herczeg}
\affiliation{Kavli Institute for Astronomy and Astrophysics, Peking University, Yiheyuan 5, Haidian Qu, 100871 Beijing, China}

\author[0000-0002-6956-0730]{Steve Mairs}
\affiliation{East Asian Observatory, 660 N. A'ohoku Place, Hilo, HI 96720, USA}

\author{Carlos Contreras-Pe\~na}
\affiliation{Physics and Astronomy, University of Exeter, Stocker Road, Exeter EX4 4QL, UK}

\author{Jennifer Hatchell}
\affiliation{Physics and Astronomy, University of Exeter, Stocker Road, Exeter EX4 4QL, UK}

\author{Tim Naylor}
\affiliation{Physics and Astronomy, University of Exeter, Stocker Road, Exeter EX4 4QL, UK}

\author[0000-0003-0438-8228]{Graham S.\ Bell}
\affiliation{East Asian Observatory, 660 N. A'ohoku Place, Hilo, HI 96720, USA}

\author[0000-0001-7491-0048]{Tyler L.\ Bourke}
\affiliation{SKA Observatory, Jodrell Bank, Lower Withington, Macclesfield, SK11 9FT, UK}
\affiliation{Jodrell Bank Centre for Astrophysics, School of Physics and Astronomy, University of Manchester, Manchester, M13 9PL, UK}

\author{Colton Broughton}
\affiliation{Department of Physics and Astronomy, University of Victoria, 3800 Finnerty Road, Elliot Building, Victoria, BC, V8P 5C2, Canada}

\author[0000-0001-8822-6327]{Logan Francis}
\affiliation{Department of Physics and Astronomy, University of Victoria, 3800 Finnerty Road, Elliot Building, Victoria, BC, V8P 5C2, Canada}
\affiliation{NRC Herzberg Astronomy and Astrophysics, 5071 West Saanich Rd, Victoria, BC, V9E 2E7, Canada}

\author[0000-0002-9959-1933]{Aashish Gupta}
\affiliation{Institute of Astronomy, National Central University, 300 Zhongda Road,, Zhongli 32001 Taoyuan, Taiwan}
\affiliation{Institute of Astronomy and Astrophysics, Academia Sinica, No. 1, Sec. 4, Roosevelt Road, Taipei 10617, Taiwan, R. O. C.}

\author{Daniel Harsono}
\affiliation{Institute of Astronomy and Astrophysics, Academia Sinica, No. 1, Sec. 4, Roosevelt Road, Taipei 10617, Taiwan, R. O. C.}

\author{Sheng-Yuan Liu}
\affiliation{Institute of Astronomy and Astrophysics, Academia Sinica, No. 1, Sec. 4, Roosevelt Road, Taipei 10617, Taiwan, R. O. C.}

\author{Geumsook Park}
\affiliation{Korea Astronomy and Space Science Institute, 776 Daedeokdae-ro, Yuseong-gu, Daejeon 34055, Republic of Korea}

\author{Spencer Plovie}
\affiliation{Department of Physics and Astronomy, University of Victoria, 3800 Finnerty Road, Elliot Building, Victoria, BC, V8P 5C2, Canada}

\author{Gerald H.\ Moriarty-Schieven}
\affiliation{NRC Herzberg Astronomy and Astrophysics, 5071 West Saanich Rd, Victoria, BC, V9E 2E7, Canada}

\author{Aleks Scholz}
\affiliation{SUPA, School of Physics \& Astronomy, University of St Andrews, North Haugh, St Andrews, KY16 9SS, United Kingdom}

\author{Tanvi Sharma}
\affiliation{Institute of Astronomy, National Central University, 300 Zhongda Road,, Zhongli 32001 Taoyuan, Taiwan}

\author{Paula Stella Teixeira}
\affiliation{SUPA, School of Physics \& Astronomy, University of St Andrews, North Haugh, St Andrews, KY16 9SS, United Kingdom}

\author{Yao-Te Wang}
\affiliation{Graduate Institute of Astrophysics, National Taiwan University, No. 1, Sec. 4, Roosevelt Road, Taipei 10617, Taiwan, R. O. C.}

\author{Yuri Aikawa}
\affiliation{Department of Astronomy, University of Tokyo, 7-3-1 Hongo, Bunkyo-ku, Tokyo 113-0033, Japan}

\author[0000-0003-4056-9982]{Geoffrey C.\ Bower}
\affiliation{Institute of Astronomy and Astrophysics, Academia Sinica, 645 N. A'ohoku Place, Hilo, HI 96720, USA}

\author{Huei-Ru Vivien Chen}
\affiliation{Department of Physics and Institute of Astronomy, National Tsing Hua University, Taiwan}


\author[0000-0001-7258-770X]{Jaehan Bae}
\affiliation{Earth and Planets Laboratory, Carnegie Institution for Science, 5241 Broad Branch Road NW, Washington, DC 20015, USA}

\author{Giseon Baek}
\affiliation{School of Space Research and Institute of Natural Sciences, Kyung Hee University, 1732 Deogyeong-daero, Giheung-gu, Yongin-si, Gyeonggi-do 446-701, Korea}

\author{Scott Chapman}
\affiliation{Department of Physics and Atmospheric Science, Dalhousie University, Halifax, NS, B3H 4R2, Canada}

\author[0000-0003-0262-272X]{Wen Ping Chen}
\affiliation{Institute of Astronomy, National Central University, 300 Zhongda Road,, Zhongli 32001 Taoyuan, Taiwan}

\author{Fujun Du}
\affiliation{Purple Mountain Observatory and Key Laboratory of Radio Astronomy, Chinese Academy of Sciences, 10 Yuanhua Road, Qixia District, Nanjing 210033, PR China}

\author{Somnath Dutta}
\affiliation{Institute of Astronomy and Astrophysics, Academia Sinica, No. 1, Sec. 4, Roosevelt Road, Taipei 10617, Taiwan, R. O. C.}

\author{Jan Forbrich}
\affiliation{Centre for Astrophysics Research, University of Hertfordshire, Hatfield AL10 9AB, UK}
\affiliation{Center for Astrophysics, Harvard \& Smithsonian, 60 Garden St, MS 
72, Cambridge, MA 02138, USA}

\author{Zhen Guo}
\affiliation{Centre for Astrophysics Research, University of Hertfordshire, Hatfield AL10 9AB, UK}

\author[0000-0003-4366-6518]{Shu-ichiro Inutsuka}
\affiliation{Department of Physics, Graduate School of Science, Nagoya University, Furo-cho, Chikusa-ku, Nagoya 464-8602, Japan}

\author{Miju Kang}
\affiliation{Korea Astronomy and Space Science Institute, 776 Daedeokdae-ro, Yuseong-gu, Daejeon 34055, Republic of Korea}

\author{Helen Kirk}
\affiliation{NRC Herzberg Astronomy and Astrophysics, 5071 West Saanich Rd, Victoria, BC, V9E 2E7, Canada}
\affiliation{Department of Physics and Astronomy, University of Victoria, 3800 Finnerty Road, Elliot Building, Victoria, BC, V8P 5C2, Canada}

\author{Yi-Jehng Kuan}
\affiliation{Department of Earth Sciences, National Taiwan Normal University, Taipei 116, Taiwan}
\affiliation{Institute of Astronomy and Astrophysics, Academia Sinica, No. 1, Sec. 4, Roosevelt Road, Taipei 10617, Taiwan, R. O. C.}

\author{Woojin Kwon}
\affiliation{Department of Earth Science Education, Seoul National University, 1 Gwanak-ro, Gwanak-gu, Seoul 08826, Republic of Korea}
\affiliation{SNU Astronomy Research Center, Seoul National University, 1 Gwanak-ro, Gwanak-gu, Seoul 08826, Republic of Korea}

\author{Shih-Ping Lai}
\affiliation{Institute of Astronomy and Dept.\ of Physics, National Tsing Hua University, 101 Section 2 Kuang Fu Road, 30013 Hsinchu, Taiwan}
\affiliation{Institute of Astronomy and Astrophysics, Academia Sinica, No. 1, Sec. 4, Roosevelt Road, Taipei 10617, Taiwan, R. O. C.}

\author{Bhavana Lalchand}
\affiliation{Institute of Astronomy, National Central University, 300 Zhongda Road,, Zhongli 32001 Taoyuan, Taiwan}

\author[0000-0001-8472-6404]{James M.\ M.\ Lane}
\affiliation{Department of Physics and Astronomy, University of Victoria, 3800 Finnerty Road, Elliot Building, Victoria, BC, V8P 5C2, Canada}.

\author{Chin-Fei Lee}
\affiliation{Institute of Astronomy and Astrophysics, Academia Sinica, No. 1, Sec. 4, Roosevelt Road, Taipei 10617, Taiwan, R. O. C.}

\author{Tie Liu}
\affiliation{Shanghai Astronomical Observatory, Chinese Academy of Sciences, 80 Nandan Road, Shanghai 200030, People's Republic of China}

\author{Oscar Morata}
\affiliation{European Southern Observatory, Karl-Schwarzchild-Str. 2, D-85748, Garching, Germany}

\author{Samuel Pearson}
\affiliation{SUPA, School of Physics \& Astronomy, University of St Andrews, North Haugh, St Andrews, KY16 9SS, United Kingdom}

\author{Andy Pon}
\affiliation{Department of Physics and Astronomy, The University of Western Ontario, 1151 Richmond Street, London, N6A 3K7, Canada}

\author[0000-0002-4393-3463]{Dipen Sahu}
\affiliation{Institute of Astronomy and Astrophysics, Academia Sinica, No. 1, Sec. 4, Roosevelt Road, Taipei 10617, Taiwan, R. O. C.}

\author[0000-0001-8385-9838]{Hsien Shang
}
\affiliation{Institute of Astronomy and Astrophysics, Academia Sinica, No. 1, Sec. 4, Roosevelt Road, Taipei 10617, Taiwan, R. O. C.}

\author[0000-0002-4502-8344]{Dimitris Stamatellos}
\affiliation{Jeremiah Horrocks Institute for Mathematics, Physics \& Astronomy, University of Central Lancashire, Preston, PR1 2HE, UK}

\author[0000-0003-4247-1401]{Shih-Yun Tang}
\affiliation{Department of Physics, National Central University, 300 Zhongda Road, Zhongli, Taoyuan 32001, Taiwan}
\affiliation{Lowell Observatory, 1400 W. Mars Hill Road, Flagstaff, AZ 86001, USA}
\affiliation{Department of Astronomy and Planetary Sciences, Northern Arizona University, Flagstaff, AZ 86011, USA}

\author{Ziyan Xu}
\affiliation{Kavli Institute for Astronomy and Astrophysics, Peking University, Yiheyuan 5, Haidian Qu, 100871 Beijing, China}

\author[0000-0002-8578-1728]{Hyunju Yoo}
\affiliation{Korea Astronomy and Space Science Institute, 776 Daedeokdae-ro, Yuseong-gu, Daejeon 34055, Republic of Korea}


\begin{abstract}
We present the four-year survey results of monthly submillimeter monitoring of eight nearby ($< 500\,$pc) star-forming regions by the JCMT Transient Survey. We apply the Lomb-Scargle Periodogram technique to search for and characterize variability on 295 submillimeter peaks brighter than 0.14 Jy\,beam$^{-1}$, including 22 disk sources (Class\,II), 83 protostars (Class\,0/I), and 190 starless sources.
We uncover 18 secular variables, all of them protostars. No 
single-epoch burst or drop events and no inherently stochastic sources are observed. We classify the secular variables by their timescales into three groups: Periodic, Curved, and Linear.  
For the Curved and Periodic cases, the detectable fractional amplitude, with respect to mean peak brightness, is $\sim4$\% for sources brighter than $\sim$ 0.5 Jy\,beam$^{-1}$. Limiting our sample to only these bright sources, the observed variable fraction is 37\% (16 out of 43). 
Considering source evolution, we find a similar fraction of bright variables for both Class\,0 and Class\,I.
Using an empirically motivated conversion from submillimeter variability to variation in mass accretion rate, six sources (7\% of our full sample) are predicted to have years-long accretion events during which the excess mass accreted reaches more than 40\% above the total quiescently accreted mass: two previously known eruptive Class\,I sources, V1647\,Ori and EC\,53 (V371\,Ser), and four Class\,0 sources, HOPS\,356, HOPS\,373, HOPS\,383, and West\,40. Considering the full protostellar ensemble, the importance of episodic accretion on few years timescale is negligible, only a few percent of the assembled mass. However, given that this accretion is dominated by events of order the observing time-window, it remains uncertain as to whether the importance of episodic events will continue to rise with decades-long monitoring.  
\end{abstract}


\section{Introduction}\label{sec:introduction}

Mass accretion is the key process of star formation; however, the time dependence is not yet well understood \citep[see review by][]{dunham14}.
A potential discrepancy between observed luminosities of YSOs and the luminosities expected from theory, the Luminosity Problem, was uncovered by \citet{kenyon90}, and updated by \citet{evans09, dunham10}, following observations by the {\it Spitzer Space Telescope}.
The episodic accretion model for protostellar assembly, which has quiescent-accretion phases interspersed with burst-accretion phases, has been suggested as a possible solution to this discrepancy.  Other solutions require that most of the accretion occurs very early in the protostellar evolution \citep[e.g.][]{offner11,fischer19}

Some support for the episodic accretion hypothesis can be found among young eruptive variable stars, which show a significant increase in their optical and near-IR brightness due to abrupt changes in the accretion rate \citep{hartmann96}. 
Based on photometric \citep{hartmann98, scholz13, hillenbrand15, contreras17, contreras19, fischer19}, chemical \citep{jorgensen15, hsieh19}, and outflow \citep{bontemps96} surveys, outburst events appear to occur more frequently during the earliest evolutionary stage of formation, in agreement with theoretical \citep{machida11, bae14, vorobyov15} studies.
Furthermore, during this early stage, while the protostar remains deeply embedded in its natal envelope, most of the stellar mass is gained. Therefore, carefully studying the variability of protostars which are densely enshrouded is essential to understanding the effects of mass accretion variability during the star formation process.

Searches for variability of YSOs using optical and infrared data have been effective for discovering variable accretion in the later stages of star formation.
Deeply embedded protostars (Class 0), however, are difficult to detect at optical and near-IR because of their thick envelope, which absorbs the short-wavelength radiation from the central protostar and re-emits it thermally, at much longer wavelengths.
\citet{Johnstone13} modeled the effect of enhanced accretion on dust emission from the heated envelope and found that the 
 thermal response timescale to accretion luminosity changes at the protostar should be weeks to months, limited only by the envelope light-crossing time as the dust is quick to thermally equilibrate. 

The possibility of detecting submillimeter variability on months-long timescales motivated the JCMT-Transient Survey \citep{herczeg17}. We are monitoring eight nearby star-forming regions at 450 and 850 $\micron$ at an approximately monthly cadence using the Submillimetre Common-User Bolometer Array 2 \citep[SCUBA-2; ][]{holland13} on the James Clerk Maxwell Telescope (JCMT).  Roughly 100 protostars in these fields are bright enough for evaluating variability, with a relative flux accuracy of $\sim$2\% relative achieved by careful calibration \citep{mairs17a}.

In our previous paper, \citet{johnstone18} investigated the stochastic and linear secular variability of 150 submillimeter emission peaks, both protostellar and pre-stellar, over the first 18 months of the monitoring survey. In that paper we uncovered six robust protostellar variables by fitting linear trends to the submillimeter light curves and measuring the observed epoch to epoch submillimeter brightness standard deviations. Five out of 50 protostars were found to be secularly (linearly) variable.  One source, \mbox{EC 53} (\mbox{V371 Ser}) in the Serpens Main region \citep[see also,][]{yoo17, mairs17b}, was found to have a significant stochasticity. \mbox{EC 53} had previously been observed to vary periodically in the near-IR \citep{hodapp12}, with a timescale of 1.5 yr.
More detailed analysis of EC 53 \citep{lee20} reveals a similar periodic variability in submillimeter brightness as well as for the near-IR color, the latter produced by the combination of variable brightness and extinction associated with the underlying accretion process \citep{lee20}. The secular variability found in EC 53 allows for a quantitative determination of accretion-related amplitudes and timescales and a qualitative discussion of the physical instabilities driving the observed periodicity. These results also motivate a detailed search for similar periodic trends in the submillimeter light curves toward our full sample of embedded protostars.

In this paper, we examine protostellar variability beyond the linear trend by applying the Lomb-Scargle Periodogram (LSP) \citep{lomb76, scargle89, vanderplas18} analysis to the first four years of monitoring measurements obtained by the JCMT Transient Survey. 
In Section \ref{sec:observation}, we summarize the JCMT Transient Survey observing methods and describe the data reduction process including precision relative pointing and flux calibration. 
In Section \ref{sec:V}, we present the analysis methods that are applied to find protostellar variability from the submillimeter light curves. We separate the secular variability, extracted from our LSP and linear analysis (Section \ref{sec:V_Sec}), and the stochastic variability, found after subtracting any secular variability (Section \ref{sec:V_STCH}). 
In Section \ref{sec:C}, we determine the completeness of our secular variability analysis as a function of source brightness.
In Section \ref{sec:Disc}, we discuss the ensemble variability statistics with respect to individual regions and to protostellar evolutionary stage (Section \ref{sec:Region} and \ref{sec:Evol}), and compare the submillimeter light curve behavior of sources identified at shorter wavelengths as either bursting (Section \ref{sec:FUors}) or appearing subluminous (Section \ref{sec:VeLLOs}). 
In Section \ref{sec:Concl}, we summarize our results.
\section{The JCMT Transient Survey}\label{sec:observation}

The JCMT Transient Survey \citep{herczeg17} uses 
the Sub-millimeter Common User Bolometer Array 2 
(SCUBA-2; \citealt{holland13}) to observe 
continuum emission simultaneously at 
$450$ and $850~\micron$ with effective 
beam sizes of $9.8 \arcsec$ and $14.6 \arcsec$, 
respectively \citep{dempsey13}. Since 2015 
December, the survey has monitored eight 
star-forming regions within \mbox{500 pc} of the 
Sun: \mbox{IC 348}, \mbox{NGC 1333}, 
\mbox{NGC 2024}, \mbox{NGC 2068}, 
Orion Molecular Cloud \mbox{(OMC) 2/3}, 
\mbox{Ophiuchus}, \mbox{Serpens Main}, 
and \mbox{Serpens South}. 
All but \mbox{Serpens Main} are monitored at an 
approximately monthly cadence as weather 
conditions and target observability allow, while 
\mbox{Serpens Main} is monitored twice per month 
to provide additional information on the 
accretion variability of deeply embedded 
protostar \mbox{EC 53} \citep{yoo17,lee20}. In this paper we  
include almost all monitoring data obtained from the beginning of 
the survey through 2020 January.
Three observations of \mbox{NGC 
1333} on 2016 February 5, 2017 December 8, and 
2018 July 07 and one observation of 
\mbox{OMC 2/3} observed on 2018 March 8 (less than 2\% of our observations)  are excluded due to 
anomalously high uncertainty in the relative spatial 
alignment of the final maps (see 
Table~\ref{tab:obssummary}). 

The observations employ the \textsc{Pong1800} 
\citep{kackley10} scan pattern to produce a 
circular field $\sim30 \arcmin$ in diameter with 
uniform background RMS noise. During each 
observation, the telescope scans across the sky 
at a rate of $~400\arcsec\,$s$^{-1}$ such that each 
part of the target field is observed from 
multiple position angles. This strategy allows 
for the separation of telluric contributions and 
astronomical source contributions to the flux in 
the data reduction process. The observations are 
performed up to a maximum zenith opacity at 
$225\mathrm{\:GHz}$ ($\tau_{225}$) of 0.12, 
corresponding to precipitable water vapor of 
\mbox{$<2.58$ mm}, as measured by the JCMT 
line-of-sight \mbox{$183$ GHz} water vapour 
radiometer \citep{dempsey13}. Depending upon the 
sky conditions, the integration time is between 
$20-40$\,minutes, in order to yield a consistent 
sensitivity of $\sim14$\,mJy\,beam$^{-1}$ at 850\,$\micron$ \citep{mairs17a}. The atmospheric transmission in the \mbox{450\,$\micron$} band, however, has a much stronger dependence on precipitable water vapor, yielding more than an order of magnitude variation in RMS noise and thus analyis of the \mbox{450\,$\micron$} survey data are deferred to a future publication. 

The data are reduced 
using the iterative {\sc{makemap}} 
\citep{chapin13} software, found in 
{\sc{starlink}}’s \citep{currie14} Submillimetre 
User Reduction Facility ({\sc{smurf}}) package 
\citep{jenness13}. No $^{12}$CO subtraction was 
performed at \mbox{850\,$\micron$}. 
The standard JCMT observing scheme and data 
reduction pipeline yields a pointing uncertainty 
of $\sim2$ or $3\arcsec$ and a flux calibration 
uncertainty of $\sim5$ to $10\%$ \citep{dempsey13}. 
All observations presented in this work are then post-processed to improve spatial 
alignment and reduce the flux calibration 
uncertainty in a relative sense (epoch to epoch).
Using the methods described in \cite{mairs17a}, the 
locations of bright, compact point sources are 
measured from epoch to epoch to derive pointing
corrections in order to align the images to 
better than $1\arcsec$. In addition, by 
monitoring the peak fluxes of a set of bright, 
non-varying sources over time, relative flux 
calibration factors are derived for each epoch, 
reducing the flux calibration uncertainty at 
\mbox{850 $\micron$} to $2\%$\footnote{The fluxes presented 
throughout this work have been normalized to the 
average images of each field calculated over the 
first $\sim6\mathrm{\:months}$ of Transient 
Survey observations.}. 

\begin{deluxetable*}{ccccccc}[htb]
\tablecaption{Observation Summary by Region \label{tab:obssummary}}
\tablehead{
\colhead{Region} & 
\colhead{Start date} & 
\colhead{Last date} &
\colhead{\# of epochs} &
\colhead{N$_{\rm submm}$\tablenotemark{a}} &
\colhead{N$_{\rm disk}$} &
\colhead{N$_{\rm protostar}$}
}
\startdata
IC 348 & 2015 Dec 22 & 2020 Jan 24 & 35 & 9 & 0 & 5 \\
NGC 1333 & 2015 Dec 22 & 2020 Jan 24 & 34\tablenotemark{b} & 31 & 1 & 16 \\
NGC 2024 & 2015 Dec 26 & 2020 Jan 24 & 35 & 36 & 3 & 3 \\
NGC 2068 & 2015 Dec 26 & 2020 Jan 24 & 36 & 29 & 0 & 15 \\
OMC 2/3 & 2015 Dec 26 & 2020 Jan 24 &  33\tablenotemark{b} & 95 & 13 & 14 \\
Ophiuchus& 2016 Jan 15 & 2020 Jan 18 & 28 & 33 & 5 & 9 \\
Serpens Main & 2016 Feb 02 & 2019 Oct 19 &  50 & 16 & 0 & 8 \\
Serpens South & 2016 Feb 02 & 2019 Nov 02 &  35 & 46 & 0 & 13 \\
\enddata
\tablenotetext{a}{Total number of detected submillimeter  sources (> 0.14 Jy beam$^{-1}$), including disks, protostars, and prestellar cores.}
\tablenotetext{b}{Three epochs of NGC\,1333 and one epoch of OMC\,2/3 have been excluded from the analysis because of poor telescope pointing.}
\end{deluxetable*}


We investigate all eight Transient Survey fields, limiting our analysis to those sources within 18$\arcmin$ of each mapping center where the map noise properties are uniform.
Among 1665 submillimeter peaks identified through clump-finding \citep{johnstone18} with the FellWalker algorithm \citep{berry15}, we analyse 295 sources with the mean peak brightness greater than 0.14 \mbox{Jy beam$^{-1}$}, corresponding to a signal to noise per epoch of $\sim$10.
 
Our final sample contains submillimeter peaks coincident to within 10 arcseconds of 22 disk (Class II) objects and 83 protostars (Class 0/I), which have previously been classified through their SEDs \citep{dunham10,stutz13,megeath16}. Table \ref{tab:obssummary} provides the numbers of submillimeter sources, protostars, and disks per star-forming region, along with the number and date range of the epochs observed.





\section{Analysis}\label{sec:V}

Following the methodology by \citet{johnstone18}, we separate our variability analysis into two investigations dependent on the light curve properties. We call those sources with light curves that can be modelled as smoothly varying over the 4.5 years of monitoring secular, and those with light curves more likely due to either singular or on-going discrete events stochastic. We note that the labels secular and stochastic implicitly depend on a timescale and are sometimes defined differently. For example, \citet{dunham14} used the term, {\it stochastic} to refer to any variability arising from random processes over a wide variety of timescales, while the term, {\it secular} to refer to very slow, long-term changes in accretion rates that occur over the full lifetime of the embedded phase. 
When contemplated over such longer times, the secular variations uncovered here may be interpreted as stochastic, with time constants of at least several years.

\subsection{Secular Variability}\label{sec:V_Sec}

We first examine secular variability, changes that take place across the entire time coverage. We determine the amplitudes and timescales associated with the observed brightness fluctuations, and then use these measurements to classify the types of variability. Furthermore, these estimates for the timescales involved in the observed variations provide a direct link to the underlying physical process responsible for the variations (e.g.\ viscously evolving disks, Keplerian orbit times).

We apply a Lomb-Scargle Periodogram (LSP) \citep{lomb76,scargle89,vanderplas18} analysis in order to uncover and quantify the relevant timescales and amplitudes of secular variability observed in the submillimeter. The LSP measures the likelihood that an observed light curve is well fit by a sinusoidal function with a set of parameters: mean brightness, amplitude, frequency, and phase. The statistical power measures the quality of the best sinusoidal fit as a function of frequency and builds the periodogram of the observed light curve. We perform the LSP analysis using the \textit{LombScargle} task in the \textit{timeseries} package of \textit{astropy} \citep{astropycollabo13} and analytically modify the output to fit our purpose. The details of the LSP output and our modifications are described below.

In the LSP analysis, the statistical power $P_{\rm A}$ of the ``A" hypothesis is defined as the fractional improvement over the non-varying (or static) ``N" hypothesis:
\begin{align}\label{eq:Power}
    P_{\rm A} &=\frac{\chi_{N}^{2}-\chi_{\rm A}^{2}}{\chi_{N}^{2}},
\end{align}
where $\chi^{2}_{X}$ denotes the chi-square value of the subscripted hypothesis. Thus, the value of $P_A$ rises as the goodness-of-fit of the sinusoidal function improves on the non-varying hypothesis; that is $\chi^2_\mathrm{A}$ reduces with respect to $\chi^2_{N}$.
Following standard practice, we consider the frequency at which $P_{\rm A}$ reaches its maximum value to denote the best fit. 

We next determine the False Alarm Probability (FAP) of this best fit to validate the detection. The FAP measures the probability that the sinusoidal fit is not real.
The FAP is calculated using the best-fit $\chi^{2}$ \citep{baluev08}.
That is, the FAP of the best-fit ``A" hypothesis, $FAP_{\rm A}$, is defined as
\begin{align}\label{eq:FAP_A}
    FAP_{\rm A} &=(1-P_{\rm A})^{N_{\rm f}/2}.
\end{align}
The FAP decreases with increasing degrees of freedom in the fit $N_{f}= (N_{epoch} - D_{\rm A})$,  where $N_f$ is determined from the number of epochs observed, $N_{\rm epoch}$, minus the number of parameters in the model, $D_{\rm A}$.  The $FAP_\mathrm{A}$ also reduces as the fractional improvement of the sinusoidal fit $P_\mathrm{A}$ increases.

Following the traditional LSP approach, the FAP of the best hypothesis must also take into account the total number of hypotheses tested. 
Thus, we define $FAP_{\rm LSP}$ as  
\begin{align}\label{eq:FAP_allh}
    FAP_{\rm LSP} &=1-(1-FAP_{\rm A})^{N_{\rm freq}},
\end{align}
where $N_{\rm freq}$ is the total number of frequencies tested. Furthermore, since we are only interested in cases where $FAP_{\rm A} \ll 1$ this can be significantly simplified to
\begin{align}\label{eq:FAP_allh2}
    FAP_{\rm LSP} & \cong  FAP_{\rm A} \times {N_{\rm freq}}.
\end{align}

Most protostellar sources are best fit by low frequency, long timescale, sinusoids.  Very short period sinusoids {\it always} result in poor fits to these protostellar light curves. 
The tested frequency range and step is calculated using the time baseline and the number of observed epochs. Using our 4 yr time baseline and 35 epochs, the tested frequencies are between 0.025 yr$^{-1}$ (once in 40 yr) and 22 yr$^{-1}$ (almost once per two weeks) with 0.05 yr$^{-1}$ steps.
The traditional $FAP_{\rm LSP}$ calculated over the full tested frequency range, which includes very high frequencies, will provide a systematic overestimate of the false alarm probability. We therefore modify the $FAP_{\rm LSP}$ to consider only longer periods than the best-fit period, with $FAP_{\rm Mod}$ given by
\begin{align}\label{eq:mod_sin}
    FAP_{\rm Mod} &\cong FAP_{\rm A} \times N_{\rm <freq},
\end{align}
where $N_{\rm <freq}$ is the number of tested periods that are longer than or equal to the best-fit period (lower than the best-fit frequency).
$FAP_{\rm Mod}$ becomes equivalent to $FAP_{\rm LSP}$ for sources with best-fit periods similar to the highest frequency tested.

Given that our LSP analysis uncovers primarily long period sinusoids, often with periods greater than the 4-year observing window, we also apply a linear least-squares fitting method to quantify the best-fit linear solution. The statistical power, $P_{\rm Lin}$, and the FAP of the linear best fit, $FAP_{\rm Lin}$, are calculated following Equations \ref{eq:Power} and \ref{eq:FAP_A}.

\begin{figure}[htp]
	\centering
	{\includegraphics[trim={0.6cm 0.0cm 1.3cm 0.9cm},clip,width=0.98\columnwidth]{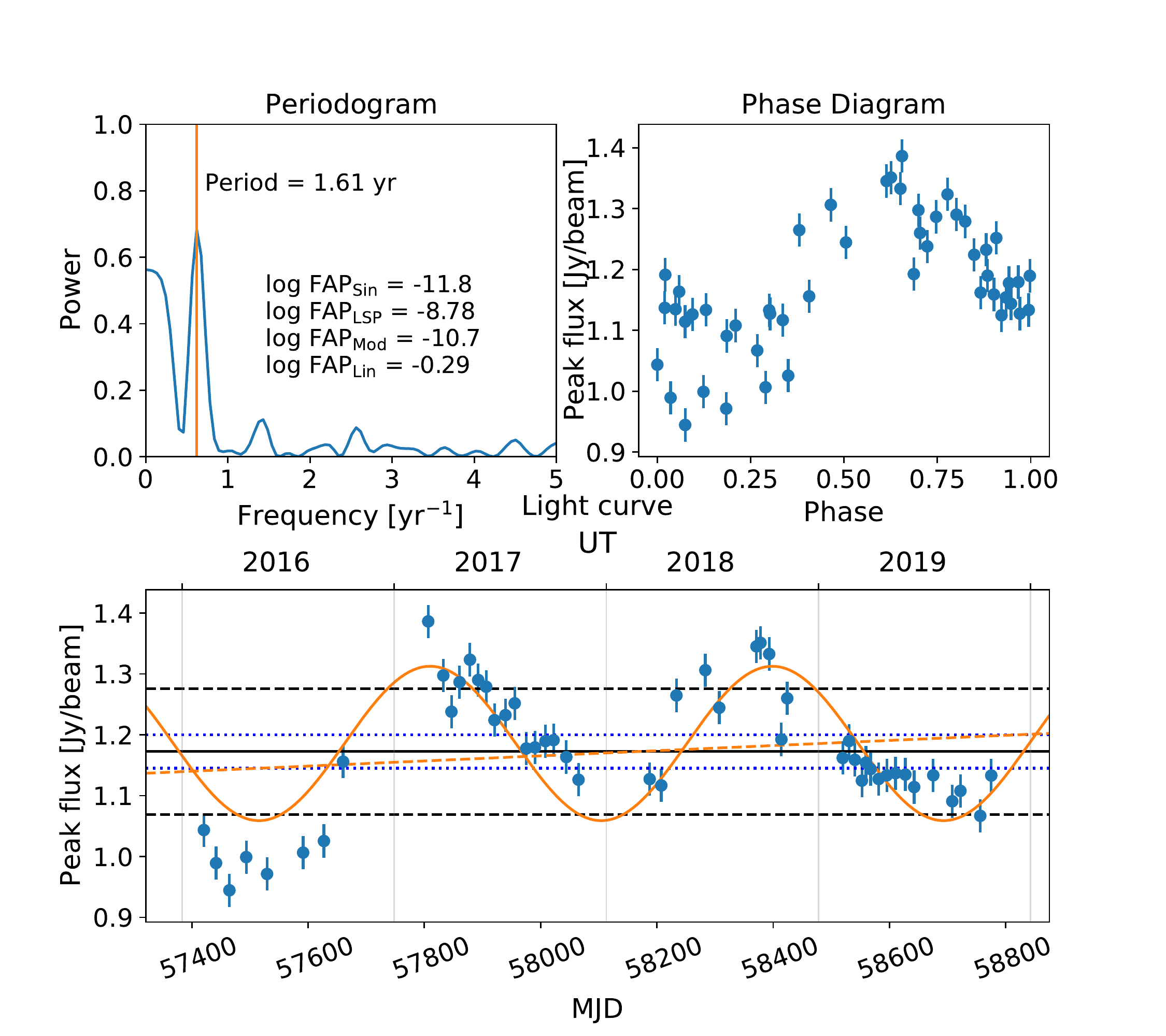}}
	\caption{ Periodogram, phase diagram, and light curve for known variable source EC 53 in Serpens Main (see also \citealt{lee20}). Upper left panel: Periodogram for EC 53. The solid orange vertical line indicates the frequency of the peak statistical power. The FAPs shown in the panel are calculated following Equations \ref{eq:FAP_A} - \ref{eq:mod_sin}. Upper right panel: Phase diagram for EC 53. The period of 1.61 yr found by the periodogram analysis is used for folding the light curve. Lower panel: light curve for EC 53. The black solid line indicates the mean peak brightness over the full time window, in \mbox{Jy beam$^{-1}$}. The black dashed lines and blue dotted lines indicate the observed standard deviation and fiducial uncertainty around the mean peak brightness, respectively. The solid orange curve denotes the best fit sinusoid recovered from the LSP analysis . The dashed orange straight line denotes the best fit linear function. MJD is JD--2400000.5.}
\label{fig:A_example}
\end{figure}

\begin{figure}[htp]
	\centering
	{\includegraphics[trim={0.0cm 1.8cm 0cm 2.7cm},clip,width=1.08\columnwidth]{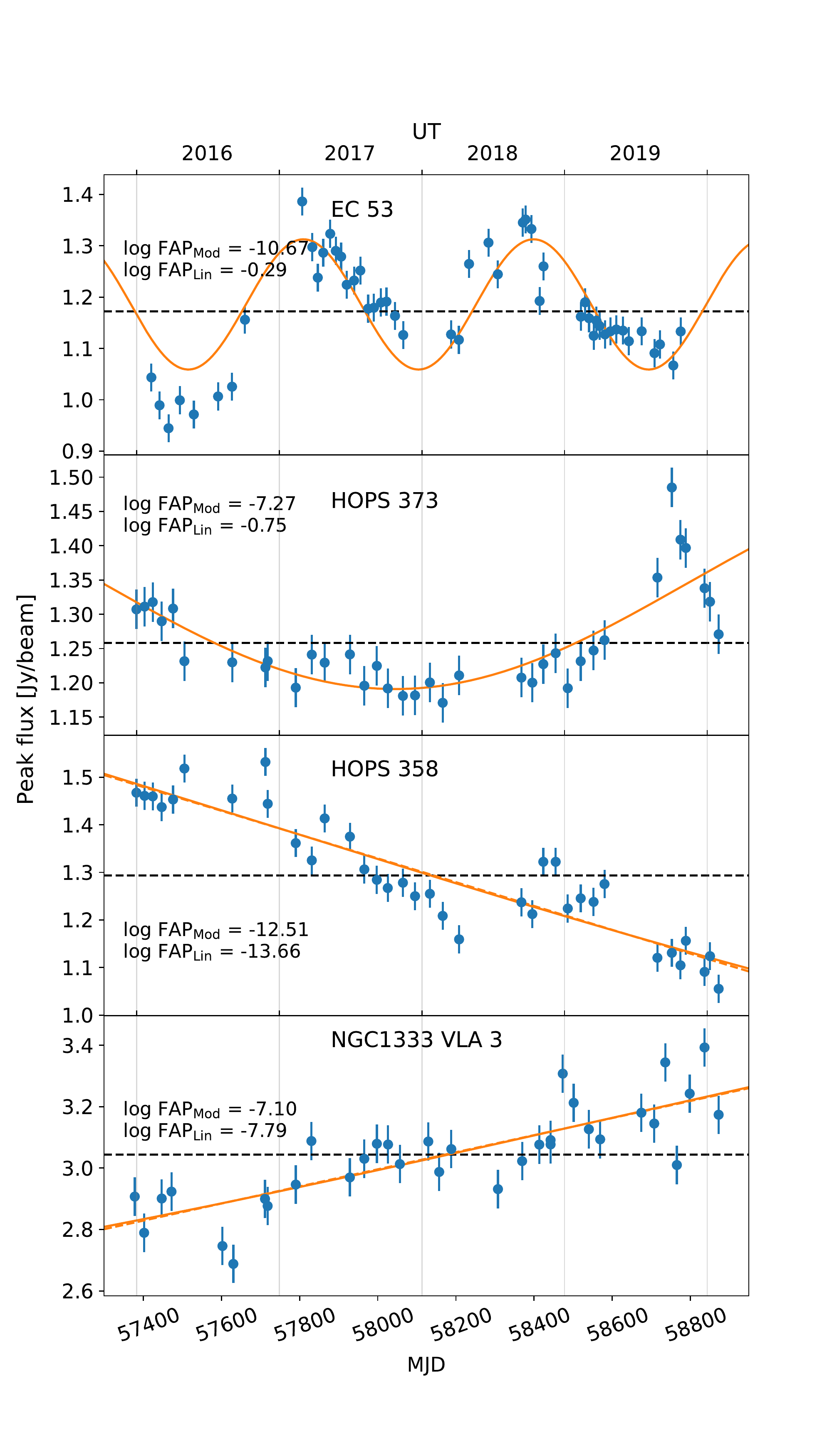}}
	\caption{ Representative light curves found in this study. The top two panels show Periodic and Curved light curves, while the lower two panels are increasing and decreasing Linear light curves. In all cases, the black horizontal line indicates the mean peak flux of the source over all epochs. The orange solid line shows the sinusoidal best-fit of the light curve, and in the bottom two panels the orange dashed line shows the linear best-fit. }
\label{fig:LC_examples}
\end{figure}

We demonstrate our dual secular analysis procedure in Figure \ref{fig:A_example}, where the bottom panel plots the 4-year light curve for a known variable (EC 53 in Serpens Main; see also \citealt{lee20}). We first perform the LSP analysis and reproduce the periodogram in the upper left panel. We next diagnose the various FAPs derived at the peak of the periodogram, denoted by the vertical orange line. For this source the periodogram peak is found at  a frequency of 0.0017 day$^{-1}$, equivalent to a 1.61 yr period, the single frequency false alarm probability is less than 10$^{-11}$, and both the LSP and modified FAPs are less than $10^{-8}$, denoting that the sinusoidal fitting is highly faithful. Finally, we also determine the best-fit linear form (dashed orange line in upper panel) and calculate a large linear FAP, greater than 0.5, denoting the linear fit has a 50\% likelihood of being a false alarm.

In Figure \ref{fig:LC_examples} we show examples of sources with light curves fit robustly by either periodic or linear secular functions. The top panel shows a periodic light curve with short period compared to our time coverage (4 yr). The second panel from the top shows a light curve robustly fit by a sinusoid with a period comparable to 4 yr, but not by a linear function. The bottom two panels show light curves well fit by both sinusoidal and linear functions, where the sinusoidal period is much longer than the time baseline.  We specify the classification of the found variables and analyze their properties in Section \ref{sec:V_rst}. 

\begin{figure}[htp]
	\centering
	{\includegraphics[trim={0.0cm 0.5cm 0.4cm 0cm},clip,width=0.98\columnwidth]{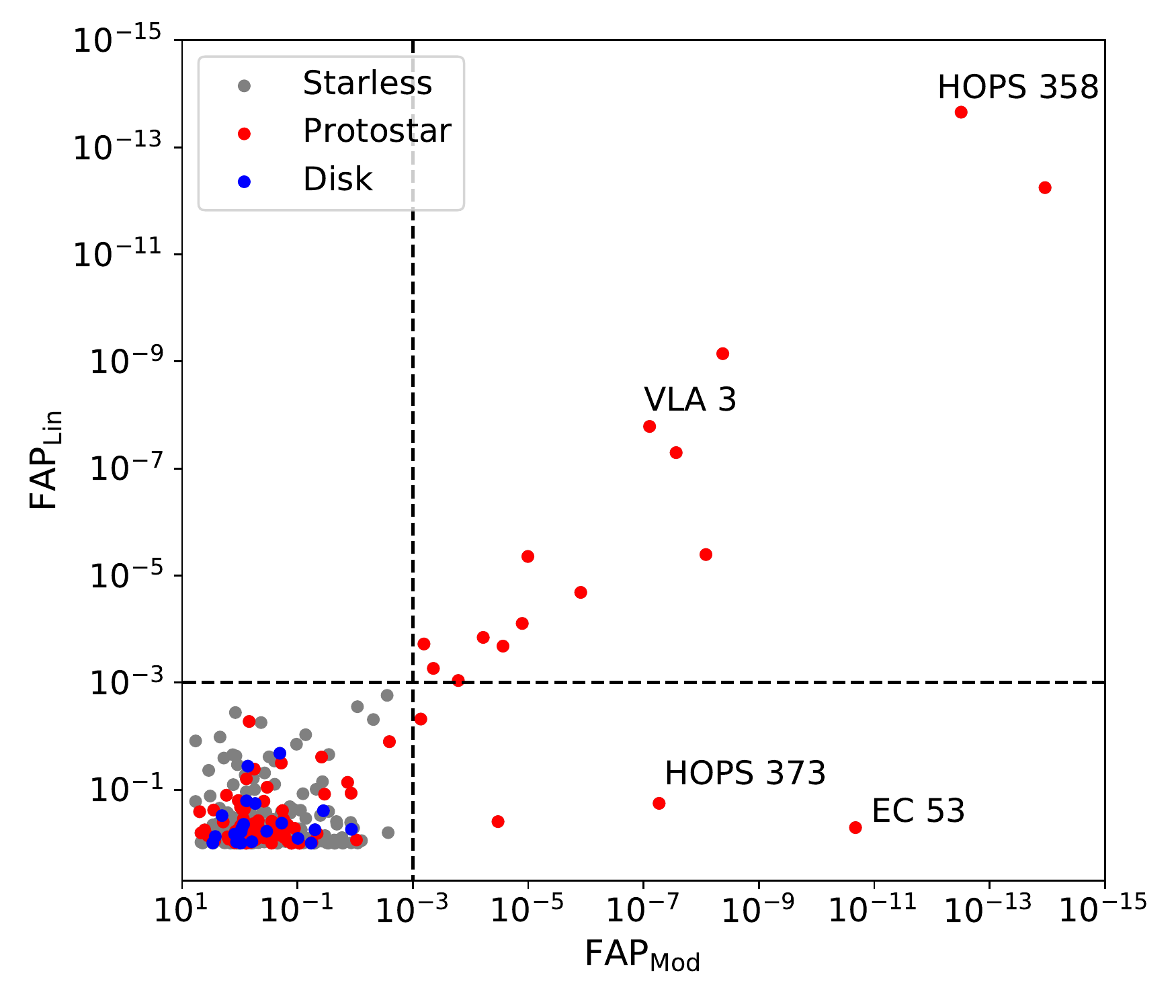}}
	\caption{Comparison of the sinusoidal, $FAP_{\rm Mod}$, and linear, $FAP_{\rm Lin}$, false alarm probabilities for all 295 submillimeter sources in our sample. The Black dashed lines correspond to FAPs of $0.1 \%$ along each axis. Sources are color coded by type as shown in the inset legend.}
\label{fig:FAPl_FAPs}
\end{figure}

To confirm the robustness of our method, in Figure \ref{fig:FAPl_FAPs} we compare the FAPs of the two fitting functions, sinusoidal and linear, for each of the 295 submillimeter sources in our sample. In the figure we color code the sources by type, protostar (red), disk (blue), and starless (grey).  The starless cores and disk sources occupy the lower left corner of the plot, where there is no robust evidence for secular variability (i.e.\ FAPs $> 0.1$\%).

Eighteen protostellar sources are found significantly outside the lower left corner of Figure \ref{fig:FAPl_FAPs}. Furthermore, all 14 protostellar sources which have $FAP_{\rm Lin} < 0.001$  also have $FAP_{\rm Mod} < 0.001$ and lie along a diagonal line in the figure. This is expected given that long period sinusoids are almost indistinguishable from linear functions over limited timescales.  In the figure, however, there are an additional four protostellar sources with $FAP_{\rm Mod} < 0.001$ for which a linear fit is poor due to the observed periodicity or curvature of the light curve. While EC 53 (see Figure \ref{fig:A_example}) shows clear periodic variability, the other three sources, HOPS 315, HOPS 373, and SMM 10, reveal clear curved trends with little overall slope and have best-fit periods of $< 8$ yr; see the light curves of HOPS 315 (Figure B.9), HOPS 373 (Figure \ref{fig:LC_examples}), and SMM 10 (Figure B.5). We discuss further all of the revealed secular variables in Section \ref{sec:V_rst}.

 

It is important to note that the LSP determination of a robust sinusoid fit does not necessarily imply that the underlying source variability is truly periodic, especially for the majority of our uncovered sources (16 out of 18) requiring long periods, of order or greater than the 4-year observing time window.  Nevertheless, the best fit parameters provide a useful, quantifiable estimate of the underlying timescale and amplitude for observed light curve variations.
We stress, however, that for periods longer than 4 years these are extrapolated estimates of the period and amplitude of the observed light curve event and not an indication of underlying periodicity.
Additionally, for the longest robust sinusoidal fits, the periods the amplitudes are extremely uncertain and the linear slopes are more appropriate quantifiable measures of the variability. Thus, in general we will use sinusoidal fits only for sources with derived periods less than $\sim$15 years.

\subsection{Stochastic Variability}\label{sec:V_STCH}

Along with the long term secular variability, sources may vary irregularly, with timescales of a few months or shorter. Two types of stochasticity are anticipated: (1) long-term chaotic brightness variations resulting in an increase in the epoch to epoch brightness variability without an appreciable linear or periodic trend, perhaps due to short timescale instabilities near the disk/star interface, (2) individual epoch rare brightening or dimming events that robustly stand out above the noise. We explore both of these stochastic variability possibilities across all observed epochs for each of the 295 submillimeter sources in our sample.

We first consider the rare, individual epoch events. The expected uncertainty of each peak brightness measurement of a given submillimeter source at 850 $\micron$ is defined as the fiducial uncertainty, $\sigma_{\rm fid}$, and calculated as 
\begin{align}\label{eq:fid}
    \sigma_{\rm fid} &= \sqrt{0.014^{2} + (0.02\times\Bar{F})^{2}}\ {\rm Jy\,beam}^{-1},
\end{align}
where $\Bar{F}$ is the mean peak brightness of the source over all observations. The other two terms in Equation \ref{eq:fid} are the relative flux calibration error for a given map epoch, $2\%$, and the typical 850 $\micron$ RMS noise, 0.014 \mbox{Jy beam$^{-1}$}. These values have been empirically measured for the JCMT Transient Survey by \citet{mairs17a} and the fiducial uncertainty is described in detail by \citet{johnstone18}.
To mitigate the effect of outlier measurements on the mean peak brightness, we determine the mean peak brightness after excluding the brightest and faintest 10\% of measurements from each source, $\bar{F}_{\rm 80\%}$, and we use this value in place of $\Bar{F}$ in Equation \ref{eq:fid}. Furthermore, for the 18 secular variables uncovered in Section \ref{sec:V_Sec}, we subtract the best fit sinusoidal function from their light curves before determining $\bar{F}_{\rm 80\%}$.


\begin{figure}[htp]
	\centering
	{\includegraphics[trim={0.0cm 0.4cm 0.0cm 0.2cm},clip,width=1.03\columnwidth]{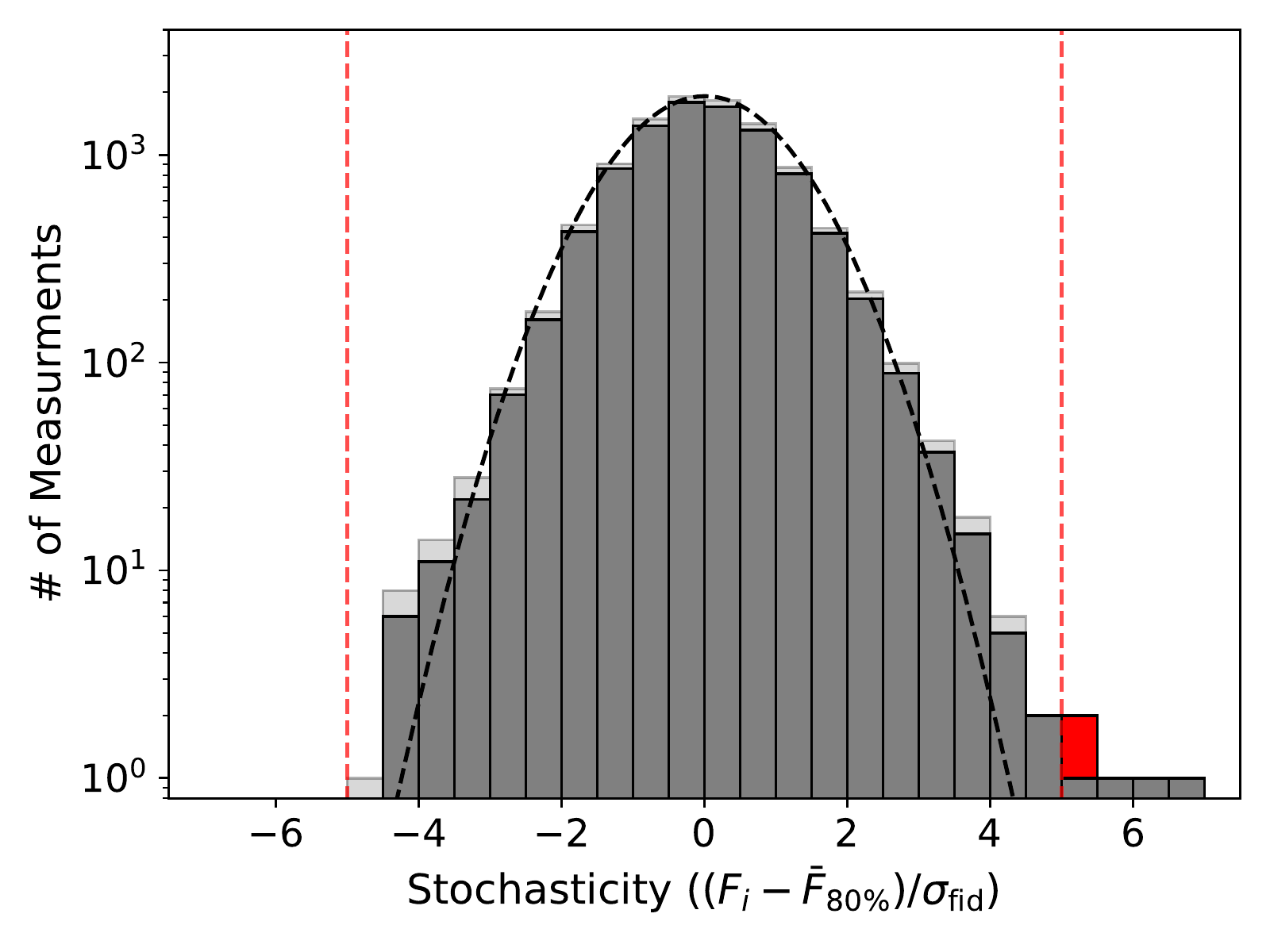}}
	\caption{Histogram of the stochasticity, $(F_{i} - \Bar{F}_{\rm 80\%}) / \sigma_{\rm fid}$, of each measurement of all 295 submillimeter sources in our sample (light), and after excluding the 18 secular variables (dark). Where necessary, the stochasticity is obtained after subtracting the best fit sinusoidal function. The black dashed line overlays the Gaussian function obtained by the standard deviation ($\sigma = 1.10$) and mean (0.01) of the stochasticity. 
	The red lines show where the stochasticity equals to $\pm 5$. The red histogram denotes the outlier detected in HOPS 373.}
\label{fig:Hist_sig}
\end{figure}

We first analyse the light curves of all 295 submillimeter sources in our sample for individual outlier events. For each measurement of each source, including the brightest and faintest 10\%, we determine the residual peak brightness after subtracting $\bar{F}_{\rm 80\%}$. We then bring these residual measurements to a standard scale by dividing by the expected measurement uncertainty $\sigma_{\rm fid}$. Figure \ref{fig:Hist_sig} presents a histogram over all these $\sigma_{\rm fid}$ scaled residual fluxes (light) and after excluding the 18 secular variables (dark).

The shape of the distribution follows well the expected normal distribution with an standard deviation fit close to unity, $\sigma = 1.10$, indicating that our uncertainty estimates are appropriate. The residuals with absolute values greater than 3, 
however, 
are somewhat overpopulated, perhaps indicating slight non-Gaussian shape in the noise characteristics. 
In addition, there are six highly stochastic events with absolute residuals greater than 5, 
indicated by the red vertical lines in the figure. These six candidate stochastic events (Table \ref{tab:stchev}) are measured across six different sources, five of which are known protostars,

Two of the six candidate stochastic events are associated with faint submillimeter sources, $\bar{F}_{\rm 80\%} \sim 0.3$ \mbox{Jy beam$^{-1}$}, for which the fiducial uncertainty is dominated by measurement error (Equation \ref{eq:fid}). This sample includes the only non-protostellar potential variable source in our sample, and also the only source to apparently dim for a single epoch. Careful examination of the residual images, after subtracting the co-add over all epochs from the epoch in which the event occurred, reveals that for both these candidates the residual emission is extended significantly beyond that of a point-source. These results strongly suggest that for these two epochs the image reconstruction procedure has introduced spurious large-scale features, not unexpected in submillimeter map reconstructions \citep[for details see,][]{mairs15, mairs17a}. We therefore drop these two candidates as potential stochastic events.

Three of our candidate stochastic events are bright, $\bar{F}_{\rm 80\%} \sim 2$ \mbox{Jy beam$^{-1}$}, but reside nearby, $r < 30\arcsec$, extremely much brighter sources, 
$\bar{F}_{\rm 80\%} > 5$ \mbox{Jy beam$^{-1}$}. For each of these candidates, the residual images for the epochs of interest show clear structure in the residual beam pattern of all the brighter sources within the map, indicating slight focus issues during the observation. The focus issue is seen to clearly produce excess emission within $r \sim 30\arcsec$, around the brighter sources and thus the measurement uncertainties for the three candidate stochastic events are significantly underestimated at these particular epochs. We therefore drop these three candidates as potential stochastic events.

\begin{deluxetable*}{cccccclr}
\tablecaption{Candidate Stochastic Events \label{tab:stchev}}
\tablehead{
\colhead{Region} &
\colhead{ID} &
\colhead{Known Name}& 
\colhead{$\mathrm{\Bar{F}}_{\rm 80\%}$($\mathrm{\Bar{F}}$)\tablenotemark{a}} &
\colhead{$\sigma_{\rm max}$} & 
F$_{\rm max}$/$\mathrm{\bar{F}_{\rm 80\%}}$ &
\colhead{Date} &
\colhead{Robust?}\\
& & & & [Jy beam$^{-1}$] & [$\sigma_{\rm fid}$] &  & (light curve)
}
\startdata
NGC 2068 & J054631.0-000232 & 
HOPS 373 & 1.25 (1.26)   & 5.34\tablenotemark{b} & 1.12 & 2019 Sep 26 & Yes (Figure \ref{fig:LC_examples})\\
OMC2/3 &  J053522.4-050111 & 
HOPS 88 & 2.50 (2.50) & 6.80 & 1.14  & 2016 Feb 29 & No/Focus\\
Serpens south &  J183002.6-020248 & 
CARMA 3 & 2.19 (2.20)  & 6.43 & 1.13 & 2019 Nov 2 & No/Focus\\
OMC2/3 &  J053527.4-050929 & 
HOPS 370 & 2.62 (2.64)  & 5.79 & 1.11 & 2017 Apr 21& No/Focus\\
NGC 2068 & J054645.6+000719 & 
LBS 11 & 0.32 (0.32) & 5.48 & 1.26 & 2016 Apr 27& No/Extended\\
OMC2/3 & J053524.2-050932\tablenotemark{c}&
starless& 0.34 (0.34)  & -5.01 &  &  2018 Nov 2 &No/Extended\\
\enddata
\tablenotetext{a}{The values in the bracket denote the mean peak brightness using the whole data points (see Section \ref{sec:V_STCH}.)}
\tablenotetext{b}{Value obtained after subtracting the best-fit sinusoidal function.}

\end{deluxetable*}


Finally, one of our protostellar sources, HOPS 373 (see Figure \ref{fig:LC_examples} and Table \ref{tab:stchev}), is detected in both the secular and stochastic variability analyses. We therefore note that the stochastic event detected in HOPS 373 appears to be part of a longer timescale burst-like event. 
Comparison of the residual map for the epoch in which the $> 5\,\sigma$ stochastic event occurred, reveals a strong point-like significant peak at the location of HOPS 373, assuring the robustness of the detection. Given that this brightening event occurs over more than one epoch,\footnote{We note that on detecting the rise, the cadence for the NGC\,2068 field, in which HOPS\,373 is located, was increased to bi-weekly. Thus, the time resolution for the decay of the burst is twice the nominal resolution of our survey.} we categorize this source as a secular variable in the rest of this work. 


 \citet {mairs19} performed a separate single-epoch transient source analysis for those areas within our monitored star-forming regions where there are no bright submillimeter sources. That investigation uncovered a prominent stochastic submillimeter variable, coincident with JW 566  in OMC2/3. JW 566 is a T Tauri star and shows a brightening by $500\,$mJy in a single epoch and a $\sim$50$\%$ decline in its peak brightness during the half hour observation. The mean brightness of JW 566 over all epochs lies below the threshold used for this paper, however, and therefore JW 566 is not identified by, or included in, our analysis.

\begin{figure*}[htp]
	\centering
	{\includegraphics[trim={0.0cm 1.0cm 0.0cm 1.5cm},clip,width=180mm]{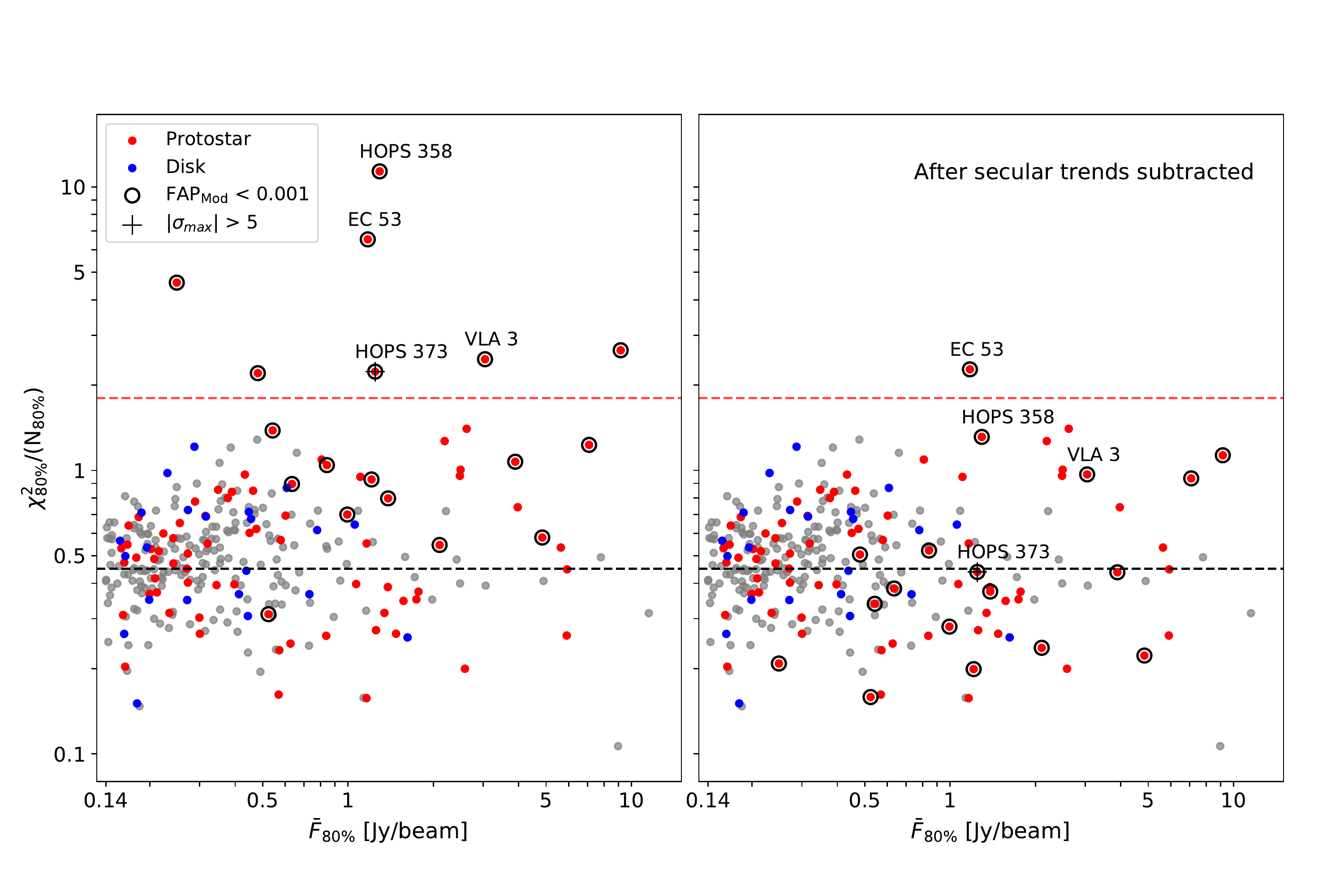}}
	\caption{The distribution of $\chi^{2}_{\rm 80\%}$ divided by the number of epochs (N$_{\rm 80\%}$) with respect to the mean peak brightness. Here we used only the middle 80 $\%$ data points, excluding each 10$\%$ of the brightest and faintest fluxes from the light curves. Circles denote sources with robust secular variability. The cross denotes the source (HOPS 373) with individual epoch stochastic variability. Right panel is the same as the left panel, except that the $\chi^{2}_{\rm 80\%}$ is calculated after subtracting secular trends from the light curves of robust secular variables. The black and red dashed line indicates $\chi^{2}_{\rm 80\%}/N_{\rm 80\%} =$ 0.45 and 1.8, which corresponds to $\chi^{2}/N =$ 1 and 4, respectively assuming a Gaussian distribution. }
\label{fig:scatter_log}
\end{figure*}

Having found no robust rare, individual epoch events within our sample of bright sources, we next consider the measured peak brightness uncertainty over the full light curve, normalized by the expected fiducial uncertainty, as a determination of the long-term stochastic nature of each of our 295 submillimeter sources. Here we are searching for sources with a significantly larger spread in brightness measurements compared with the known noise properties of the measurements. Figure \ref{fig:scatter_log} plots 
this dispersion against mean peak brightness, with color again indicating source type. The y-axis plots $\chi^{2}_{80\%}$, the chi-square obtained by using only the flux measurements that went into calculating $\Bar{F}_{80\%}$. Thus, any rare significant outlier events are not included in this analysis. The eighteen known secular variables are marked with black circles, and HOPS 373 with the individual stochastic event is
marked with cross. The left panel presents the calculation before removing the best fit sinusoid from the secular variables whereas the right panel shows the result after subtracting the secular trend. 

Two results are immediately evident in Figure \ref{fig:scatter_log}. First, all sources in the left panel with anomalously large variations in the measured peak brightness over their full light curves, $\chi^2_{80\%}/N_{80\%} > 1.8$ (corresponding to an expected $\chi^{2}/N \sim 4$ when measured over a full Gaussian distribution), are secular variables for which the large deviations are significantly removed by subtracting the secular fit (right panel). 
Thus, given the present survey sensitivity there are {\it no} sources with on-going random brightness variations significantly larger than the measurement uncertainty.
Second, the typical peak brightness of the robustly determined variables is shifted to the right (brighter), compared with the full sample even when only considering protostars (red dots). This is an expected occurrence due to the increased signal to noise for bright sources and its carryover effect on the completeness of our analyses. We return to this discussion in Section \ref{sec:C}.


\subsection{Properties of the Variables}\label{sec:V_rst}

In this section we consider the quantifiable properties of the robustly recovered variables in our sample in order to determine an approximate classification scheme. Using a false alarm threshold criterion of 0.1\% and a stochastic threshold of $\pm 5\sigma$ we recover only 18 secular variables 
among the 295 bright submillimeter sources in our sample. 

\begin{figure}[htp]
	\centering
	{\includegraphics[trim={0.0cm 0.5cm 0.4cm 0.0cm},clip,width=0.98\columnwidth]{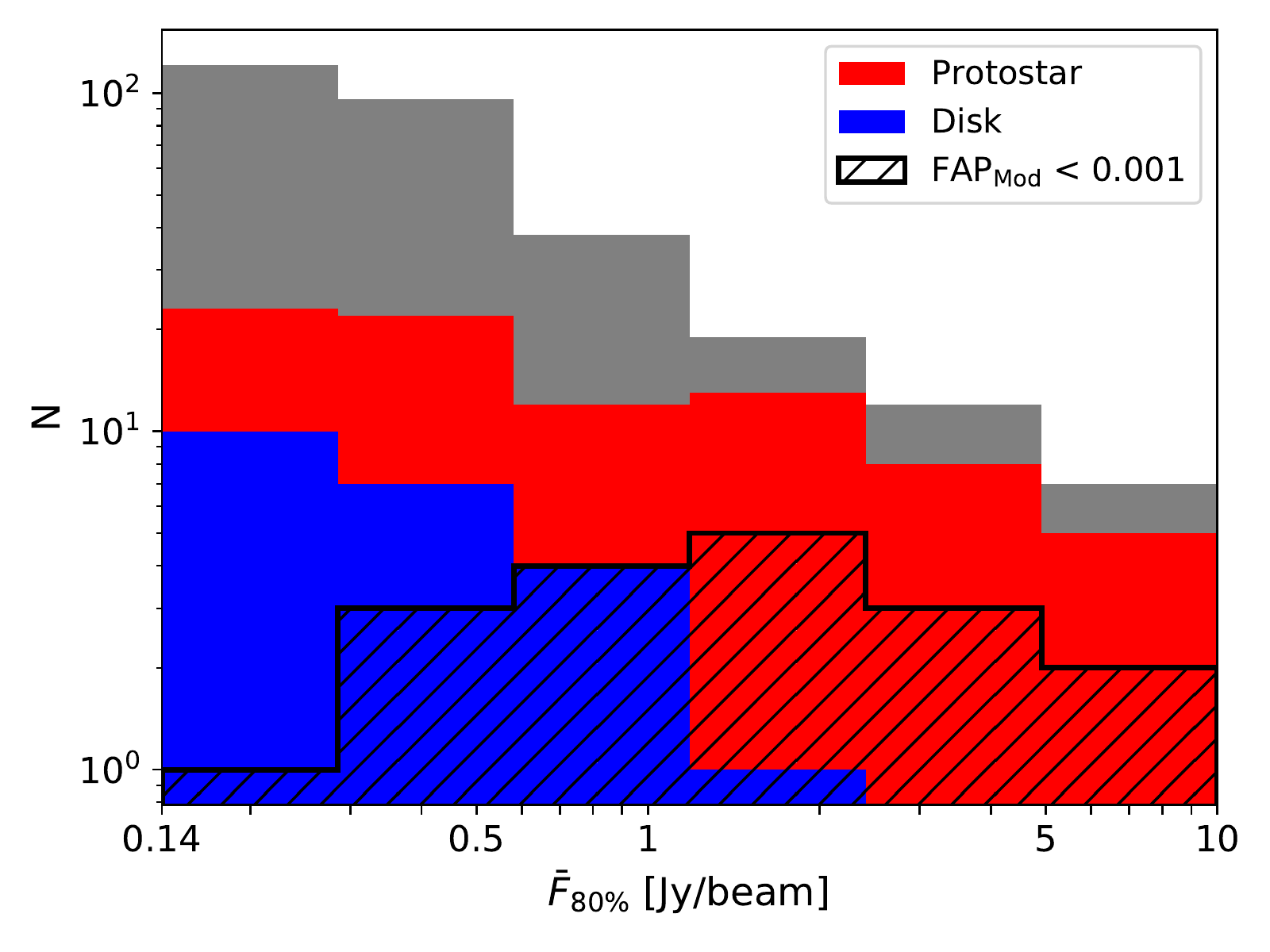}}
	\caption{Histograms of $\bar{F}_{\rm 80\%}$ for all sources with peak brightnesses $\ge$ 0.14 \mbox{Jy beam$^{-1}$} (Grey). The red histogram shows the number of protostars and the blue histogram shows the number of disks. The black hatched region shows the histogram for the secular variables. The disk sources are distributed at lower mean peak brightness than the protostars. Furthermore, the secular variables generally have higher $\bar{F}_{\rm 80\%}$ compared to the whole sample (see Section \ref{sec:C} for a discussion of completeness). }
\label{fig:hist_Fmean80}
\end{figure}

All the robust variables within our survey are known protostars. None of the known disk sources show any variability in the submillimeter using our LSP, linear, and stochasticity analyses. 
{This absence is at least in part due to the low peak brightness of the disk sources in submillimeter (see Figure \ref{fig:hist_Fmean80}), and the brightness-dependent completeness discussed in Section \ref{sec:C}. We further discuss the variability depending on evolutionary stage in Section \ref{sec:Evol}.}
Limiting our sample only to protostars, we find that $\sim 22\%$ (18 out of 83) are observed to be secular variables and 
none show evidence of a single stochastic outlier event.

\begin{deluxetable*}{ccccccccc}
\tablecaption{Physical Properties of Robust Secular Variables \label{tab:secularvar1}}
\small
\tablehead{
 \colhead{Region} & \colhead{ID} & \colhead{Figure}\tablenotemark{*} & \colhead{Known Name} & \colhead{$\mathrm{\Bar{F}}$}  & \colhead{$\Delta$F/$\sigma_{\rm fid}$} & \colhead{$\Delta$F/$\mathrm{\Bar{F}}$} & L$_{\rm bol}$ & T$_{\rm bol}$ \\
& & & & [Jy bm$^{-1}$]  & & [$\%$] & [L$_{\odot}$] & [K]
}
\startdata
IC 348 & J034356.5+320050 & B.1 & IC348 MMS 1 & 1.39 & 4.57 & 10.2 & 1.6 & 23\tablenotemark{a} \\
Serpens main & J182951.2+011638 & B.2 & EC 53 & 1.17  & 16.2 & 37.7 & 5.9 & 161\tablenotemark{a} \\
NGC 2068 & J054647.4+000028 & B.3 & HOPS 389 & 0.99  & 5.09 & 12.4 & 6.0 & 43\tablenotemark{b} \\
NGC 1333 & J032910.4+311331 & B.4 & IRAS 4A & 9.16 & 7.84 & 15.7 & 8.3 & 31\tablenotemark{a} \\
Serpens main  & J182952.0+011550 & B.5 & Serpens SMM 10 & 0.84 & 6.21 & 16.1 & 8.3 & 70\tablenotemark{a} \\
Ophiuchus & J162626.8-242431 & B.6 & VLA 1623-243 & 3.89  & 5.42 & 11.0 & 0.5& 45\tablenotemark{c}\\
NGC 2068 & J054613.2-000602 & B.7 & V1647 Ori & 0.25  & 7.93 & 47.3 & 27 & 322\tablenotemark{b} \\
IC 348 & J034357.0+320305 & B.8 & HH 211 & 1.21  & 6.59 & 15.2 & 1.4 & 23\tablenotemark{a} \\
NGC 2068 & J054603.6-001447 & B.9 & HOPS 315 & 0.52  & 3.06 & 10.2 & 6.2 & 180\tablenotemark{b} \\
NGC 2068 & J054631.0-000232 & B.10 & HOPS 373 &1.26  & 10.9 & 25.0 & 5.3 & 37\tablenotemark{b} \\
OMC2/3  & J053529.8-045944 & B.11 & HOPS 383 & 0.54  & 5.56 & 18.1 & 7.8 & 46\tablenotemark{b} \\
NGC 1333 & J032903.8+311449 & B.12 & West 40 & 0.48  & 7.86 & 27.8 & 0.7 & 18\tablenotemark{a} \\
Serpens south  & J182937.8-015103 & B.13 & IRAS 18270-0153 & 0.63  & 5.79 & 17.2 & 6.9 & 57\tablenotemark{a} \\
Serpens main  & J182949.8+011520 & B.14 & Serpens SMM 1 & 7.08 & 9.35 & 18.8 & 70 & 13\tablenotemark{a} \\
Serpens main  & J182948.2+011644 & B.15 & SH 2-68 N& 2.10  & 3.20 & 6.76 & 14 & 31\tablenotemark{a} \\
Serpens south  & J183004.0-020306 & B.16 & CARMA 7 & 4.85  & 5.31 & 10.7 & 15 & 35\tablenotemark{d}  \\ 
NGC 2068 & J054607.2-001332 & B.17 & HOPS 358 & 1.29  & 16.2 & 37 & 25 & 42\tablenotemark{b} \\
NGC 1333 & J032903.4+311558 & B.18 & NGC1333 VLA 3 & 3.04  & 11.3 & 23 & 38 & 220\tablenotemark{a} \\
\enddata
\tablenotetext{*}{The reference number of the light curve in Appendix B.}
\tablenotetext{a}{For these sources the L$_{\rm bol}$, T$_{\rm bol}$ values are taken from \citealt{Mowat:2018}. See also Section \ref{sec:Region} for more details.}
\tablenotetext{b}{For these sources the L$_{\rm bol}$, T$_{\rm bol}$ values are taken from \citealt{furlan16}.}
\tablenotetext{c}{For this source the L$_{\rm bol}$, T$_{\rm bol}$ values are taken from \citealt{mur18}}
\tablenotetext{d}{For this source the L$_{\rm bol}$, T$_{\rm bol}$ values are taken from \citealt{maury11}.}
\end{deluxetable*}

\begin{deluxetable*}{cccccccccccc}
\tablecaption{Light Curve Properties of Robust Secular Variables \label{tab:secularvar2}}
\small
\tablehead{
 \colhead{Known Name} & \colhead{Period\tablenotemark{a} } & \colhead{$A$/$\sigma_{\rm fid}$} & \colhead{A/$\mathrm{\Bar{F}}$}& \colhead{Slope/$\mathrm{\Bar{F}}$} & \colhead{Grp\tablenotemark{b}} & \colhead{log FAP$_{\rm Mod}$} & \colhead{log FAP$_{\rm Lin}$}\\
 & [yr] & & [$\%$] & [$\%$ yr$^{-1}$] & 
}
\startdata
 MMS 1 & 1.05 & 1.78 & 2.70 & -1.3 & P & -3.36 & -3.27\\
 EC 53  & 1.61 & 8.77 & 10.8 & & P & -10.7 & -0.29\\
 HOPS 389  & 4.53 & 2.69 & 3.17 & -1.5 & C & -4.90 & -5.39\\
 IRAS 4A  & 4.54 & 0.46 & 4.1 & -2.0 & C & -3.79 & -3.04\\
 SMM 10  & 5.30 & 4.50 & 4.61 & & C & -4.48 & -0.40\\
 VLA 1623-243  & 5.72 & 0.668 & 2.57 & -1.6 & C & -3.19 & -3.72\\
 V1647 Ori  & 5.83 & 30.0 & 18.1 & -10.4 & C & -14.0 & -12.2\\
 HH 211  & 5.84 & 2.86 & 3.81 &  2.15 & C & -7.56 & -7.30\\
 HOPS 315 & 8.16 & 3.79 & 3.26 &  & C & -3.14 & -2.32\\
 HOPS 373 & 8.16 & 8.94 & 12.2 &  & C & -7.27 & -0.75\\
 HOPS 383 & 8.16 & 9.56 & 8.11 &  -2.80 & C & -8.08 & -5.39\\
 West 40 & 8.18 & 11.0 & 8.55 & 4.84 & C & -8.37 & -9.15\\
 IRAS 18270-0153 & 12.5 & 12.1 & 11.1 & -2.0 & C & -4.56 & -3.68\\
 SMM 1 & 37.1 & &  & 2.07 & L & -5.91 & -4.69\\
 SH 2-68 N & 37.1 &  & & -0.96 & L & -4.22 & -3.85\\
 CARMA 7 & 37.5 &  &  & 1.29 & L & -4.99 & -5.36\\ 
 HOPS 358 & 40.8 &  &  & -7.06 & L & -12.5 & -13.7\\
 VLA 3 & 40.9 &  &  & 3.34 & L & -7.10 & -7.79 \\
\enddata
\tablenotetext{a}{The best-fit period from the LSP method.}
\tablenotetext{b}{P: Periodic Group. C: Curved Group. L: Linear Group.}
\end{deluxetable*}

Tables \ref{tab:secularvar1} and \ref{tab:secularvar2} present the quantitative results for individual secular variables. The observed amplitude ($\Delta$F) is defined as F$_{\rm max}$-F$_{\rm min}$ where F$_{\rm max}$ and F$_{\rm min}$ are the brightest and faintest peak brightness among the measured fluxes of a source. The LSP-derived periods of the detected variables vary from 1 year to 40 years. We categorize the variables with their best-fit periods compared to the 4-year observing window of our survey: Periodic ($< 4$ yrs: Group P), Curved ($4$ -- $15$ yrs: Group C), and Linear ($>15$ yrs: Group L). 

We find two protostars in the Periodic Group:  the known periodic infrared and submillimeter variable source EC 53 in Serpens Main \citep{hodapp12, yoo17, lee20} with $\sim$1.5 yr period, and the protostellar source MMS1 in Perseus \mbox{IC 348}, with $\sim$ 1 yr period. MMS1 is an intriguing source because it is also detected through our linear analysis (see the light curve presented in Figure B.1). 
The majority of the recovered secular variables, $61\%$, belong to the Curved Group, for which the observing window is comparable to the derived source periodicity. 
As mentioned earlier, for these sources the quantified period is more likely a measure of variability timescale rather than a determination of a sinusoidal nature. These sources require dedicated long-term, many year, monitoring to further unravel their episodic nature.
Finally, the Linear Group contains five protostars, $28\%$ of the secular variables, that show predominantly a linear trend in their light curves but with the possibility of slight curvature. We discuss each of these individual sources in the Appendix.

\begin{figure}[htp]
	\centering
	{\includegraphics[trim={0.0cm 0.0cm 0.0cm 0.0cm},clip,width=1.03\columnwidth]{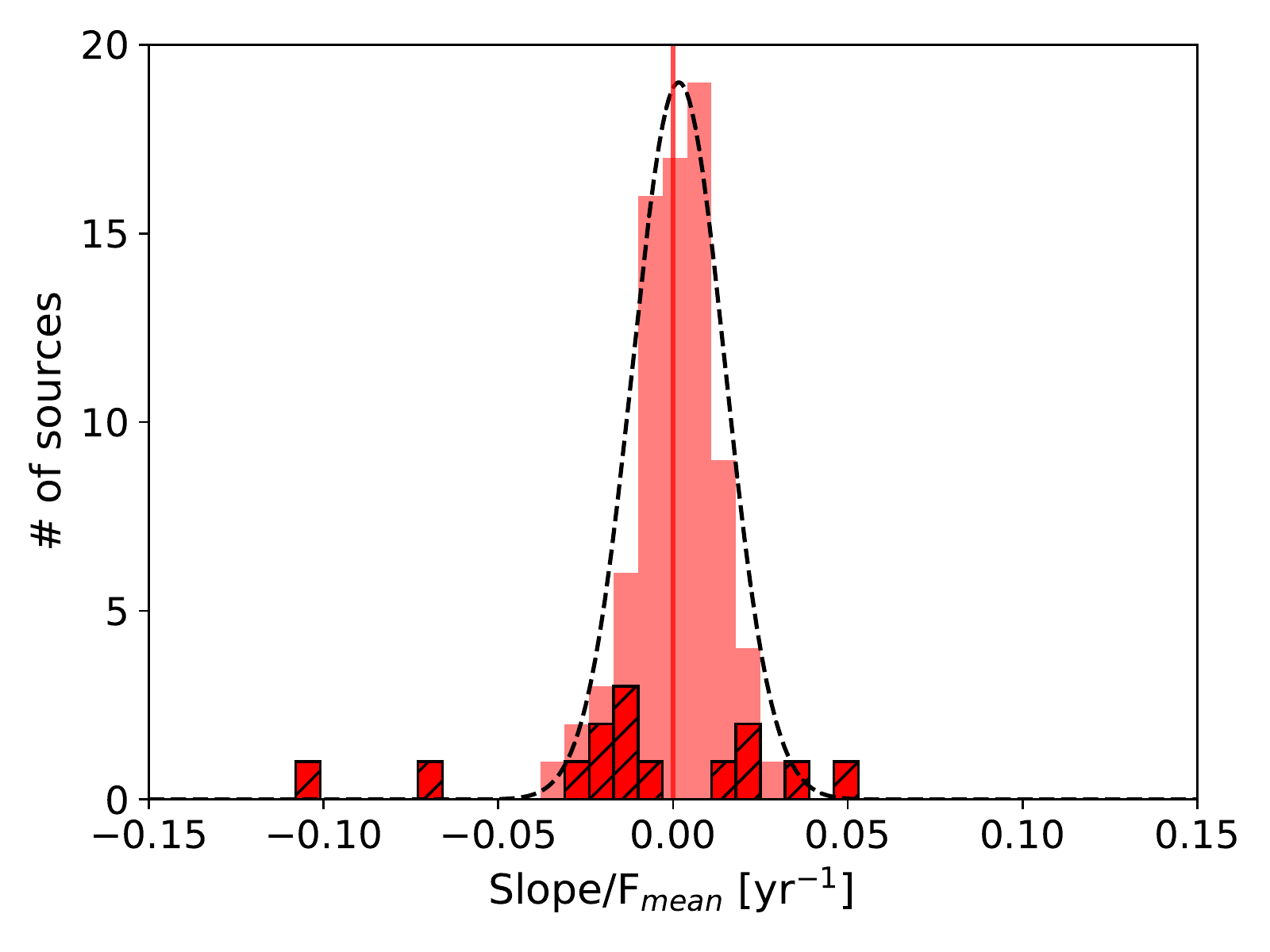}}
	\caption{Histogram of the fractional slopes determined by the linear fitting. The light red histogram shows the slopes of every protostar. The red hatched histogram shows only the robust linear slopes. 	
	The red vertical line marks the origin while the overlaid black dashed Gaussian fits the
	mean (0.0016 yr$^{-1}$) and standard deviation (0.013 yr$^{-1}$) of all the measured slopes.} 
\label{fig:Hist_slope}
\end{figure}

Alternatively to studying periodicity timescales, we can also consider the distribution of the linear fits to the light curves to determine if sources are in a brightening or dimming phase. 
The distribution of slopes, in units of fractional change per year, is obtained from the least-square linear fitting to the light curves of all protostars and is shown in Figure \ref{fig:Hist_slope} (light red) overlaid with the 14 robust linear fits (red hatched). For the protostars with robust fits, 9 are declining in brightness while only 5 are rising.  Furthermore, the typical strength of the decline or brightening is similar, although there are two significant outliers with large negative slopes. The lack of symmetry between rising and dimming submillimeter sources is intriguing but suffers from small number statistics at present.

Across the 4 years of observations analysed here, the distribution of the slopes over all investigated protostars (light red) is strongly peaked near the origin, and has a mean value of 0.0016, essentially zero,  and $\sigma_{\rm obs}(4\,$yrs$) = 0.013$, equivalent to a rising or falling slope of 1.3\% per year. Individually these slopes are not considered robust precisely because the measured slope is smaller than, or similar to, the uncertainty in the slope determination. The ensemble, however, contains useful information on the allowable range of these individually uncertain slopes. Previously, on the basis of the first 18 months of observations, \citet{johnstone18} fit a Gaussian distribution to a similarly derived set of  light curve slopes and measured $\sigma_{\rm obs}(1.5\,$yrs$) = 0.023$, of which at least half of the spread was calculated to be due to the uncertainty in fitting the slopes, an uncertainty that decreases with longer observing timelines. Recognizing that measurement uncertainty contributed to their observed spread in slopes, \citet{johnstone18} determined that the underlying intrinsic distribution of light curve slopes is  $\sigma_{\rm int} \lesssim 0.01$ and predicted that after three years of observations the observed width of the distribution should shrink to be at most $\sigma_{\rm obs}(3\,$yrs$) \sim 0.01$, with the upper limit occurring if $\sigma_{\rm int} = 0.01$. Here, when considering all the protostars in our sample, we obtain that upper limit. We therefore predict that submillimeter monitoring surveys that can reach a relative brightness calibration significantly below 1\% will find that a majority of the protostellar sources are robust secular variables.

\citet{johnstone18} recovered five robust secular variables within the JCMT Transient Survey sample after the first 18 months of the survey. All of those sources continue to be identified as  robust secular variables by the present study. Furthermore, the lengthening of the observing window to 4 yrs has increased the number of recovered secular variables by greater than a factor of three.

Finally, as mentioned above, the robust secular variables are biased toward the brighter protostars. In Figure \ref{fig:hist_Fmean80} we present histograms of the peak brightness of our full source sample as well as subsets for protostars, disks, and secular variables.  We discuss the issue of  completeness within our secular variability analysis in the next section.

\section{Completeness of the Survey}\label{sec:C}




Understanding the completeness of our secular variability analysis as a function of source brightness (see Figure \ref{fig:hist_Fmean80}) is important as many trends that we wish to examine rely on determinations of sub-samples for which the underlying distribution of source brightness is intrinsically biased. For example, due to envelope-clearing older protostars are expected to be fainter in the submillimeter than the youngest, most deeply embedded sources.

\begin{deluxetable}{cccccc}
\tablecaption{Variability Detection by Source Brightness \label{tab:Fdep}}
\tablehead{
\colhead{Condition} & \colhead{S/N} & \colhead{N$_{\rm submm}$}  &\colhead{N$_{\rm protostar}$} & \colhead{N$_{\rm secular}$} & \colhead{P$_{\rm sec}$\tablenotemark{a}}\\
(Jy bm$^{-1}$)& & & & &
}
\startdata
$\ge$ 0.14 & 10 & 295 & 83 & 18 & 0.22 \\
$\ge$ 0.35 & 22 & 141 & 51 & 17 & 0.33 \\
$\ge$ 0.5 & 29 & 95 & 43 & 16 & 0.37 \\
$\ge$ 1.0 & 41 & 48 & 31 & 11 & 0.35 \\
$\ge$ 2.0 & 47 & 45 & 15 & 6 & 0.40
\enddata
\tablenotetext{a}{Fraction of secular variables(N$_{\rm secular}$/N$_{\rm protostar}$).}
\end{deluxetable}

In Appendix \ref{sec:A_Comp} we derive the minimum amplitude required to satisfy our FAP threshold via the LSP analysis.
For our observed short-period sinusoids (the Periodic Group), the secular variability is detectable as long as the amplitude $A$ satisfies
\begin{align}\label{eq:Amp1}
    A &> \left[ 2(\alpha-1) \right]^{1/2}\,\sigma_{\rm fid},
\end{align}
where $\alpha = (10^{-3}/N_{\rm freq})^{-2/N_{\rm f}}$. For our short period detections the number of frequencies requiring testing, $N_{\rm freq}$, is $\sim 10$ and  the degrees of freedom, $N_{\rm f}$, is $\sim 30$; therefore, $\alpha \sim 1.85$ and the detectable amplitude is $1.30\, \sigma_{fid}$. Furthermore, the detectable amplitude decreases slowly through increasing $N_{\rm f}$, i.e. observing additional epochs.

For longer period sinusoids (Curved and Linear Groups), the formulation is more  complicated as the observing window does not fully sample the underlying period.  We used a Monte Carlo analysis to generate 10,000 hypothetical light curves with randomized noise patterns using the observational windows from the JCMT Transient Survey. 
The detectable amplitude, scaled to the underlying measurement uncertainty, depends on both the sinusoidal period and the phase over which the sinusoid is observed.
Averaging over the phase dependency for a given period, we define a practical detectable amplitude, such that the probability of a robust detection is at either the 68\% or the 99\% level. 
We discuss the result of the Monte Carlo analysis in the Appendix \ref{sec:A_Comp}.



\begin{figure}[htp]
	\centering
	{\includegraphics[trim={0.2cm 0.3cm 0.0cm 0.3cm},clip,width=0.98\columnwidth]{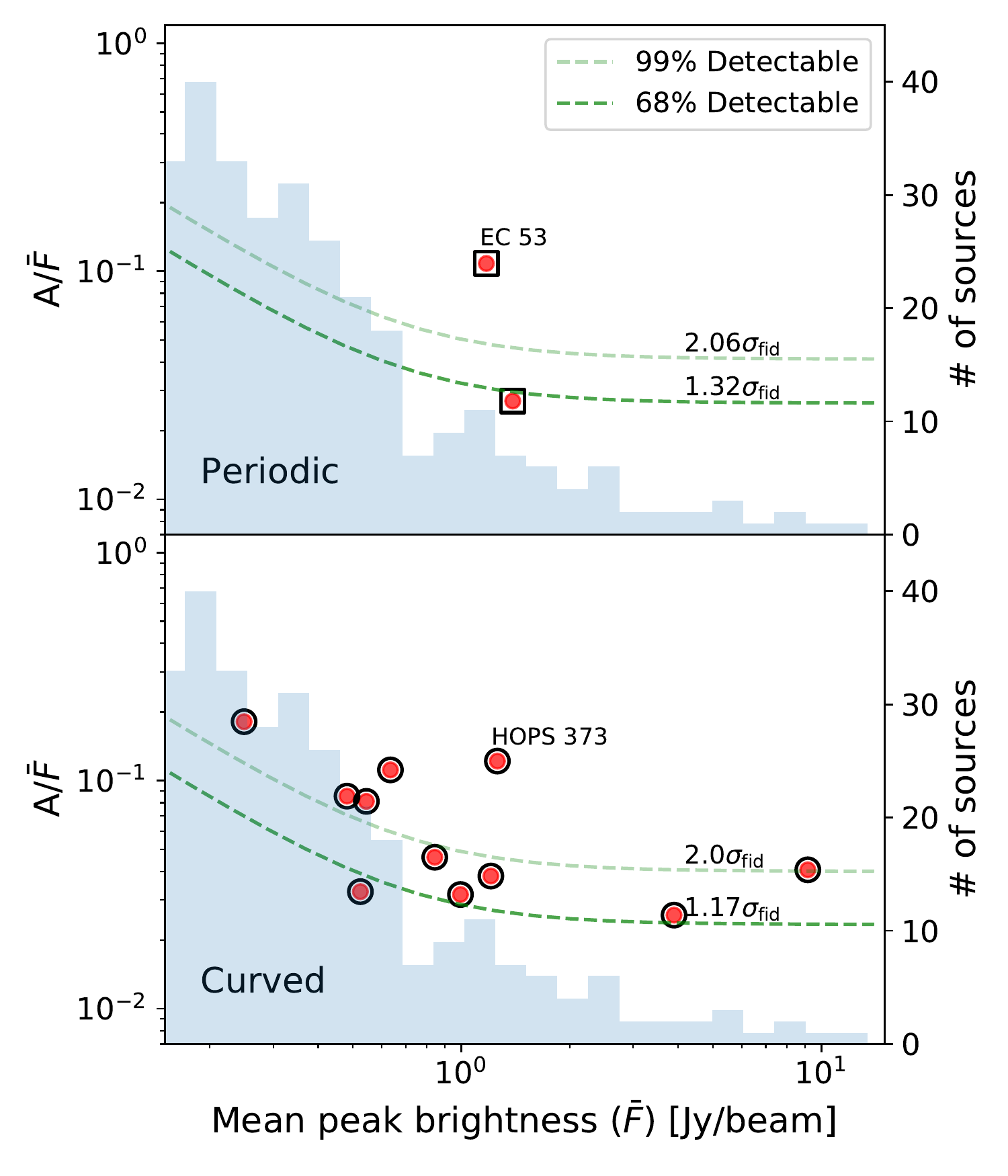}}
	\caption{Scatter plot of fractional amplitude vs. mean peak brightness (The Periodic Group: squares, and the Curved Group: circles). Background: Histogram of the mean peak brightness for the whole submillimeter sources. The green lines indicate the 99$\%$ (faint dashed), and 68$\%$ (dashed) detectable level. The corresponding values are noted above each line.} 
\label{fig:Fmean_fracamp}
\end{figure}
\begin{figure}[htp]
	\centering
	{\includegraphics[trim={0.2cm 0.2cm 0.0cm 0.2cm},clip,width=0.98\columnwidth]{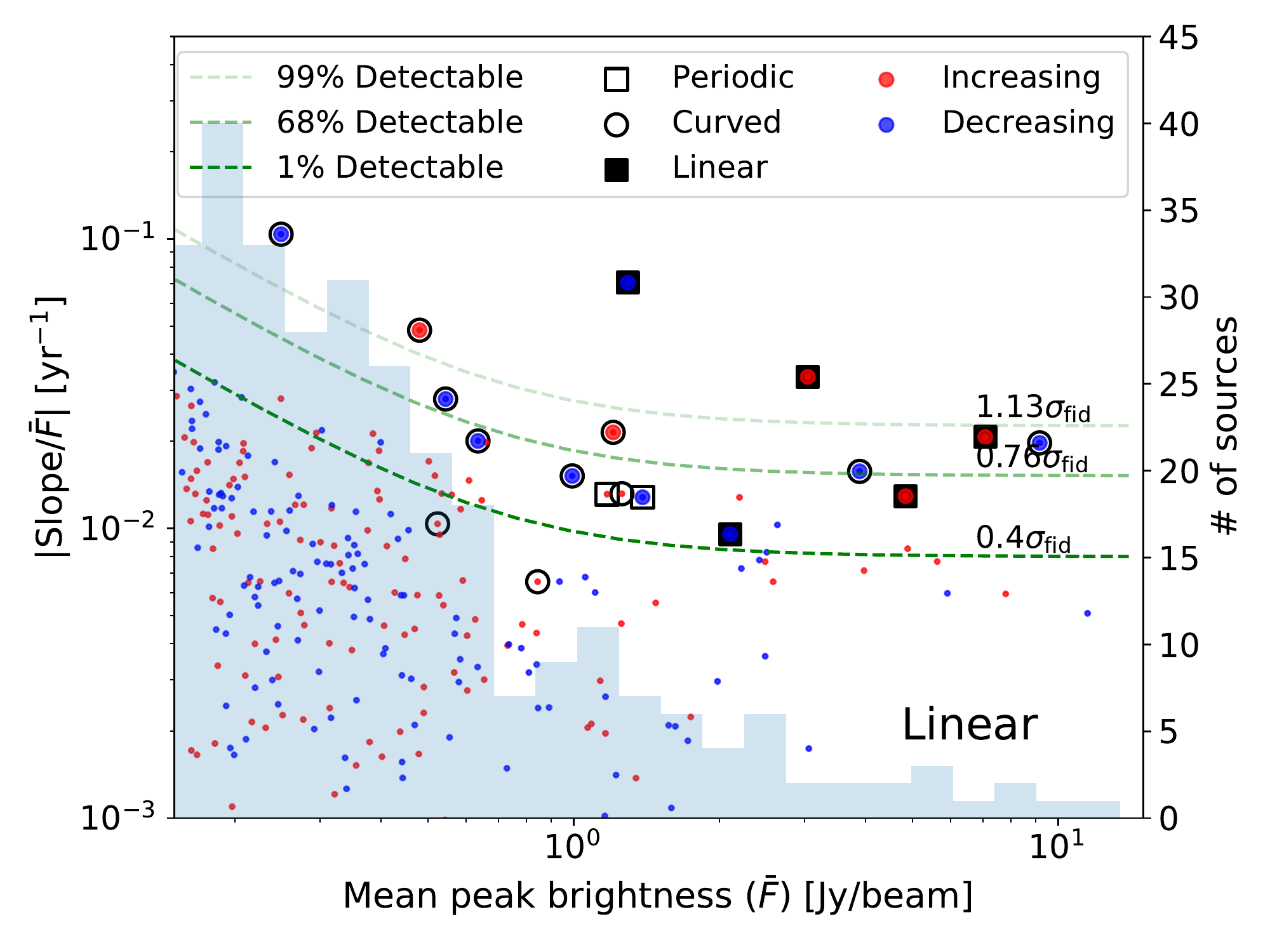}}
	\caption{Scatter plot of fractional slope vs. mean peak brightness. Background: Histogram of the mean peak brightness for the whole submillimeter sources. Small marks indicate the sources of which linear variability are not shown. Three green lines indicate the 99$\%$, 68$\%$, and 1$\%$ detectable level from top. The corresponding values are noted above each line. The red and blue markers indicate the sources with the positive, and negative best-fit slopes. Empty circles and squares denote four sources which are only found by sinusoidal fitting (see Figure \ref{fig:FAPl_FAPs}).}
\label{fig:Fmean_fracslope}
\end{figure}

Following our classification of the secular variables depending on their determined period (see Section \ref{sec:V_rst}), we present the practical detection limit determined by the above completeness analysis in Figures \ref{fig:Fmean_fracamp} and \ref{fig:Fmean_fracslope}. In each panel of Figure \ref{fig:Fmean_fracamp} we calculate the limiting fractional amplitudes, as a function of the mean peak brightness, that provide a 68\% and 99\% probability of detection. Figure \ref{fig:Fmean_fracslope} shows the practical detection limits of the linear fitting method.  Most of the non-variables (small marks) are below 1\% detection line.  These estimated detectable limits are the averages over all eight star-forming regions and thus some robustly detected sources lie below the limiting curves (see discussion below). 

As expected, the limiting detectable fractional amplitude varies with source brightness. For the brighter sources, where the measurement uncertainty is dominated by the flux calibration within the map and thus the measurement uncertainty is roughly proportional to the source brightness, the limiting detectable fractional amplitude becomes constant. Once the mean peak submillimeter brightness becomes less than $\sim$0.5 \mbox{Jy beam$^{-1}$}, the limiting fractional amplitude increases quickly. Given that the majority of secular variable detections are found within a factor of a few around the limiting threshold while the numbers of protostars increases toward the faint end (see Figure \ref{fig:hist_Fmean80}), it is clear that the completeness of the survey drops significantly below this brightness limit.
The completeness transition can be clearly seen in Table \ref{tab:Fdep}. The fraction of secular variables remains about 40\% when we limit the sample to bright sources. Once we include sources fainter than $\sim 0.5$ \mbox{Jy beam$^{-1}$}, however, the fraction of secular variables drops. 

Finally, we caution that the detectable limit varies with region due to the different number of epochs for each region (see Table \ref{tab:obssummary}). In general, this should introduce only minor differences between regions; however, we anticipate a somewhat higher completeness for Serpens Main, where the number of observed epochs is 50\% larger than the average over the other regions. 


\section{Discussion}\label{sec:Disc}

In Section \ref{sec:V} we found 18 secular variable protostars, out of 83 monitored by the JCMT Transient Survey, and quantified their variability timescales and amplitudes. Next, in Section \ref{sec:C} we established the completeness of our variable sample as a function of source brightness and variability amplitude, determining that for 43 protostars with peak brightness $> 0.5$ \mbox{Jy beam$^{-1}$} our sample has a uniform fractional amplitude detection threshold. In this section, we investigate the regional dependency of detected variability, and compare the submillimeter behavior to the known properties of the variables in the literature.

\subsection{Variability Across Star-Forming Regions}\label{sec:Region}

Although our survey reveals only a small number of variables, which limits the statistic significance of results from sub-samples, we consider possible regional variations in the numbers of secular variables uncovered. Table \ref{tab:numperregions} shows the ratio ($P_{\rm sec}$) between the numbers of secular variables and protostars in each region to examine any regional dependency. For the full survey, we find 22\% of protostars varying, while across the eight individual star-forming regions we find a wider range, 0 -- 50\%. To account for the completeness (Section \ref{sec:C}), we repeat our analysis using only those sources with mean peak brightness greater than 0.5 Jy beam$^{-1}$ and find that 37\% of protostars vary across the entire survey. 
 
We compare the regional variability ratios by calculating the probability ($p$) that each region is drawn from an underlying sample with the mean value. Each region is compared against the entire sample, excluding the region of interest.
When taking the samples to include all protostars, there is no clear evidence of regional dependence.   Serpens Main is somewhat overabundant in variables and has $p = 0.04$ ($ P < 0.05$ is typically used as a dissimilarity threshold), but has also been most frequently observed and should
have a slightly higher completeness threshold than the other regions. 
Considering only protostars brighter than 0.5 Jy beam$^{-1}$, where the completeness is more uniform, the \mbox{OMC 2/3} region becomes somewhat exceptional in harbouring fractionally few variables, 1 out of 12 protostars, with $p = 0.01$. There is no obvious correlation in these numbers with distance to each region. The full set of results is presented in Table \ref{tab:numperregions}.


\begin{deluxetable*}{ccccccccccc}
\tablecaption{Detected Variables by Region \label{tab:numperregions}}
\tablehead{
\colhead{Region} & 
\colhead{D [pc]} &
\colhead{N$_{\rm proto}$} & 
\colhead{N$_{\rm stch}$} & 
\colhead{N$_{\rm secular}$ } & 
\colhead{P$_{\rm sec}$} & 
\colhead{$p$\tablenotemark{a}} & 
\colhead{N$_{\rm proto}$\tablenotemark{*}} & 
\colhead{N$_{\rm secular}$\tablenotemark{*}} & 
\colhead{P$_{\rm sec}$\tablenotemark{*}}  &
\colhead{$p$\tablenotemark{*}}\\
& & & & & & & & & &
}
\startdata
NGC 1333 & 293\tablenotemark{b} & 16 & 0 & 3 & 0.19 & 0.75 & 7 & 2 & 0.29 & 0.62  \\
OMC2/3 & 388\tablenotemark{c} & 14 & 0 & 1 & 0.07 & 0.15 & 12 & 1 & 0.08 & 0.01 \\
NGC 2068 & 388\tablenotemark{c} &  15 & 0 & 5 & 0.36 & 0.16 & 8 & 4 & 0.50 & 0.42 \\
Serpens south & 436\tablenotemark{c} & 13 & 1 & 2 & 0.15 & 0.55 & 3 & 2 & 0.67  & 0.28\\
Ophiuchus cores & 137\tablenotemark{c} & 9 & 0  & 1 & 0.11 & 0.42 & 1 & 1 & 1.00 & 0.37\\
Serpens main & 436\tablenotemark{c} & 8 & 0  & 4 & 0.50 & 0.04 & 7 & 4 & 0.57 & 0.24 \\
IC 348 & 321\tablenotemark{c} & 5 &0  & 2 & 0.40 &  0.40 & 3 & 2 & 0.67 & 0.28 \\
NGC 2024 & 423\tablenotemark{c} & 3 & 0  & 0  & 0.00 & 0.36 & 2 & 0 &  0.00 & 0.28 \\ \hline
Whole survey & & 83 & 1 & 18 & 0.22 & & 43 & 16 & 0.37 &\\
\enddata
\tablenotetext{*}{Results when we include only those protostars brighter than 0.5 \mbox{Jy beam$^{-1}$}.}
\tablenotetext{a}{Region dependent $p$-values, obtained from Student's $T$ Test using the \textit{ttest\_ind} task in the \textit{stats} package of \textit{scipy}, comparing the fraction of protostellar secular variables in each region against the same fraction over all other regions combined. The test results are null except for Serpens Main, when using all protostars, and OMC\,2/3, when limiting to bright protostars ($> 0.5$ \mbox{Jy beam$^{-1}$}).}
\tablenotetext{b}{Parallaxes from the Gaia DR2 catalog \citep{ortizleon18}}
\tablenotetext{c}{Parallaxes from the VLBA GOBELINS program \citep{kounkel17,ortizleon17a,ortizleon17b,ortizleon18}}
\end{deluxetable*}



\subsection{Variability Across Evolutionary Stage}\label{sec:Evol}

Protostars are expected to be more variable at earlier evolutionary stages. 
For this analysis, we use the bolometric temperature, corrected for extinction, as a proxy for evolutionary stage.
For the three Orion fields, these values were directly obtained from the literature \citep{furlan16}. 
For Serpens, Perseus and Ophiuchus, bolometric temperature and luminosity values were updated by \citet{Mowat:2018} based on the SED-fitting methodology in \citet{dunham15} using SCUBA-2 fluxes at 450\micron\ and 850\micron\ from the JCMT Gould Belt survey  \citep{Pattle:2015,Chen:2016} to improve the submillimeter coverage.\footnote{In addition to the SCUBA-2 and Spitzer fluxes, data from many additional surveys also included by \citet{dunham15} were used to determine the SEDs:  J, H and K-band photometry from the Two Micron All Sky Survey \citep[2MASS,][]{Skrutskie:2006,Cutri:2012yCat}; 12\micron\ and 22\micron\ from the Wide-field Infrared Survey Explorer \citep[WISE, ][]{WISE,Cutri:2012yCat}; millimetre fluxes from 1100\micron\ observations with Bolocam \citep{bolocam,Enoch:07}; and for some sources, 350\micron\ observations with SHARC-{\sc II} \citep{dowell03,JWu:07,Suresh:2015}.
Extinction corrections were applied to the literature fluxes for the calculations of $T_\mathrm{bol}$. Following \citet{c2d} and \citet{dunham15}, the \citet{WeingartnerDraine:2001} $R_\mathrm{V} = 5.5$ extinction law was used to de-redden the fluxes for each candidate protostar. Extinction was corrected for all YSOs with the same values of visual extinction $A_\mathrm{V}$ used in the Spitzer catalogue. The trapezoidal rule was used to calculate $L_\mathrm{bol}$ following \citet{Dunham:2013}.}

\begin{figure*}[htp]
	\centering
	{\includegraphics[trim={0.0cm 0.0cm 0.0cm 0.0cm},clip,width=180mm]{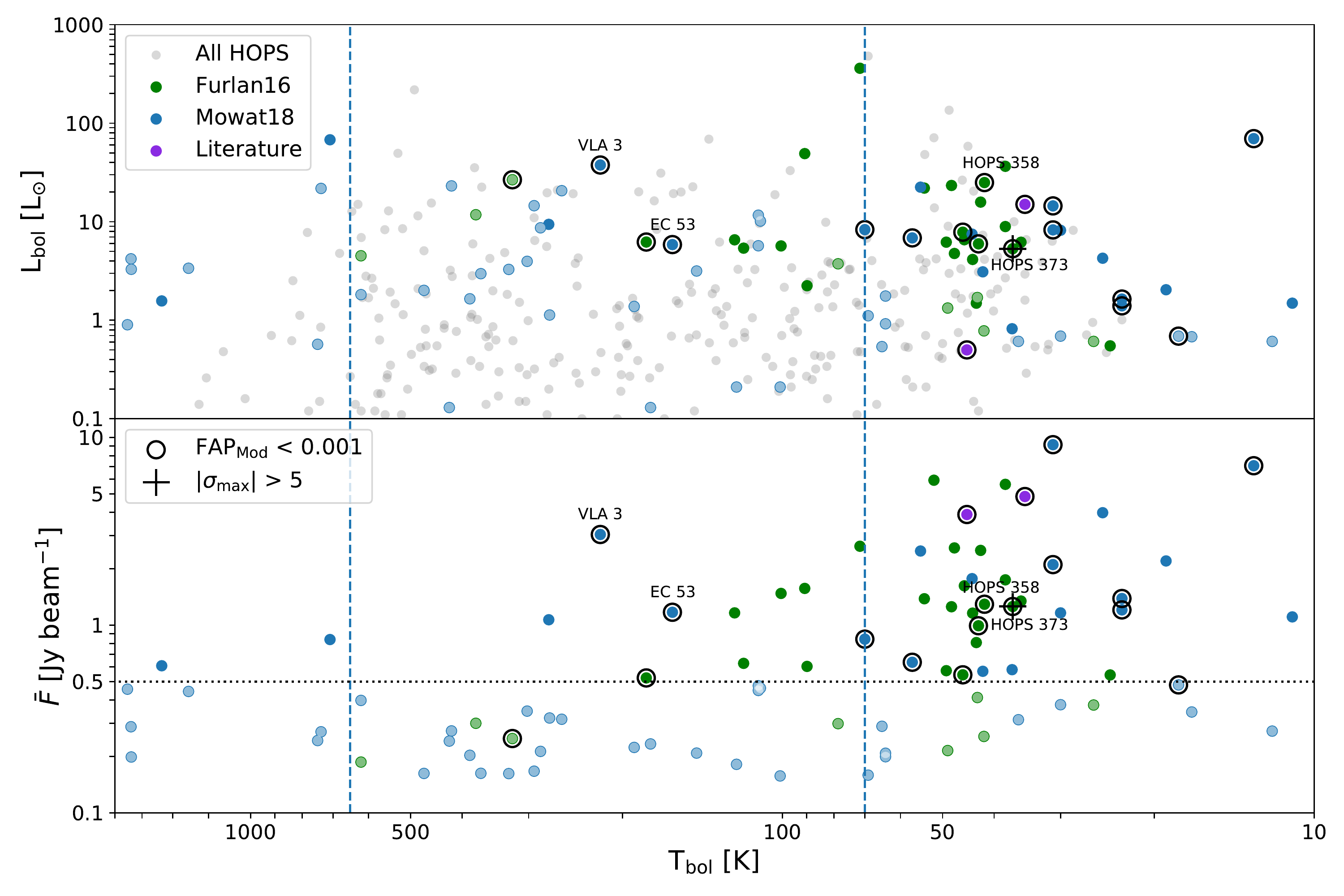}}
	\caption{ Bolometric luminosity (L$_{\rm bol}$, top panel) and mean peak brightness at 850 $\micron$ ($\Bar{F}$, bottom panel) on bolometric temperature (T$_{\rm bol}$) space. Green and blue markers denote JCMT Transient Survey sources with physical parameters obtained from the HOPS catalogue \citep{furlan16} and {\it Spitzer} Space telescope Gould Belt Survey catalogues \citep{dunham15,Mowat:2018}, respectively. Purple circles denote two secular variables for which the physical properties were found via a literature search, VLA 1623-243 \citep{mur18} and CARMA 7 \citep{maury11}. Gray markers denote all the sources from HOPS \citep{furlan16}. Faint markers denote the sources that are fainter than 0.5 \mbox{Jy beam$^{-1}$} (dotted line) at 850 $\micron$. Black circles denote the secular variables, and the black cross denotes the stochastic variable. }
\label{fig:TbolLbol}
\end{figure*}

In Figure \ref{fig:TbolLbol} we plot both the extinction corrected bolometric luminosities (upper panel) and the peak submillimeter brightnesses (lower panel) of our sample against the extinction corrected bolometric temperature. We also include the larger Orion HOPS sample \citep{furlan16} in the luminosity panel.

Considering just the luminosity versus temperature (upper panel), the protostellar variables are more luminous and cooler than the underlying HOPS sample, but not significantly more luminous than the typical protostar in the JCMT Transient Survey sample. However, in the bottom panel it is clear that peak submillimeter brightness is a strong function of evolutionary class, as expected.  Care must be taken before inferring changes in the variability fraction with evolutionary stage due to changing completeness limits as a function of source brightness. 

Following the results of Section \ref{sec:C}, we consider only those sources for which the mean brightness is greater than 0.5 \mbox{Jy beam$^{-1}$}, for which the completeness to secular variability is reasonably flat. We thus find 0 (out of 2) Class\,II secular variables, 3 (out of 10, 30\%) Class\,I secular variables, and 13 (out of 34, 38\%) Class\,0 secular variables. 
Although the variability ratio is higher among Class 0 than Class I, the difference between the ratios of Class 0 and Class I is insignificant ($p = 0.61$).

\subsection{Variability of Known Eruptive YSOs}\label{sec:FUors}

We conducted a literature search for known eruptive variables within the eight JCMT Transient Survey fields. Six such stars were identified across four regions, of which three are coincident with submillimeter peaks. All three of the coincident submillimeter protostars, \mbox{EC 53} in \mbox{Serpens Main} \citep[also known as \mbox{V371 Serpentis,}][]{hodapp12}, \mbox{V1647 Ori} in \mbox{NGC 2068} \citep{aspin09}, and \mbox{HOPS 383} in \mbox{OMC 2/3} \citep{safron15} are observed to be varying (see Table \ref{tab:secularvar2}). Based on bolometric temperature (see Section \ref{sec:Evol}), the first two are classified as a Class I  while \mbox{HOPS 383} is a Class 0.  The three known eruptive stars for which there is no coincident submillimeter peak are \mbox{OO Serpentis}\footnote{OO Serpentis is classified as a young Class I source \citep{kospal07} and resides at a location of enhanced submillimeter emission in our Serpens Main map; however, the peak does not significantly stand out from the larger scale structure.} and \mbox{V370 Serpentis}, both in Serpens Main, as well as \mbox{IRAS 18270-0153W}\footnote{\citet{johnstone18} incorrectly identified the linearly varying submillimeter peak associated with IRAS 18270-0153 as being associated with the FU Ori candidate IRAS 18270-0153W \citep{connelley10}; however, the submillimeter peak is offset by $>15\arcsec$ from the FU Ori source location. In this paper, we refer to the submillimeter source, which is still seen to secularly vary (see Table \ref{tab:secularvar2}), without the ``W" designation and consider it a separate entity.} in Serpens South. 


The two optical/IR eruptive (FUOr/EXOr) stars that are found to vary in our submillimeter sample, V1647\,Ori and EC\,53, are both observed to have episodic accretion outbursts. \mbox{EC 53} has a short term, $\sim 1.5$ yr, periodicity, observed at near-IR and submillimeter wavelengths \citep[see,][]{lee20}.  \mbox{V1647 Ori} has undergone multiple eruptions, with times between bursts in the 2 to 5 year range \citep{aspin06, acosta-pulido07, ninan13}. Most recently, 2018, V1647\,Ori was observed in a quiescent phase \citep{giannini18}. The decrease during 2018 is observed in our survey (see Figure B.7).
The submillimeter light curve for this source is fit with a $\sim 6$ yr period, placing it in the Curved Group. The light curve itself suggests that it may be reaching a minimum (Figure B.7).

These two known optical/IR eruptive Class I sources are among the most extreme submillimeter variables in our sample in terms of fractional flux change across the observing window. Only \mbox{HOPS 383}, our mid-infrared identified eruptive Class\,0 variable, along with the three Class\,0 sources \mbox{HOPS 358}, \mbox{HOPS 373} and \mbox{West 40} reveal a similarly large brightness range (see also Section \ref{sec:V_mass}). 
Furthermore, similar to \mbox{HOPS 383}, the three strongly varying Class 0 sources have extrapolated timescales (periods) greater than $8\,$yrs. Interestingly, \mbox{HOPS 383} has a somewhat lower variability amplitude across the last 4 years, compared with the other three Class\,0 variables. This may be due to the fact that its submillimeter brightness decay has recently slowed, after the well-known outburst event between 2004 and 2010 \citep{safron15}.

Additional details on each of these protostellar variables can be found in the Appendix.


\subsection{Variability of Known Subluminous YSOs}\label{sec:VeLLOs}
For embedded protostars, their luminosity is dominated by accretion luminosity. Therefore, the low-luminosity sources, which are classified as YSOs with $\leq$ 1 L$_{\odot}$ \citep{dunham08}, should have low mass accretion rates. The most extreme low luminosity sources, called Very Low-Luminosity Objects (VeLLOs), were discovered by the {\it Spitzer} survey \citep{youngCH04}. 
By definition, VeLLOs are embedded protostars with the internal luminosity $\leq$ 0.1 L$_{\odot}$, and are thus considered as YSOs in the most quiescent phase of the episodic accretion process. We investigate the variability of low-luminosity sources and VeLLOs revealed in submillimeter.

First, we cross-matched those low-luminosity sources from \citet{dunham08} with the source list in this study. We note that the \mbox{SMM 1} was classified as low-luminosity source by \citet{dunham08} using 2MASS and {\it Spitzer} dataset (1.25-70 $\micron$).\footnote{SMM 1 shows a bolometric luminosity higher than 1 L$_{\odot}$ with the peak of SED at > 100 $\micron$ \citep{dunham15}. We therefore exclude \mbox{SMM 1} from our low-luminosity source list in the following discussion.}

There are 28 low-luminosity sources located within our coverage, with 7 LLSs brighter than $>0.14$ \mbox{Jy beam$^{-1}$} at 850 $\micron$: 2 in \mbox{IC 348}, 4 in \mbox{NGC 1333}, and 1 in \mbox{Ophiuchus}. The two low-luminosity sources in \mbox{IC 348}, \mbox{MMS 1} and \mbox{HH 211}, are secular variables (Figure B.1 \& B.8).
Thus, 2 out of 7 low-luminosity sources ($\sim$30\%) show secular variability in the submillimeter. Comparing this variability fraction against the entire sample of monitored protostars (see Table \ref{tab:numperregions}), we obtain a p-value of 0.69. For the sources F$_{80\%}$ > 0.5 \mbox{Jy beam$^{-1}$}, the number of variables from the sample becomes 2 out of 4 (50\%) of which p-value is 0.59. Thus, we don't detect any clear differentiation in the secular variability fraction of low-luminosity sources, with any conclusions limited by the small sample size.

As an alternate sample, we compare our source list to the VeLLO list from \citet{kim16}. \citet{kim16} added four complementary criteria (a ratio of the 1.65 $\micron$ flux to the 70 $\micron$ flux less than 2.8, 4.5 $\micron$ magnitude brighter than 15.3, not registered as galaxies in known databases, and a color index [8]-[24] > 2.2) from \citet{dunham08} for identifying VeLLOs. Only 4 VeLLOs from this sample, all in the Serpens South region, are detected in the submillimeter, although 14 VeLLOs (1 in \mbox{IC 348}, 2 in \mbox{NGC 1333}, 2 in \mbox{Ophiuchus}, and 9 in \mbox{Serpens South}) are within our coverage. None of the matched VeLLOs shows secular variability in our analysis. Despite the null detection of variability among these 4 observed VeLLOs, there is no evidence that the small VeLLO sample is different than the larger survey sample. That is, comparing the fractional variability within the VeLLO sample against the larger survey sample yields a p-value $\approx$ 0.14, well above the 0.05 typically used to justify differences in the samples.

\subsection{Comparison with NEOWISE Variables}\label{sec:WISE}

According to our results, submillimeter continuum emission traces only variability in protostellar luminosity, which appears as a temperature change in the thick envelope.
Optical and near-IR brightness variations are sensitive to the luminosity, but may also be caused by extinction changes within the small inner region close to the central object, (i.e., optical depth to the stellar photosphere and inner disk). As a corollary, the variability of YSOs at later evolutionary stages, without much remaining envelope, is primarily revealed at shorter wavelength.  Furthermore, different physical and geometrical scales associated with underlying variability processes also determine the timescales of variability. 

In this sense, the mid-infrared (MIR) covers a wider range of physical/geometrical scales, and thus, a greater variety of variability mechanisms than are expected to be traced in the submillimeter. An intensive investigation for YSO variability in the MIR has been undertaken for  $\sim\,5400$ YSOs in 20 nearby low-mass star-forming regions, using  NEOWISE W2 band (4.6 $\micron$) light curves covering $\sim$7 years (Park et al. submitted). Here, we compare the JCMT Transient Survey results to that MIR variability investigation. 


In total, 1204 NEOWISE sources are located within our eight Transient survey regions.  Of those, 49 have submillimeter counterparts: 
 39 are classified as protostars, 9 are disk sources, and 1 submillimeter peak is coincident with a Class\,III or evolved source.
For the protostars in common, 23 out of 39 (\mbox{59 $\%$})  are 
variables in the MIR over the 6.5 years monitored by NEOWISE, very similar to the 55$\%$ variability likelihood over all protostars detected by that survey (\mbox{W.~Park et al.} submitted). 
 
From our survey, 8 submillimeter protostellar variables have counterparts observed in the W2 band, yielding a 20$\%$ (8 out of 39) variability rate almost identical to the likelihood over the full submillimeter survey. 
Five of these sources are identified as variable at both wavelengths.
\mbox{West 40}, \mbox{V1647 Ori}, and \mbox{Serpens Main}-\mbox{SMM 1} are classified as Curved in both submillimeter and MIR.
\mbox{EC 53} and \mbox{Serpens Main}-\mbox{SMM 10} are varying in W2 but classified as MIR irregulars, which means no notable secular trend in the W2 light curves. 
\mbox{HOPS 389}, \mbox{NGC 1333}-\mbox{VLA 3}, and \mbox{SH 2-68-N} are not observed to vary in the MIR.


Similar to the submillimeter analysis in this work, Park et al.~(submitted) classified the YSO MIR variability largely into two categories, i.e., secular and stochastic. As here, they further divided secular variability into three types: Linear, Curved, and Periodic; however, the boundary between the groups is different because of the different cadences and coverage in the two surveys. Given the large number of MIR stochastic variables that they found, they further divided the stochastic variability into three types: Burst, Drop, and Irregular. Burst and Drop are identified by sudden brightening and dimming only in a few epochs (i.e. with short timescales) over the 7-years light curve, while Irregular is identified by the random distribution of brightness with a high standard deviation, in which no underlying timescale of variability is measured. 

The distinct difference between protostellar variability observed in the MIR versus the submillimeter is that most of the MIR variability is irregular while all variables in the submillimeter show secular trends.  This difference can be interpreted in terms of both the different cadences of observations and the different origins for the variability at these two wavelengths. 
The cadence of NEOWISE is about 6 months, but that of the JCMT Transient Survey is typically one month or shorter. As a result, periodic MIR variability with a short period is more likely to be classified as irregular variability for the NEOWISE light curves. EC 53 provides a good example; it is classified as an irregular variable in the MIR, but as a periodic variable in the submillimeter. The period of EC 53 is about 1.5 yrs. However, the periodogram analysis in the MIR \mbox{(W. Park et al.} submitted) did not find it as a periodic source because the light curve is not simply sinusoidal   \citep[see Figure \ref{fig:A_example} and][]{lee20}, and the MIR data cadence is too sparse to pull out the periodicity. 

There is an additional explanation for the significantly larger numbers of observed irregular variables in the MIR compared with the submillimeter. MIR brightness is intrinsically sensitive to the variability of the protostar while the submillimeter radiation is sensitive only to the variability of the envelope temperature, where the submillimeter emission arises. As a result, stochastic changes in the accretion luminosity or extinction events taking place close to the central source can be more easily detected in the MIR. Sudden changes in accretion luminosity, however, are smoothed by the large envelope given its long thermal relaxation time, a month or longer \citep{Johnstone13}, and the fractional change in submillimeter flux, which is proportional to the temperature change, is lowered by a factor of $\sim$5.5 compared to that in MIR \citep{contreraspena20}.

Finally, there are no epochs in which the observed submillimeter emission becomes much lower, drops, for a single epoch. In the MIR sample a non-negligible number of disk sources show such short-timescale behaviour. The lack of dips in the submillimeter is most likely due to the fact that the observed submillimeter radiation emitted by the envelope traces the time-averaged luminosity of the source, where the averaging is over the months thermal equilibrium time of the radiating envelope \citep[see][]{Johnstone13}. Thus, while it is possible that a non-thermal brightening event such as a flare \citep{mairs19} might add to this emission to produce a burst, there are no short-timescale subtractions of emission. Furthermore, the submillimeter radiation from the envelope is optically thin, and thus, responds only linearly to changes in optical depth, unlike the optical through MIR where small changes in the line of sight column density can provide very large optical depth variations and strong dimming on a variety of timescales.

\subsection{Variability as a Proxy for Episodic Mass Accretion}
\label{sec:V_mass}

A key result from four years of monitoring protostars in the submillimeter is that at least 20\% are seen to undergo secular variations in their brightness.  The fraction rises to 40\% when only the brighter protostars are included in the sample. Nevertheless, the measured fractional variability amplitudes and slopes for individual sources (see Table \ref{tab:secularvar2}, columns 6 \& 7) appear to be modest and thus one might be tempted to dismiss the observed variability as physically unimportant with respect to the mass assembly process for protostars. However, as noted by \citet{Johnstone13} the submillimeter emission reacts to the temperature of the protostellar envelope, effectively probing the Rayleigh-Jeans tail of the Planck curve, and thus converting the observed variability to the underlying protostellar brightness variations requires a significant, exponential, boost \citep[see also][]{macfarlane19a, macfarlane19b}. Through comparison of observed variables at both MIR and submillimeter wavelengths, \citet{contreraspena20}, empirically uncovered a submillimeter to protostellar luminosity exponential factor of $\sim4$, and confirmed that it matched well with radiative transfer expectations \citep{baek20}. Thus, assuming that the protostellar luminosity is a reasonably linear proxy for protostellar accretion, it is appropriate to consider the importance of the variable mass accretion uncovered by these submillimeter observations.

We begin by assuming that the twelve secular variables best fit with periodic light curves undergo bursts with duration timescales half the measured period and burst amplitudes equal to twice the amplitude, $A_{\rm burst} = 2\times\,A$ (see Table \ref{tab:secularvar2}), effectively requiring that quiescence is determined at the lowest point on the observed light curve. Furthermore, for those six variables best fit by linear slope measurements we assume that the burst takes place during half of the observing window, $\sim\,2$\,yrs, and that the burst amplitude is four times the (yearly) slope, $A_{\rm burst} = 4 \times\,$Slope (see Table \ref{tab:secularvar2}). This latter assumption is likely a significant underestimate of the importance of the burst as we are constraining the amplitude by the observing time window.

With these estimates for the submillimeter burst strength, we can now directly compute the excess amount of mass accreted during each burst after subtraction of the quiescent mass accretion, $M_{\rm burst}$,
against the total mass that is accreted at the fixed quiescent rate over the full time period, $M_{\rm quiescent}$;
\begin{equation}
    \frac{M_{\rm burst}}{M_{\rm quiescent}}
    = 0.5\,([1+A_{\rm burst}]^{4} - 1). 
\end{equation}
Thus, a null value refers to no variability, $A_{\rm burst} = 0$, whereas a very powerful burst would add many multiples of the quiescent accreted mass. For the quantities tabulated in Table \ref{tab:secularvar2}, we find that the smallest amplitude variables add $\sim\,10\%$ of the quiescent accreted mass (integrated over the full time range) through their years long burst enhancement. The six most prominent variables, however each add greater than $40\,\%$ during the burst phase. 

Using this simple model, the most powerful burst accretor in our sample is V1647\,Ori, adding an additional $125\,\%$ of the quiescently accreted mass during each $\sim\,3$\,yr burst. Furthermore, during each burst V1647\,Ori attains a peak mass accretion rate $3.5$ times the quiescent rate,
\begin{equation}
\frac{\dot M_{\rm peak}}{\dot M_{\rm quiescent}} = [1+A_{\rm burst}]^{4}.
\end{equation}
Similarly, EC\,53 (V371 Ser) adds an additional $60\,\%$ of the quiescently accreted mass during each of its $\sim\,0.75$\,yr bursts, attaining a peak burst mass accretion rate twice the quiescent rate. As noted in Section \ref{sec:FUors}, both of these known Class\,I repeaters belong to the FUOr/EXOr type of eruptive stars and have repetition timescales in the literature similar to what is determined in our analysis. 

In addition to EC\,53 and V1647\,Ori, four Class\,0 protostars in our sample have measured high amplitude variability and associated large burst accretion rates, HOPS\,356 (85\%), HOPS\,373 (70\%), West\,40 (45\%), and HOPS\,383 (40\%) (see Table \ref{tab:secularvar2}). All of these sources are associated with long timescales, periods of at least twice the 4\,yr monitoring window, and thus the measured amplitudes, and burst accretion rates, are anticipated to be lower limits. 

The simple analysis presented in this section suggests that the mass assembly of protostars can be significantly influenced by bursts on timescales of years. Roughly 7\%, 6 out of the 83 monitored protostars, show order unity mass accretion variability on this timescale. Assuming that all protostars sample the range of variability observed by our sample, we predict that for a given source, years-long bursts occur on average about every $\sim\,50$\,yrs. Thus, the importance of these accretion events for the overall mass assembly of the typical protostar is not large, accounting for only a few percent of the accreted mass. 

The results presented here are in contrast to the extreme bursting events searched for in the MIR by \citet{scholz13} and \citet{fischer19}, where less than 1\% of the protostars were seen to vary on half a decade timescales, implying an underlying random distribution of extreme bursts for a given source with $\sim\,1000$\,yrs between each event. There remains a vast unexplored time domain separating these two extremes for episodic mass accretion that warrants detailed investigation. This will require continued searches for rare brightening events as well as dedicated monitoring of protostellar samples in order to determine if the variability uncovered in this paper smoothly connects with the rarer but more extreme events or if the variability separates cleanly into  well-defined episodic timescales. Both possibilities would yield significant constraints for theoretical models of mass accretion, both steady-state and unstable, through the disk.

Finally, we note that continued monitoring in the submillimeter is essential for the youngest, most embedded, protostars. The four strongly varying Class\,0 sources identified here, with variability amplitudes similar to the known eruptive systems EC\,53 and V1647\,Ori, are hard to interpret from MIR observations alone due to the extreme extinction of the central source and disk (see for example the multi-wavelength analysis of HOPS\,373 by Yoon et al.\ in prep.). ALMA observations are also revealing complicated geometries during these very early phases of protostar evolution, such as multiple protostars and an arc-like structure within an extremely dense and compact 
core 
\citep{tokuda14} and multiple misaligned outflows from a single young protostar \citep{okoda21}. The explanation for these features is assumed to be complex gas accretion within a turbulent core. Such complex structure will inevitably produce protostellar accretion variability. It will, therefore, be important to also study the relation between submillimeter variability and the small scale structure near the protostar using high spatial resolution observations such as by ALMA.

\section{Conclusions}\label{sec:Concl}
In this paper, we have analyzed 4 years of JCMT Transient Survey \citep{herczeg17} monitoring of 295 (95) submillimeter peaks, $> 0.14$ Jy beam$^{-1}$ ($> 0.5$ Jy beam$^{-1}$), of which 83 (43) are protostellar. We analyzed the light curves by searching for and statistically quantifying single-epoch events, long term monotonic trends by fitting linear functions, and long-term non-linear trends or periodicity by fitting sinusoidal functions  with the Lomb-Scargle periodograms.  Although the light curves are more complicated and most variables are not well fit with any single function, these fits provide uniform statistical results that reasonably describe the size and timescale of the measured variability.
Eighteen of 83 protostars are variable with a secular trend and are classified as Periodic, Curved, or Linear by their best-fit periods. No robust single epoch events or sources with indefinite stochastic trends across the time coverage are detected for these bright sources. 

To evaluate sensitivity limits for our ability to detect changes with different amplitudes and timescales, we performed a completeness analysis for secular variability, taking into account source brightness, measurement uncertainty, variability timescale, and fractional amplitude of the variation. The sensitivity of the survey to variability mainly depends on the source brightness, and becomes uniform for the sources brighter than 0.5 Jy beam$^{-1}$. 
Following this result, we expect to find more secular variables from our extended survey as the time coverage extends making it easier to observe secular trends.  
Additionally, efforts to improve the relative calibration between epochs from $\sim 2$\% to $\sim 1$\%, should also increase significantly the number of variables recovered by our analysis techniques.
Across the 8 regions monitored by the survey, the sample of sources brighter than 0.5 Jy beam$^{-1}$ in \mbox{OMC 2/3} region show a statistically significant low variable fraction compared against the other regions. No
other region showed significant evidence for regional variation. The evolutionary dependency of variability only showed marginal evidence for more episodic sources in Class\,0 versus Class\,I for our sample.

We compared the variability properties of the known eruptive and sub-luminous YSOs within our sample. Three eruptive YSOs identified previously in the optical, NIR, or MIR, V1647\,Ori, EC\,53 (V371\,Ser), and HOPS 383,  robustly vary in the submillimeter. We note that an additional three Class 0 sources showing strong submillimeter variability, HOPS\,373, HOPS\,356, and West\,40, should continue to be monitored for potential powerful eruptions. Somewhat unexpectedly, the subluminous sample of YSOs shows no evidence of a different variability behaviour compared against the eruptive sources.

Finally, using a simple model to convert submillimeter variations to underlying mass accretion variability, we find that all of the secular variables uncovered by the JCMT Transient Survey require significant enhancement, greater than 10\%, in mass accreted due to bursts versus quiescence. For the six most variable sources, 7\% of our sample, we find that the accreted mass due to the burst alone is greater than 40\% and reaches more than 100\%.
When integrated over the full protostellar ensemble, the importance of episodic accretion on these few years timescale is  negligible, only a few percent of the assembled mass.  However, the measured variability is dominated by events of order the observing time-window on a relatively small sample of objects.
Continued submillimeter monitoring of these fields and of intermediate-mass star-forming regions with more ongoing star formation is needed to reveal the importance of episodic events on decades and longer timescales.


\acknowledgments
\section*{Acknowledgements}
We thank the anonymous referee for the helpful comment on our paper. The authors wish to recognise and acknowledge the very significant cultural role and reverence that the summit of Maunakea has always had within the indigenous Hawaiian community. We are most fortunate to have the opportunity to conduct observations from this mountain. 

The James Clerk Maxwell Telescope is operated by the East Asian Observatory on behalf of The National Astronomical Observatory of Japan; Academia Sinica Institute of Astronomy and Astrophysics; the Korea Astronomy and Space Science Institute; the Operation, Maintenance and Upgrading Fund for Astronomical Telescopes and Facility Instruments, budgeted from the Ministry of Finance (MOF) of China and administrated by the Chinese Academy of Sciences (CAS), as well as the National Key R\&D Program of China (No. 2017YFA0402700). Additional funding support is provided by the Science and Technology Facilities Council of the United Kingdom and participating universities in the United Kingdom and Canada. Additional funds for the construction of SCUBA-2 were provided by the Canada Foundation for Innovation. The James Clerk Maxwell Telescope has historically been operated by the Joint Astronomy Centre on behalf of the Science and Technology Facilities Council of the United Kingdom, the National Research Council of Canada and the Netherlands Organisation for Scientific Research.

This research used the facilities of the Canadian Astronomy Data Centre operated by the National Research Council of Canada with the support of the Canadian Space Agency. This research has made use of the SIMBAD database, operated at CDS, Strasbourg, France (Wenger et al. 2000).

D.J.\ is supported by NRC Canada and by an NSERC Discovery Grant. 
J.-E.L.\ is supported by the National Research Foundation of Korea (NRF) grant funded by the Korea government (MSIT) (grant number 2021R1A2C1011718). 
G.J.H.\ is supported by general grant 11773002 awarded by the National Science Foundation of China.
The contribution of C.C.P.\ was funded by a Leverhulme Trust Research Project Grant. G.B. was also supported by the National Research Foundation of Korea (NRF) Grant funded by the Korean Government (NRF-2017H1A2A1043046-Global Ph.D. Fellowship Program).
D.H.\ acknowledges support from the EACOA Fellowship from the East Asian Core Observatories Association.
G.P.\ is supported by Basic Science Research Program through the National Research Foundation of Korea (NRF) funded by the Ministry of Education (NRF-2020R1A6A3A01100208). 
P.S.T.\ acknowledges support from YSTFC grant ST/R000824/1. 
J.B.\ acknowledges support by NASA through the NASA Hubble Fellowship grant \#HST-HF2-51427.001-A awarded by the Space Telescope Science Institute, which is operated by the Association of Universities for Research in Astronomy, Incorporated, under NASA contract NAS5-26555. 
F.D.\ is supported by the National Natural Science Foundation of China through grant 11873094. 
H.S.\ acknowledges support from the Ministry of Science and Technology (MoST) of Taiwan through grant109-2112-M-001-028-. 
D.S.\ recognizes support from STFC grants ST/N504014/1 and ST/M000877/1. SPL acknowledges grants from the Ministry of
Science and Technology of Taiwan 106-2119-M-007-021-MY3 and 109-2112-M-007-010-MY3.
Y.A. acknowledges support by NAOJ ALMA Scientific Research Grant Numbers 2019-13B, Grant-in-Aid for Scientific Research (S) 18H05222, and Grant-in-Aid for Transformative Research Areas (A) 20H05844 and 20H05847.
S.-P.L.  acknowledges grants from the Ministry of
Science and Technology of Taiwan 106-2119-M-007-021-MY3 and 109-2112-M-007-010-MY3.

\software{makemap \citep{chapin13}, STARLINK \citep{currie14}, SMURF \citep{jenness13}, Astropy \citep{astropycollabo13}, SciPy \citep{scipy20}, Matplotlib \citep{hunter07}}

\clearpage

\bibliographystyle{aasjournal}
\bibliography{lee20}

\appendix

\section{Derivation of Detectable Amplitude with  $\chi^{2}$}\label{sec:A_Comp}
\setcounter{figure}{0}
\renewcommand{\thefigure}{A.\arabic{figure}}
Assume that a source varies as a sinusoid with $F(t) = A\,sin(2 \pi \omega t + \phi_{0}) + F_0$, where $A$, $\omega$,and $\phi_{0}$ are the amplitude, frequency,  and initial phase of the sinusoidal function, respectively, and $F_0$ is constant. Considering $N$ observations taken at fixed times and with fixed measurement uncertainties, the analytic definitions of $\chi^{2}$ under two hypotheses, the sinusoidal function ($\chi^{2}_{\rm sine}$) and a constant, or null, function ($\chi^{2}_{\rm null}$), are then
\begin{align}\label{eq:A1}
    \chi_{\rm sine}^{2} &= \sum_{i}^{N}{G_{i}^{2}(\sigma_{\rm fid})} \nonumber \\
    \chi_{\rm null}^{2} &= \sum_{i}^{N}{(A (sin(X_{i}) - \Bar{f}) + G_{i})^{2}},
\end{align}
where $G_{i}(\sigma_{\rm fid})$ is the Gaussian error of $i$th epoch, $X_{i} = 2 \pi \omega t_{i} + \phi_{0}$, $\Bar{f} = \Bar{F}/A$, and $\Bar{F}$ is the mean value of $F(t)$ over the observations. 

Expanding the null solution one obtains three terms with varying powers of $A$,
\begin{align}
    \chi_{\rm null}^{2} &= A^{2} \sum_{i}^{N} (sin(X_{i}) - \Bar{f})^{2} \nonumber\\ 
    &\quad + 2 A \sum_{i}^{N} (sin(X_{i}) - \Bar{f})  G_{i} \nonumber \\ 
    &\quad + \sum_{i}^{N} G_{i}^{2}.
\end{align}
Thus, $\chi^{2}_{\rm null}$ can be written in the form of a quadratic equation depending on A such that
\begin{align}\label{eq:A_quad}
    \chi_{\rm null}^{2} &= T_{1} A^{2} + 2 T_{2} A + T_{3},
\end{align}
where 
\begin{align}\label{eq:ts}
    T_{1} &= \sum_{i}^{N} (sin(X_{i}) - \Bar{f})^{2}, \nonumber\\
    T_{2} &= \sum_{i}^{N} (sin(X_{i}) - \Bar{f})  G_{i}, \nonumber\\
    T_{3} &= \sum_{i}^{N} G_{i}^{2} = \chi_{\rm sine}^{2}.
\end{align}
We further note that the last term, $T_{3}$, is equivalent to $\chi_{\rm sine}^{2}$.

The ratio of the $\chi^2$ terms determines the power of the best fitting sinusoid (Equation \ref{eq:Power}) as well as the false alarm probability (Equation \ref{eq:FAP_A}). Thus we obtain
\begin{align}\label{eq:A_FAP}
    1-P_{\rm sine} &= \frac{\chi_{\rm sine}^{2}}{\chi_{\rm null}^{2}}, \nonumber \\
    \Rightarrow FAP_{\rm Mod} &= (\frac{\chi_{\rm sine}^{2}}{\chi_{\rm null}^{2}})^{N_{f}/2} N_{\rm freq},
\end{align}
where we make the assumption that $FAP_{\rm Mod}$ is small.

\begin{figure}[htp]
	\centering
	{\includegraphics[trim={0.1cm 1.0cm 1.0cm 1.0cm},clip,width=0.48\columnwidth]{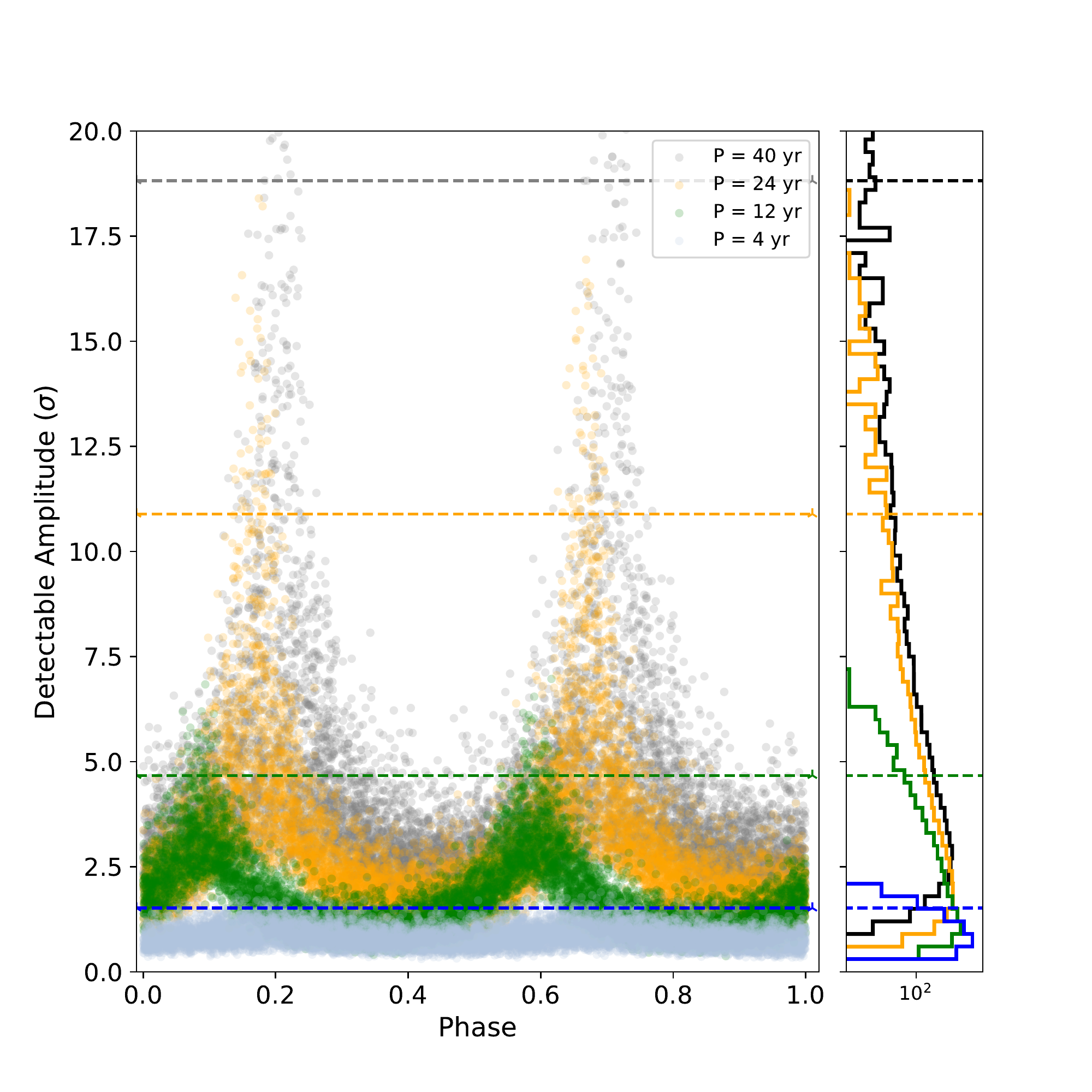}}
	\caption{The detectable amplitude in $\sigma_{\rm fid}$ level depends on phase. The different colors indicate the different periods of the hypothetical signals. For the longer periods, the spikes in a certain phase becomes sharper and higher. Right panel shows the histogram of the detectable amplitude. The colored dashed lines in both panels denote the 99$\%$ detectable level of each corresponding period.} 
\label{fig:Phase_DAmp}
\end{figure}

In our analysis, we consider a source robustly detected when $FAP_{\rm Mod} < 10^{-3}$; therefore, we can solve Equation \ref{eq:A_FAP} for $\chi^2_{\rm null}$:
\begin{align}\label{eq:ineq1}
    \chi_{\rm null}^{2} &> \chi_{\rm sine}^{2}(10^{-3}/N_{\rm freq})^{-2/N_{f}}.
\end{align}
We thus set $\alpha = (10^{-3}/N_{\rm freq})^{-2/N_{f}}$ and, remembering that $\chi^{2}_{sine}$ is equal to $T_{3}$, we substitute Equation \ref{eq:ineq1} into Equation \ref{eq:A_quad} to  obtain the inequality
\begin{align}
    A^{2} T_{1} + 2 A T_{2} + (1-\alpha)T_{3} &> 0.
\end{align}
This equation can be solved for the minimum detectable amplitude as a function of sinusoidal properties, measurement times, and measurement uncertainties.
The minimum detectable amplitude, A$_{\rm det}$, is therefore
\begin{align}\label{eq:fin_ineq}
    A_{\rm det} &= \frac{- T_{2} + \sqrt{T_{2}^{2} + (\alpha-1) T_{1} T_{3}}}{T_{1}}.
\end{align}
We use a monte-carlo approach to solve for the $T_1$, $T_2$, and $T_3$ terms in Equation \ref{eq:ts} under observing conditions similar to those in our survey and for sinusoids with fixed periods and random phases. The results are presented in Figure \ref{fig:Phase_DAmp}. Each point indicates a detectable amplitude of a hypothetical light curve. The phase dependency of detectable amplitude appears as spiked-shaped features in the left panel. In short, for sinusoids with long periods it is difficult to distinguish the underlying signal when the source is near minimum or maximum amplitude and the light curve is relatively flat over the observing window.  We sum over the various phases to determine the completeness threshold as a function of period. The period dependency of the detectable amplitude is indicated by the histogram in the right panel, which shows that the detectable amplitudes of short period sources become concentrated around the value of $\sigma$ derived above. 

Furthermore, we derive the minimum detectable amplitude under an ideal assumption of a sinusoid observed well enough in time to provide uniform phase coverage. This is most likely to occur for sources with periods shorter than the observing window. We can then treat the summation of the coefficients as $T_1 = N/2$, $T_2 = 0$, and $T_3 = N \sigma^2_{\rm fid}$, where $\sigma^2_{\rm fid}$ is the expected uncertainty for a given measurement. Under these simplifications, Equation \ref{eq:fin_ineq} reduces to
\begin{align}\label{eq:Amp2}
    A_{\rm det} &= \sqrt{2(\alpha-1)} \sigma_{\rm fid},
\end{align}
where the amplitude of the minimum detectable periodic signal is proportional to the measurement uncertainty.

We can apply the same sequence of steps to determine the required slope $a$ of a robustly detected linear function. Here the two $\chi^{2}$ are 
\begin{align}
    \chi_{\rm lin}^{2} &= \sum_{i}^{N}{G_{i}^{2}}, \nonumber\\
    \chi_{\rm null}^{2} &=
    \sum_{i}^{N}{(a\,t_{i}-\Bar{F}+G_{i})^{2}}.
\end{align}
Expanding $\chi^2_{\rm null}$ we obtain
\begin{align}
    \chi_{\rm null}^{2} &= a^{2}\sum_{i}^{N}{(t_{i}-\Bar{f})^{2}} +2a\sum_{i}^{N}{(t_{i}-\Bar{f})G_{i}} 
    +\sum_{i}^{N}{G_{i}^{2}}.
\end{align}
Following Equation \ref{eq:A_FAP}, for the linear hypothesis, we obtain

\begin{align}\label{eq:A_FAP_Lin}
    1-P_{\rm lin} &= \frac{\chi_{\rm lin}^{2}}{\chi_{\rm null}^{2}}, \nonumber \\
    \Rightarrow FAP_{\rm Lin} &= (\frac{\chi_{\rm lin}^{2}}{\chi_{\rm null}^{2}})^{N_{f}/2},
\end{align}
where, for a robust detection, we require $FAP_{\rm Lin} < 10^{-3}$. Therefore,

\begin{align}
    \chi_{\rm null}^{2} > \chi_{\rm lin}^{2} (10^{-3})^{-2/N_{f}}.
\end{align}

\newpage
\section{Detected Variables as Individual Cases}\label{sec:A_LCs}
\setcounter{figure}{0} \renewcommand{\thefigure}{B.\arabic{figure}}

In this appendix we note the information derived from literature searches about each of all the identified  secular (Table \ref{tab:secularvar2}) variables with their light curves.

\subsection{Perseus IC\,348,\ MMS1: Periodic Group}

The submillimeter light curve for secular source Perseus IC\,348, MMS\,1 is presented in Figure B.1.
The JCMT source is associated with a millimeter peak with a separation of $0\farcs66$ from the Class 0 protostar IC\,348 MMS. Continuum observations by  by \citep{2013ApJ...768..110C} with the SMA established IC\,348 MMS as a multiple system, with two continuum sources, MMS1 and MMS2, embedded in a common envelope and separated by $9\farcs8$.
IC\,348 MMS1 drives molecular outflow traced in H$_2$ \citep{2003ApJ...595..259E} 
with distinct red and blue shifted lobes detected in CO \citep{2016ApJ...820L...2L}.
Additionally, \citet{2016ApJ...817L..14S} identified MMS1 as a disk-candidate in the VLA Nascent Disk and Multiplicity (VANDAM) 
Survey. The second continuum source, MMS2, is potentially a protobinary confirmed by presence of an outflow, HH797, with signatures of a rotating jet \citep{1994ApJ...436L.189M,2012ApJ...751...78P}.

\begin{figure}[htp]
	\centering
	{\includegraphics[trim={0.2cm 0.0cm 0.0cm 0.0cm},clip,width=0.90\columnwidth]{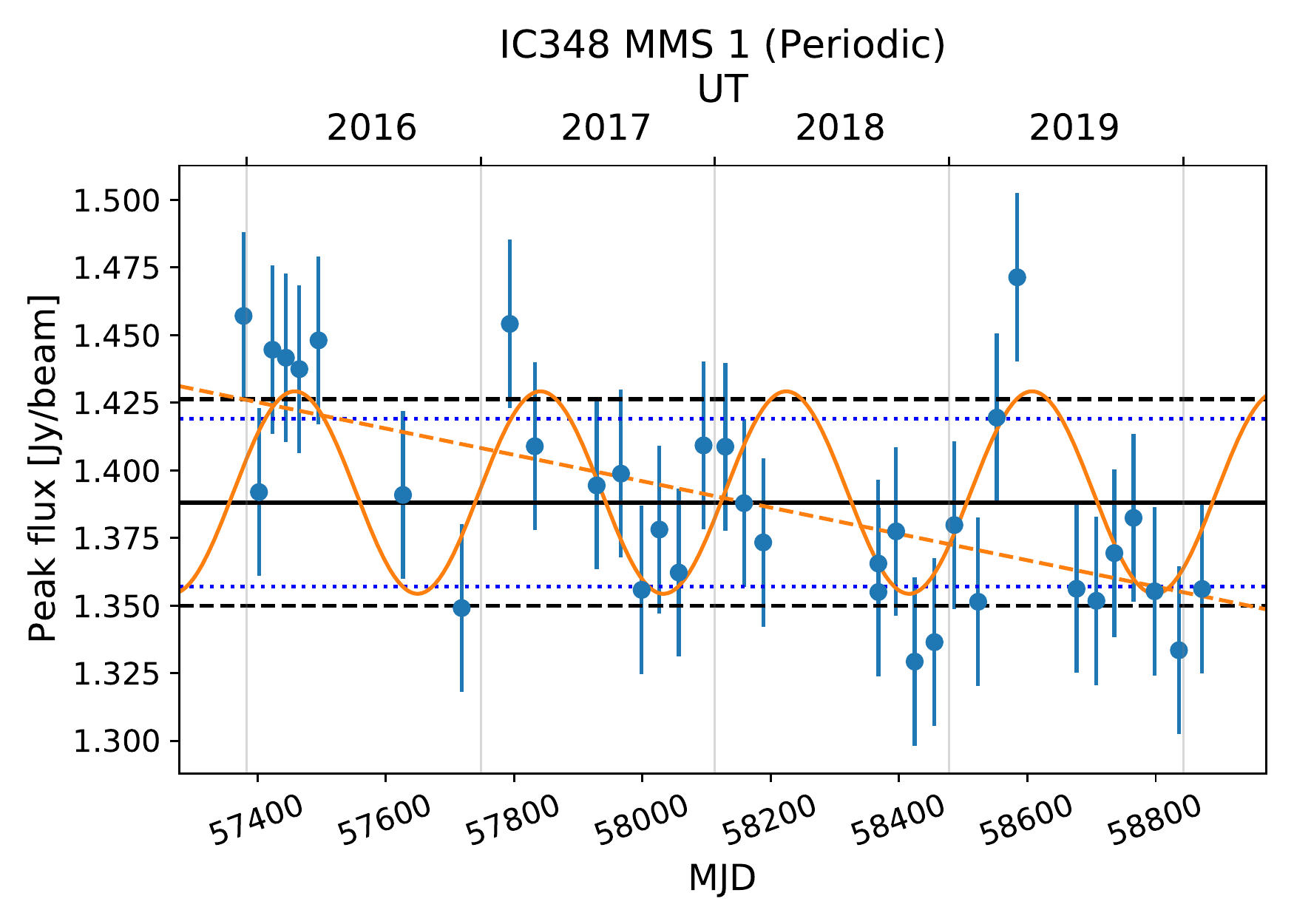}}
	\caption{Light curve of the identified variable source, MMS 1 in IC 348. Black solid line indicates the mean peak brightness in \mbox{Jy beam$^{-1}$}. Black dashed line and blue dotted line indicate the observed standard deviation and fiducial error from the mean peak brightness, respectively. The orange solid line denotes the sinusoidal fitting provided by LSP while the orange dashed line denotes the linear fitting result. Both the linear and sinusoidal fits are robust, FAP $< 0.1$\%. See Tables \ref{tab:secularvar1} and \ref{tab:secularvar2} for details.}
\label{fig:LC_IC348_MMS1}
\end{figure}

\newpage
\subsection{Serpens Main EC\,53 (V371 Serp): Periodic Group}

The submillimeter light curve for secular source Serpens Main EC\,53 (V371 Serp) is presented in Figure B.2. Based on its bolometric temperature and spectral index, EC 53 is a Class I source \citep{evans09, dunham15}. However, the envelope contains $\sim 6\,$M$_\odot$ \citep{baek20} while the protostar and disk masses are only, $0.3\,$M$_\odot$ and $0.07\,$M$_\odot$, respectively \citep{leex20}, suggestive of a much younger evolutionary state. Periodic variability has been observed previously in the near-IR \citep{hodapp99, hodapp12}, MIR \citep{contreraspena20}, and the submillimeter \citep{mairs17b, yoo17, johnstone18}. Deep CO and H$_2$O absorption lines, indicative of viscous disk heating and characteristic of FUOrs, are also observed \citep{connelley18}. \citet{lee20} combine these observations across wavelengths to analyse the physical process responsible for each burst.

\begin{figure}[htp]
	\centering
	{\includegraphics[trim={0.2cm 0.0cm 0.0cm 0.0cm},clip,width=0.90\columnwidth]{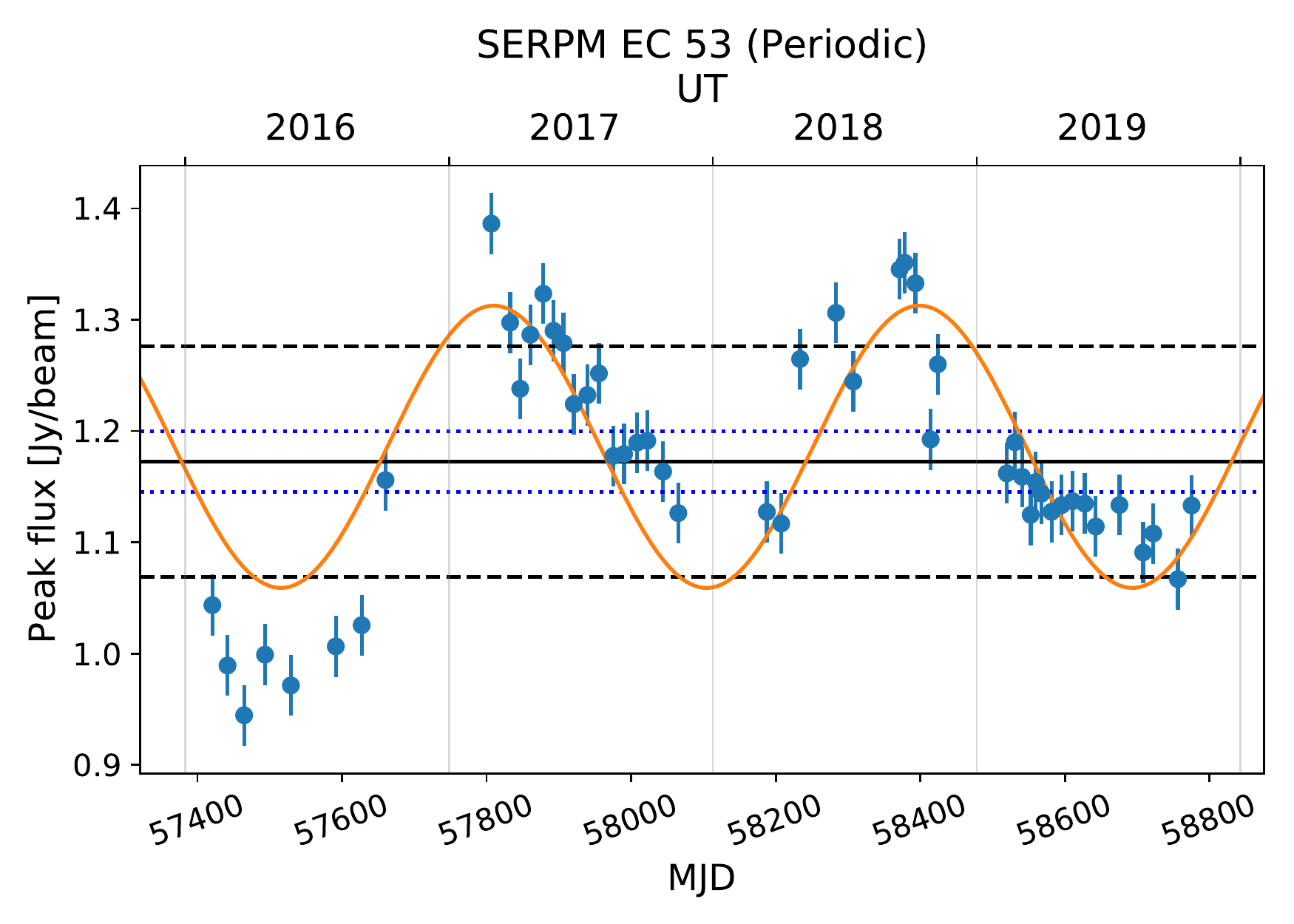}}
	\caption{Same as Figure \ref{fig:LC_IC348_MMS1} but for EC 53 in Serpens Main. The orange solid line denotes the robust, FAP $< 0.1$\%, sinusoidal fitting provided by LSP.} 
\label{fig:LC_SERPM_EC53}
\end{figure}

\newpage
\subsection{Orion NGC\,2068, HOPS\,389: Curved Group}

The submillimeter light curve for secular source Orion NGC\,2068, HOPS\,389 is presented in Figure B.3.
\citet{1996A&A...312..569L} observed this source in 1.3\,mm continuum using the IRAM 30\,m telescope,
finding it to have a compact condensation with more extended envelope.
The presence of two embedded infrared sources led to its identification
as a self-luminous protostellar core (Class 0/I).
\citet{2001ApJ...556..215M} mapped the source using SCUBA
and weakly detected an associated outflow in $^{12}$CO 3-2.
\citet{2013MNRAS.429.3252W} also observed patches of red- and blue-shifted $^{13}$CO 3-2 emission
in the vicinity of this source.
\citet{2002ApJ...571..356M}, using the SCUBA polarimeter,
found the polarization towards the source to be consistent with that of its surroundings.

\citet{2015A&A...581A.140S} list one source for these coordinates,
classified as Class I based on the SED slope between 2.2 and 24\,$\micron$.
\citet{furlan16} find two closely associated sources, HOPS 323 and 389, classifying HOPS 323 as Class I and HOPS 389 as Class 0. Here we identify the submillimeter source with HOPS 389, although HOPS 323 is somewhat more luminous. Both sources are listed as protostars in the
Spitzer survey of YSOs in Orion \citep{megeath12}.
These two objects, together with HOPS 322, are described as a group of protostars.

The VLA/ALMA Nascent Disk and Multiplicity (VANDAM) Survey of Orion Protostars \citep{tobin20}
detected four localized 870\,$\micron$ sources, presumed to be disks:
HOPS-323-A at 05:46:47.697 +00:00:25.27 (44.475 $\pm$ 0.754 mJy) and
HOPS-323-B at 05:46:47.667 +00:00:24.81 (81.351 $\pm$ 1.484 mJy)
both Class I, and
HOPS-389-A at 05:46:47.019 +00:00:27.07 (19.722 $\pm$ 0.668 mJy) and
HOPS-389-B at 05:46:46.604 +00:00:29.52 (4.785 $\pm$ 0.325 mJy)
both Class 0.

\begin{figure}[htp]
	\centering
	{\includegraphics[trim={0.2cm 0.0cm 0.0cm 0.0cm},clip,width=0.90\columnwidth]{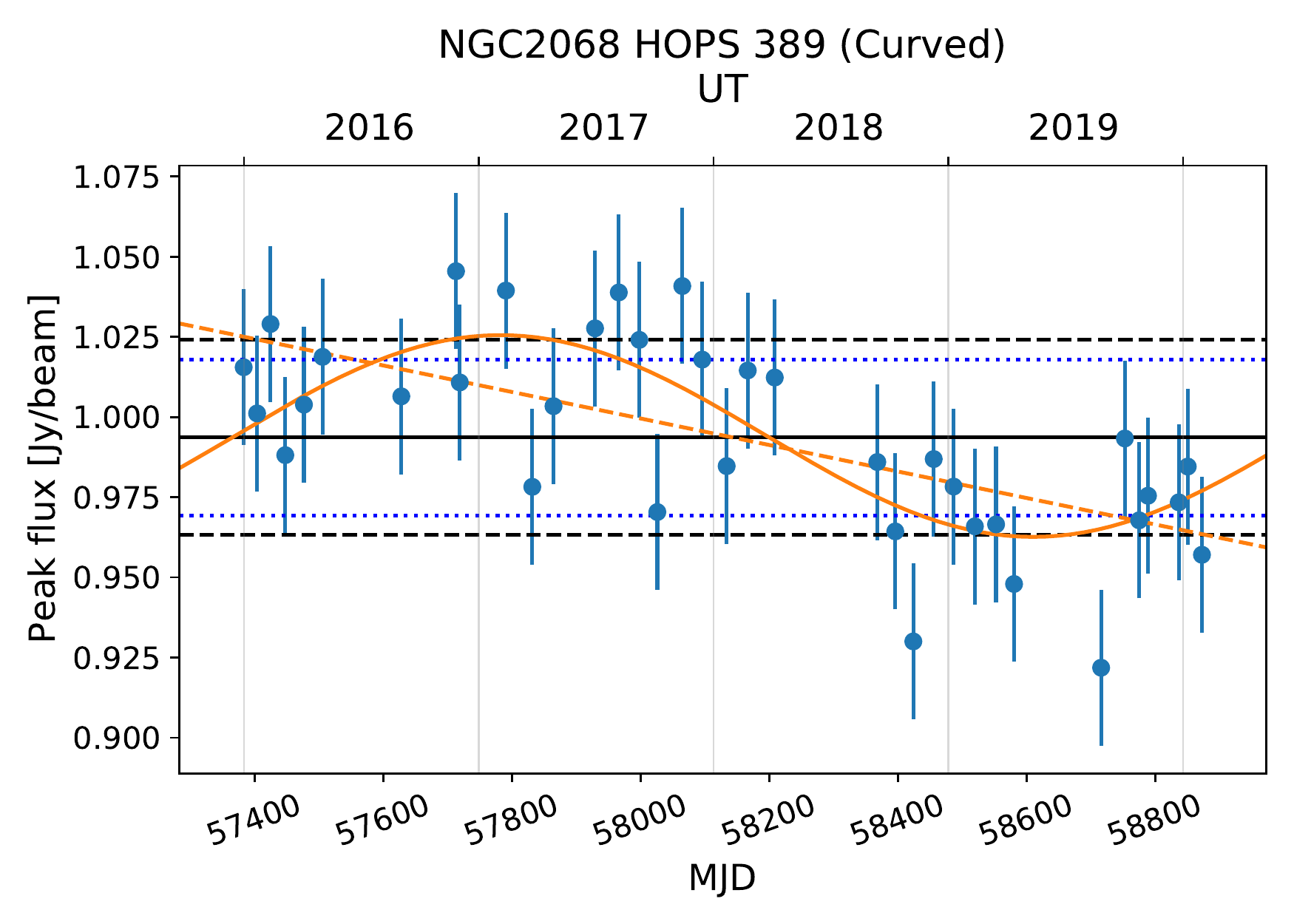}}
	\caption{Same as Figure \ref{fig:LC_IC348_MMS1} but for HOPS 389 in NGC 2068. The orange solid line denotes the sinusoidal fitting provided by LSP while the orange dashed line denotes the linear fitting result. Both the linear and sinusoidal fits are robust, FAP $< 0.1$\%. } 
\label{fig:LC_NGC2068_HOPS389}
\end{figure}

\newpage
\subsection{Perseus NGC\,1333, IRAS\,4A: Curved Group}

The submillimeter light curve for secular source Perseus NGC\,1333, IRAS\,4A is presented in Figure B.4.
The bright infrared \citep{dunham15} through submillimeter \citep{kirk06, enoch09} source is classified as Class 0. IRAS\,4A is resolved as a double source by \citet{tobin16} and appears to harbour a massive disk $\sim 1$\,M$_{\odot}$ 
\citep{tychoniec18} and a large scale bipolar molecular outflow \citep{choi01}. The source was monitored by Spitzer at 3.6 and 4.5$\mu$m for $\sim$35 days as part of YSOVAR \citep{rebull15}, but was not found to be variable in the mid-infrared.

\begin{figure}[htp]
	\centering
	{\includegraphics[trim={0.2cm 0.0cm 0.0cm 0.0cm},clip,width=0.90\columnwidth]{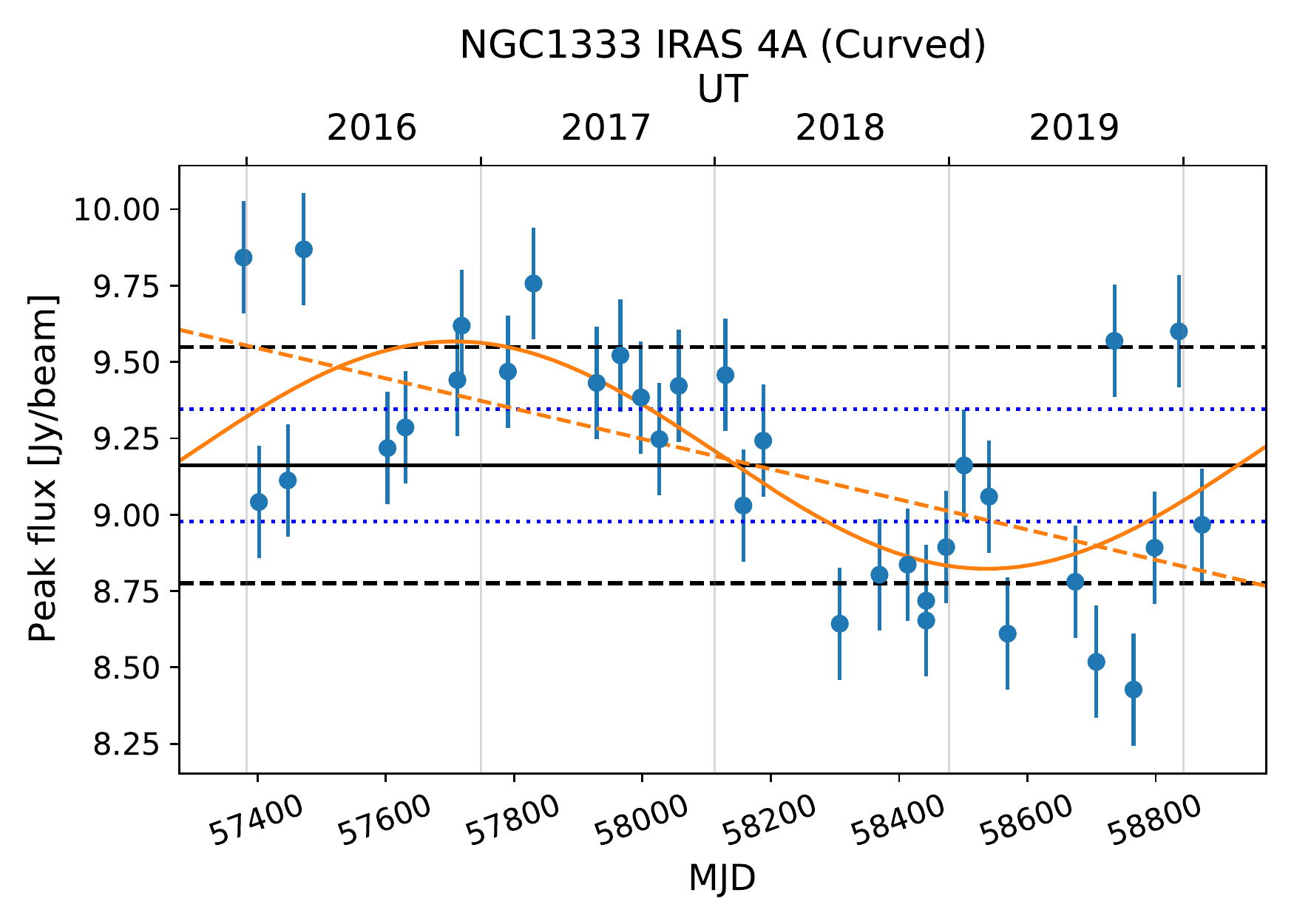}}
	\caption{Same as Figure \ref{fig:LC_IC348_MMS1} but for IRAS 4A in NGC 1333. The orange solid line denotes the sinusoidal fitting provided by LSP while the orange dashed line denotes the linear fitting result. Both the linear and sinusoidal fits are robust, FAP $< 0.1$\%.} 
\label{fig:LC_NGC1333_IRAS4A}
\end{figure}

\newpage
\subsection{Serpens Main, SMM\,10: Curved Group}

The submillimeter light curve for secular source Serpens Main, SMM\,10 is presented in Figure B.5.
The
source has been identified as Class\,I \citep{kryukova12}, and is also known as Ser-emb-12 \citep{enoch09}. Near-infrared K band observations have identified a
stellar counterpart to the submillimeter source, SMM\,10\,IR, which
is found to vary by about 2 mag in K band, and is associated with
a fan-shaped nebulosity \citep{hodapp99}. 
Correlated secular variability on
year-to-year timescales is found for both mid-infrared NEOWISE and
Transient Survey submillimeter observations of the source \citep{contreraspena20}.  ALMA 12m array band\,6 observations (Project ID:
2013.1.00618.S, ALMA source name: 211) of the dust continuum at
0.94\arcsec\ resolution reveal the source to be a binary, with a
separation of 4.25\arcsec, hosting a complex outflow structure in the $^{12}$CO line.

\begin{figure}[htp]
	\centering
	{\includegraphics[trim={0.2cm 0.0cm 0.0cm 0.0cm},clip,width=0.90\columnwidth]{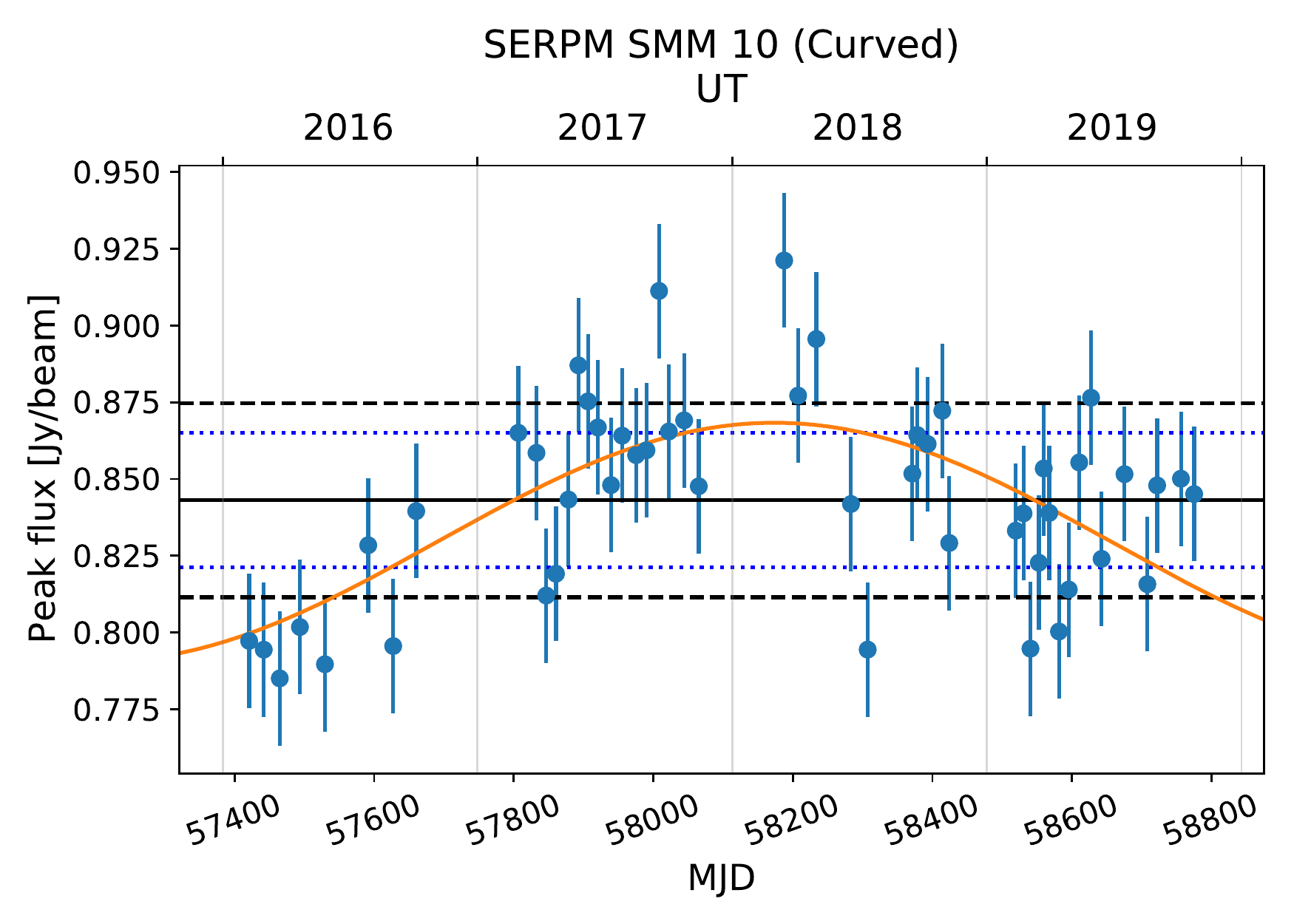}}
	\caption{Same as Figure \ref{fig:LC_IC348_MMS1} but for SMM 10 in Serpens Main. The orange solid line denotes the robust, FAP $< 0.1$\%, sinusoidal fitting provided by LSP.} 
\label{fig:LC_SERPM_SMM10}
\end{figure}

\newpage
\subsection{Ophiuchus Core, VLA\,1623-243: Curved Group}

The submillimeter light curve for secular source Ophiuchus Core, VLA\,1623-243 is presented in Figure B.6.
VLA\,1623-243 comprises three protostars: VLA\,1623\,A, an embedded Class\,0 source 5.6" from peak of 850$\micron$ emission; VLA\,1623\,B, a Class\,0 source 6.5" from the peak; and VLA\,1623\,W, a Class\,I YSO that is much further away with projected distance of 16.0" \citep{war11,mur13}. VLA\,1623\,A and B are just 1.2" apart and thus, cannot be resolved by the JCMT. We therefore associate the binary system with the observed submillimeter light curve, although VLA\,1623\,A is brighter \citep{mur18} and thus expected to contribute most to the observed flux. VLA\,1623\,A has an envelope mass $\sim0.8$\,M$_\odot$ \citep{mur18}, a disk with an outer radius of 180\,AU traced by C$^{18}$O emission \citep{mur13A}, and a large-scale bipolar outflow \citep{san15}. VLA\,1623\,B is a weaker source with about a quarter of the envelope mass of that surrounding VLA\,1623\,A \citep{mur18}. \cite{san15} also found a smaller strongly collimated outflow associated this component. Interestingly, VLA\,1623\,B was determined to be variable at centimeter wavelengths by \citet{cou19}.

\begin{figure}[htp]
	\centering
	{\includegraphics[trim={0.2cm 0.0cm 0.0cm 0.0cm},clip,width=0.90\columnwidth]{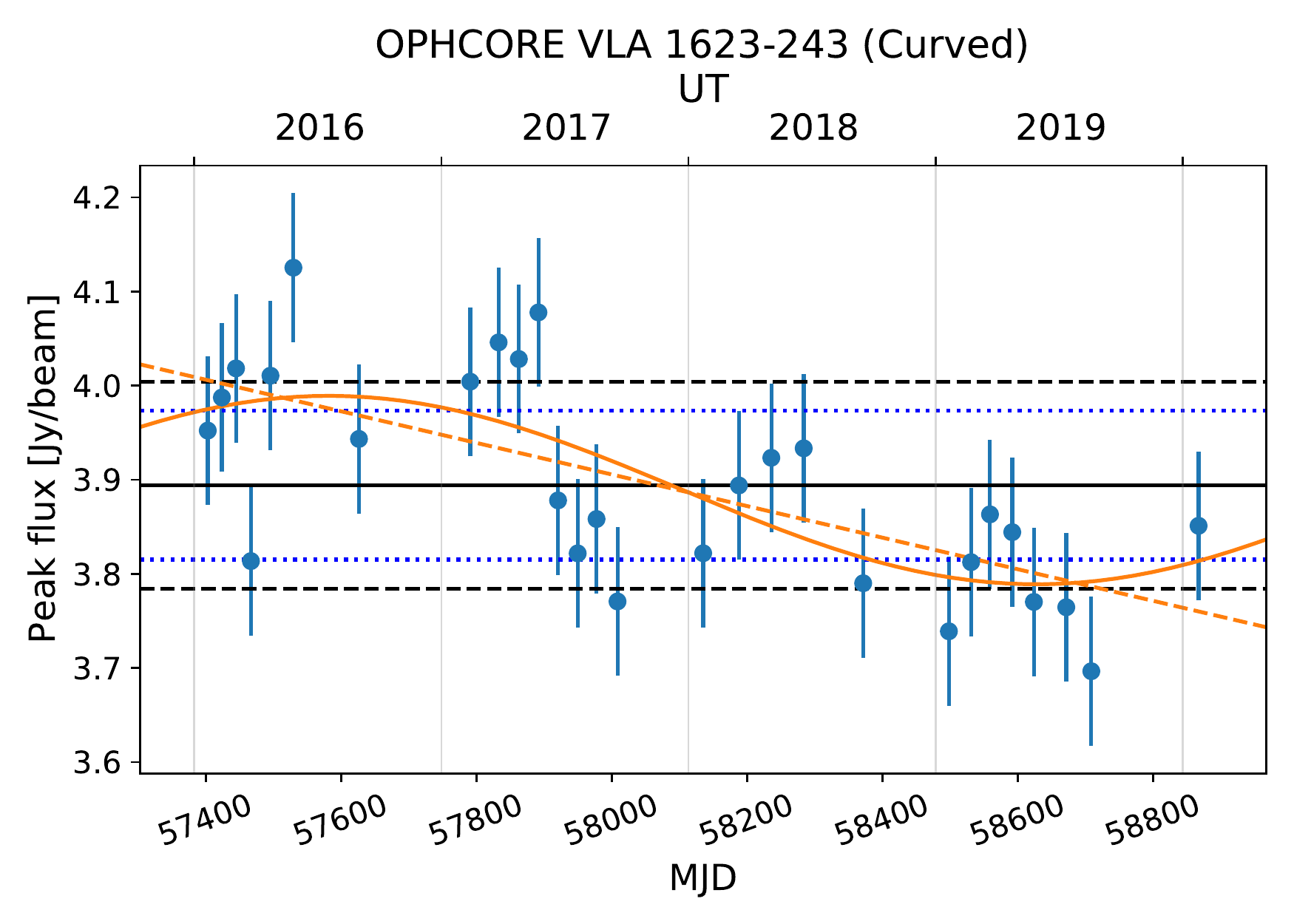}}
	\caption{Same as Figure \ref{fig:LC_IC348_MMS1} but for VLA 1623-243 in Ophiuchus cores. The orange solid line denotes the sinusoidal fitting provided by LSP while the orange dashed line denotes the linear fitting result. Both the linear and sinusoidal fits are robust, FAP $< 0.1$\%.} 
\label{fig:LC_OPHCORE_VLA1623-243}
\end{figure}

\newpage
\subsection{Orion NGC\,2068, V1647\,Ori: Curved Group}

The submillimeter light curve for secular source Orion NGC\,2068, V1647\,Ori is presented in Figure B.7.
V1647\,Ori is the only viable target that is close to the observed submillimeter peak. It is classified as FUOr-like \citep{connelley18} with multiple bursts: 2003, 2008, and 2011 
\citep{aspin06, acosta-pulido07, ninan13,giannini18}.
The source was reported to dim in July 2012 by \citet{ninan12}, four years before the start of the JCMT Transient Survey. Its classification, based on the spectral energy distribution, changes over time but the recent work \citep{furlan16} classifies it as a Class I source in transition to Class II with a flat spectrum. \citet{andrews04} reported an increase of a factor of twenty-five at 12 microns and an order of magnitude increase in bolometric luminosity up to 34\,L$_{\odot}$. In the quiescent phase, \citet{abraham04} approximate V1647\,Ori's luminosity around 5--6 times solar. 

Using VLTI, \citet{abraham06} found  no companion around V1647\,Ori. There is a possible flared disk around V1647\,Ori with a disk mass of 0.05\,M$_{\odot}$ as inferred from mid-infrared. \citet{mosoni13} modeled the MIDI observations toward this source and found that the disk+envelope pre- and post- burst structures are similar. With the recent ALMA observations, \citet{cieza18} constrain the target to a 40\,au disk whose mass is 80 Jupiter masses and with an inclination of 57 degrees. They did not include the envelope model in this case. The earliest jets observations were obtained  by \citet{eisloffel97}.  A molecular CO outflow was later detected by \citet{andrews04}. From infrared imaging, the outflow direction is north to south with respect to the source.

The long term evolution of this source up to 2006 is reported by \citet{acosta-pulido07}. The I band magnitude dropped by 5 magnitude from Feb 2004 to Jan 2006 with a rapid drop at the end of 2005. A periodicity of 56 days was determined from the long term decay that can be explained by dust structures as indicated by \citet{eiroa02}. A high cadence photometry reported by \citet{bastien11} uncovered flickers that may be due to interactions between the magnetosphere and the disk. \citet{aspin11} reported a new flare in 2011 with a total luminosity of 16\,L$_{\odot}$ and an accretion rate of $4 \times 10^{-6}$\,M$_{\odot}$\,yr$^{-1}$  which is similar to the earlier flare in 2004 \citep{fedele07}.

\begin{figure}[htp]
	\centering
	{\includegraphics[trim={0.2cm 0.0cm 0.0cm 0.0cm},clip,width=0.8\columnwidth]{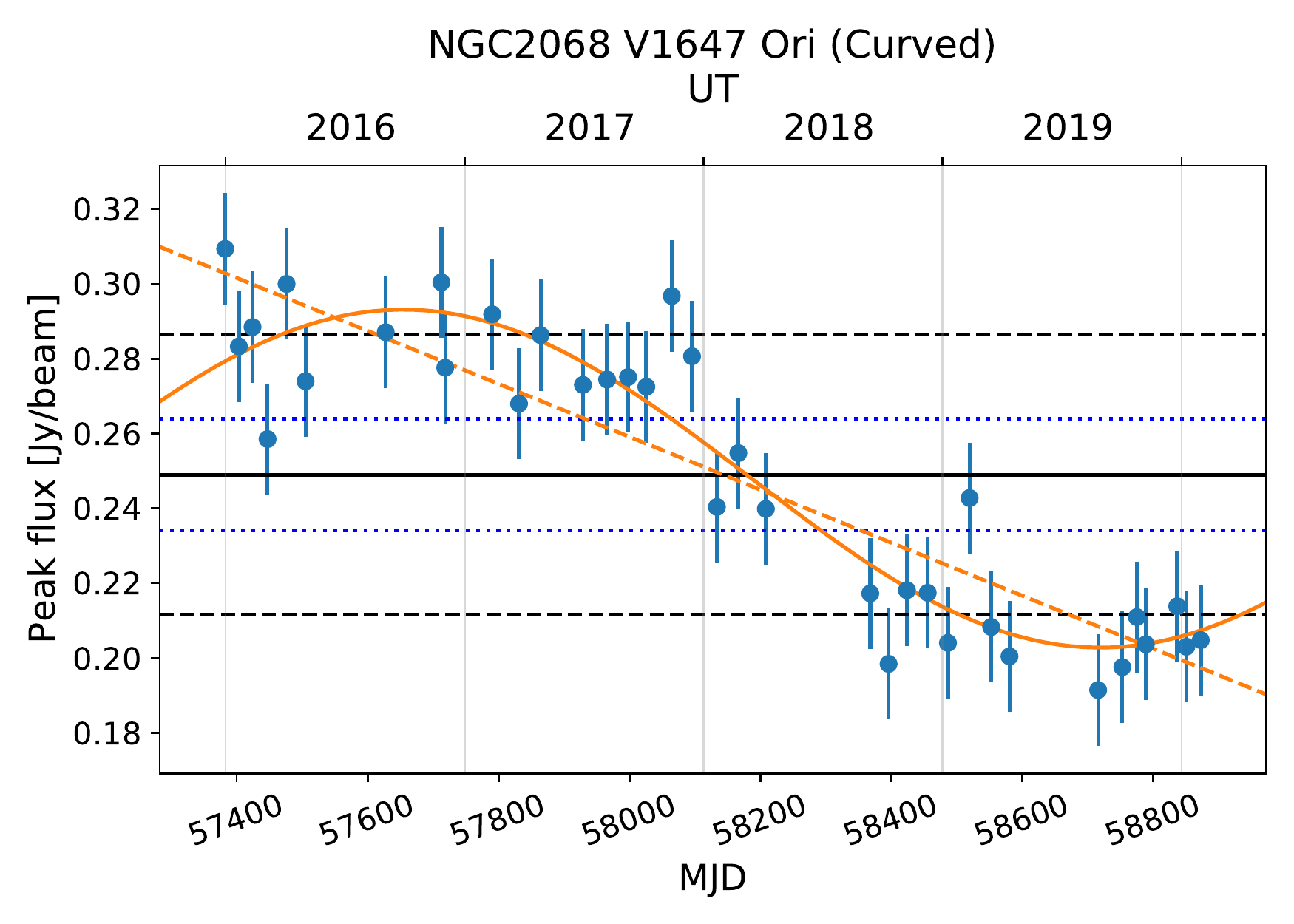}}
	\caption{Same as Figure \ref{fig:LC_IC348_MMS1} but for V1647 Ori in NGC 2068. The orange solid line denotes the sinusoidal fitting provided by LSP while the orange dashed line denotes the linear fitting result. Both the linear and sinusoidal fits are robust, FAP $< 0.1$\%.} 
\label{fig:LC_NGC2068_V1647 Ori}
\end{figure}

\newpage
\subsection{Perseus IC 348, HH\,211: Secular Source}

The submillimeter light curve for secular source Perseus IC\,348, HH\,211 is presented in Figure B.8. The JCMT peak coincides with the dense core HH\,211. This continuum source comprises two embedded sources HH\,211\,SMM1 and HH\,211\,SMM2, respectively, with an angular separation of $0\farcs31$ resolved by high resolution observations with the SMA \citep{Lee_2009,2013ApJ...768..110C}.

HH\,211\,SMM1 is an embedded young Class 0 protostar object, potentially the youngest in Perseus \citep{2008PASJ...60...37H}. It is the exciting source \citep{2001RMxAA..37..201A} for  a highly collimated jet observed in H$_{2}$  \citep{1994ApJ...436L.189M} and SiO  \citep{2006ApJ...636L.141H}.
 Furthermore, \citet{1999A&A...343..571G} detected molecular outflow activity in CO, with low-velocity CO emission tracing bipolar cavities and high-velocity emission aligned with the collimated jet-like structure closer to the central object.
 \citet{cflee20} notes that the knots of emission in the jet have estimated timescales between decades and centuries.
 The central source appears as a flattened disk-like structure, with a size $\sim80$ AU, inclined perpendicular to the jet and outflow axis \citep{1999A&A...343..571G,Lee_2009}. 

\citet{2001ASPC..235..179W} uncovered a rotating envelope in ammonia emission, normal to the outflow and very close to the central protostar. Additionally, \citet{2011ApJ...726...40T} and \citet{2011ApJ...740...45T} detected notable line broadening indicative of outflow-envelope interaction using ammonia observations. High resolution mapping with ALMA, by \citet{2018ApJ...863...94L}, confirmed the presence of a disk and found small outflow velocities indicating a rotating disk-atmosphere. 
Properties of the adjacent continuum source, SMM2, which has a lower mass than SMM1, are mostly uncertain.

\begin{figure}[htp]
	\centering
	{\includegraphics[trim={0.2cm 0.0cm 0.0cm 0.0cm},clip,width=0.90\columnwidth]{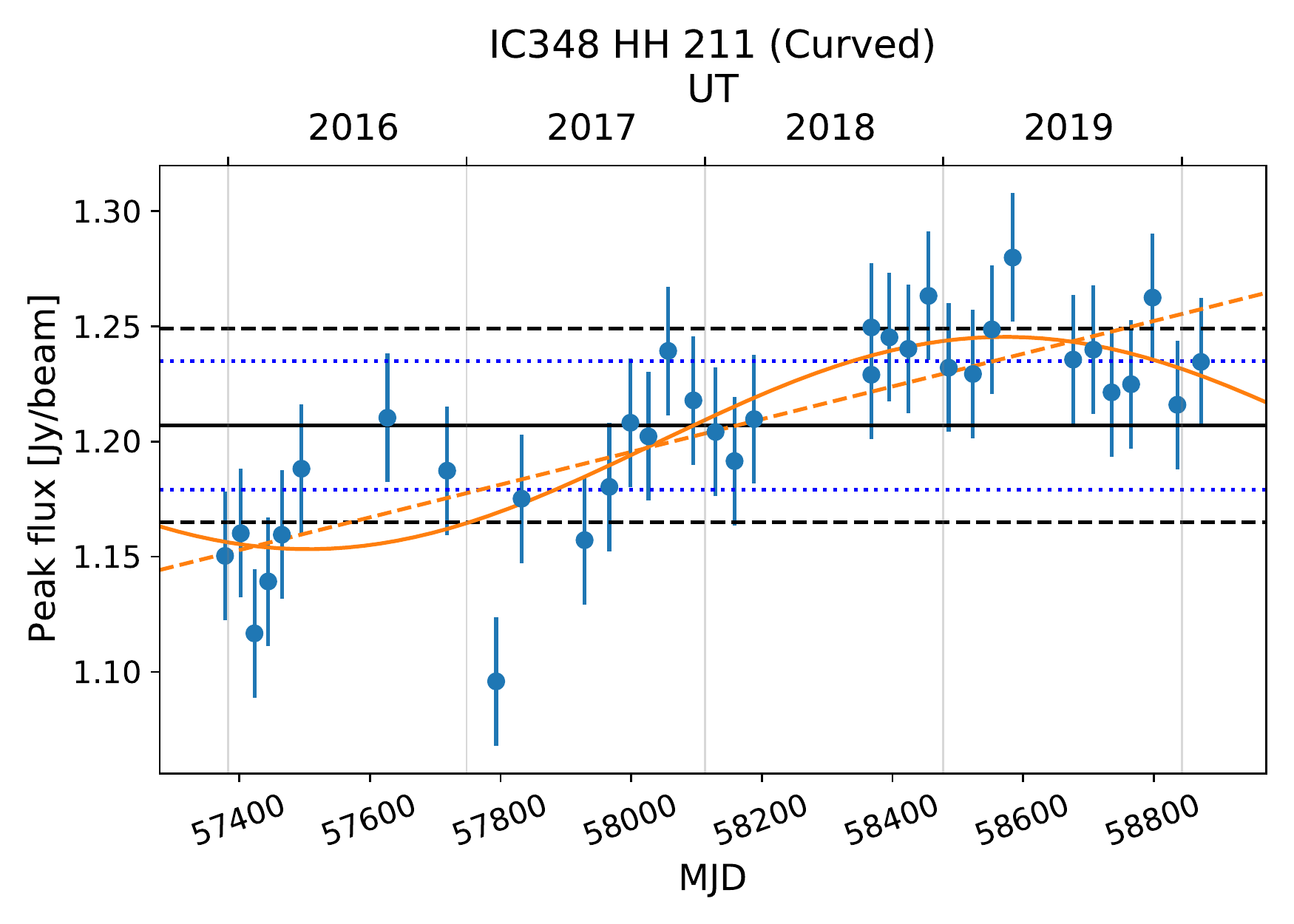}}
	\caption{Same as Figure \ref{fig:LC_IC348_MMS1} but for HH 211 in IC 348. The orange solid line denotes the sinusoidal fitting provided by LSP while the orange dashed line denotes the linear fitting result. Both the linear and sinusoidal fits are robust, FAP $< 0.1$\%.} 
\label{fig:LC_IC348_HH211}
\end{figure}

\newpage
\subsection{Orion NGC\,2068, HOPS\,315: Curved Group}

The submillimeter light curve for secular source Orion NGC\,2068, HOPS\,315 is presented in Figure B.9.
HOPS\,315 is classified as Class\,I protostar via SED fitting by \citet{furlan16}.  \citet{konyves20} identified a dense core 3.6" away from our JCMT peak using PACS and SPIRE images from the Herschel Gould Belt survey. The authors derived a critical Bonnor-Ebert mass ratio of 0.2 for the core, indicating that it is self-gravitating. Recently, \citet{tobin20} used ALMA 870\,$\micron$ observations to derive protostellar disk properties, including the dust disk radius $\sim46$\,au and dust disk mass $\sim100$ Earth masses. \citet{kounkel16} found no companions within 100-1000\,au of HOPS\, 315 using near-IR observations. ALMA observations of the source and outflow are also presented by \citet{dutta20}.

\begin{figure}[htp]
	\centering
	{\includegraphics[trim={0.2cm 0.0cm 0.0cm 0.0cm},clip,width=0.90\columnwidth]{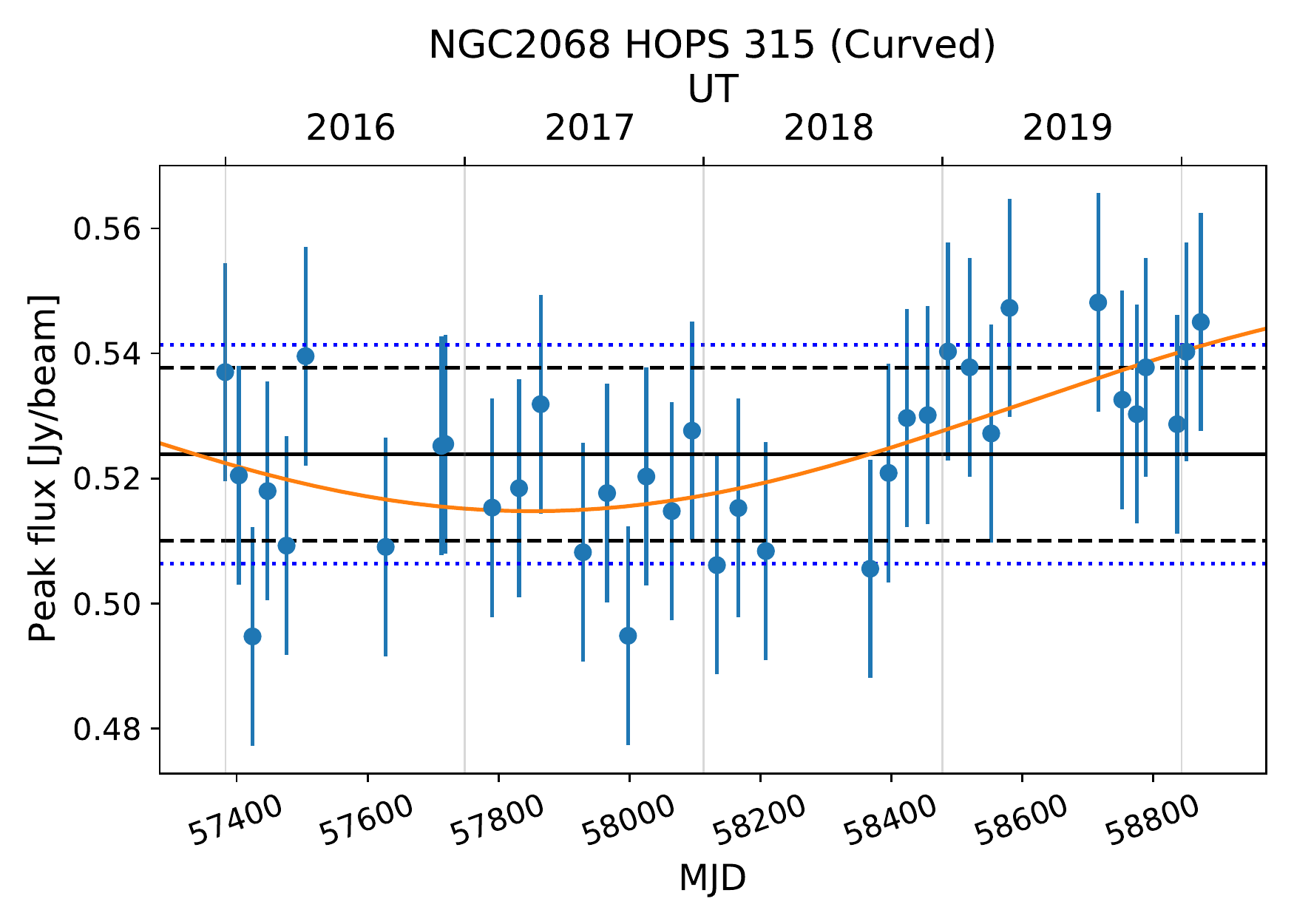}}
	\caption{Same as Figure \ref{fig:LC_IC348_MMS1} but for HOPS 315 in NGC 2068. The orange solid line denotes the robust, FAP $< 0.1$\%, sinusoidal fitting provided by LSP.} 
\label{fig:LC_NGC2068_HOPS315}
\end{figure}

\newpage
\subsection{Orion NGC\,2068, HOPS\,373: Curved Group}

The submillimeter light curve for secular source Orion NGC\,2068, HOPS\,373 is presented in Figure B.10.
\citet{2000MNRAS.313..663G} identified this source as a Class 0 protostar,
and mapped a bipolar outflow with high-velocity dense gas.
The geometric center of this outflow was close to the position of
a previously detected H20 maser \citep{1983ApJ...265..281H}.
\citet{1984ApJ...281..237E} had previously detected high-velocity molecular
gas in the vicinity of the maser.
\citet{2001MNRAS.326..927P} observed the source with SCUBA
and found its 450\,$\micron$ emission to be elongated perpendicular to the outflow direction.
In CARMA 2.9\,mm continuum imaging \citep{2015ApJ...798..128T},
the source appears to be a binary.
The outflow was further studied with CARMA \citep{2016ApJ...831...36T}
and a compact second component with the opposite orientation was seen,
maybe due to a second outflow from the binary source.
From the (Herschel PACS) O I (63.18\,$\micron$) line intensity, they estimate a mass loss rate of $1.1 \times 10^{-7}$ M$_{\odot}$ yr$^{-1}$.

The VLA/ALMA Nascent Disk and Multiplicity (VANDAM) Survey of Orion Protostars \citep{tobin20}
detected two 870\,$\micron$ sources, presumed to be disks:
HOPS-373-A at 05:46:31.099 -00:02:33.02 (103.827 $\pm$ 1.626 mJy) and
HOPS-373-B at 05:46:30.905 -00:02:35.19 (92.206 $\pm $ 0.947 mJy),
classifying both as Class 0.

The JCMT Gould Belt Survey \citep{2016ApJ...817..167K}
did not find an associated protostar in the Spitzer \citep{megeath12}
or Herschel \citep{stutz13} survey catalogues of YSOs in Orion.
However, the source is included in the second paper on the Spitzer Orion survey
\citep{megeath16} as a newly Spitzer-identified YSO (there classified as disk)
and later listed by \citet{stutz13} as a PACS Bright Red source (PBRS). 

The source was observed using the SCUBA polarimeter by
\citet{2002ApJ...571..356M} and, while polarization was seen along
the filament where it is located, no polarization was detected towards the source itself.
The authors suggest that this end of the filament may lie in the foreground
of the nebula, as do \citet{2013MNRAS.429.3252W}.

\citet{2015ApJ...814...31K} find a high deuteration ratio (HDCO/H2CO) for this
source implying that it is probably in the very earliest stage of star formation.
Its classification as a PBRS also argures for it being a very young protostar,
and this is corroborated by the models of \citet{furlan16}.
However, \citet{2015ApJ...798..128T} found rapidly declining visibility amplitudes
and suggest it may be slightly more evolved.

\begin{figure}[htp]
	\centering
	{\includegraphics[trim={0.2cm 0.0cm 0.0cm 0.0cm},clip,width=0.80\columnwidth]{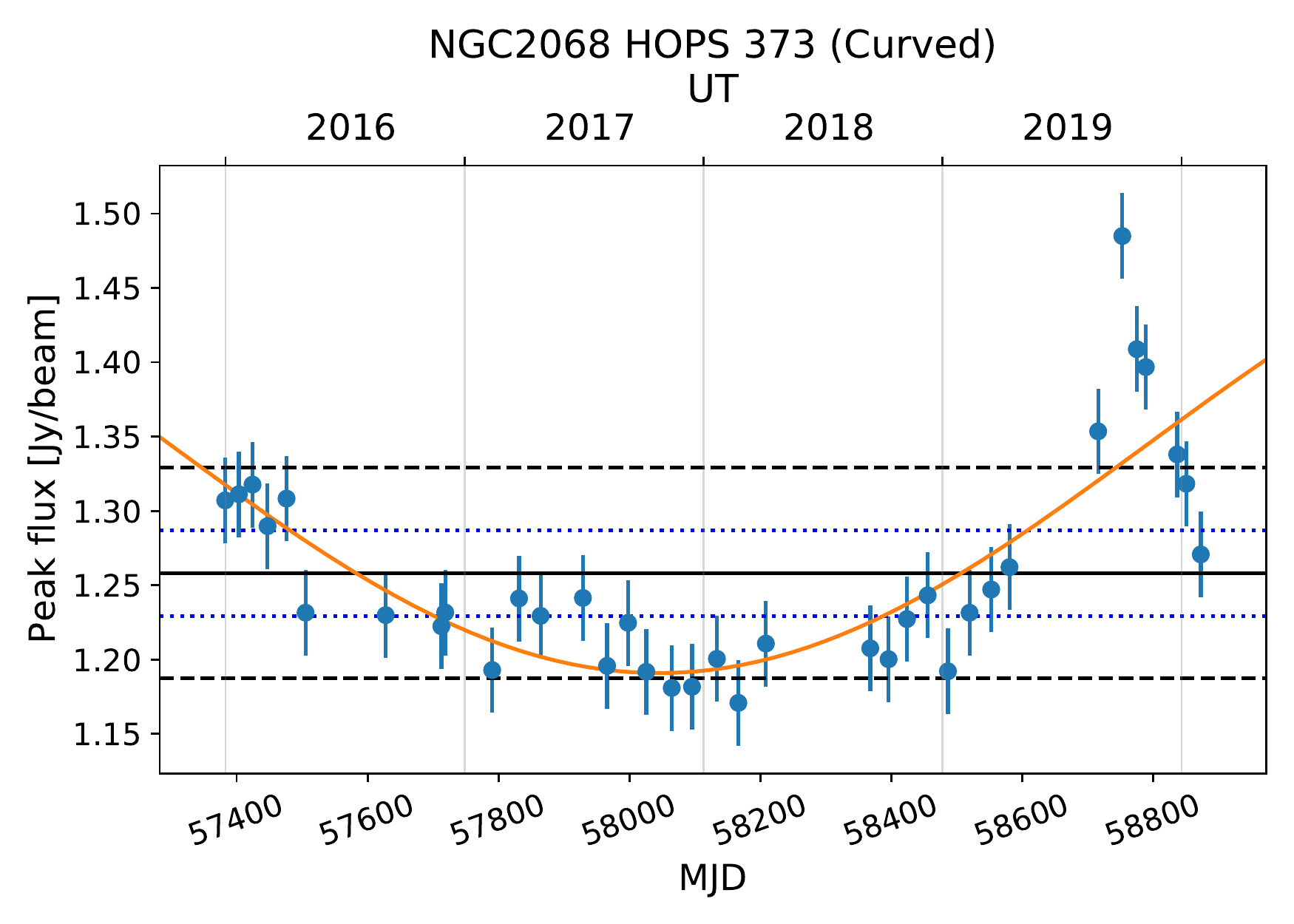}}
	\caption{Same as Figure \ref{fig:LC_IC348_MMS1} but for HOPS 373 in NGC 2068. The orange solid line denotes the robust, FAP $< 0.1$\%, sinusoidal fitting provided by LSP.} 
\label{fig:LC_NGC2068_HOPS373}
\end{figure}

\newpage
\subsection{Orion\,A OMC2/3, HOPS\,383: Curved Group}

The submillimeter light curve for secular source Orion Orion\,A OMC2/3, HOPS\,383 is presented in Figure B.11. \citet{safron15} reported an apparent MIR outburst of HOPS\,383 between 2004 and 2006. By 2008, its brightness at 24\,$\micron$ became 35 times brighter than the observation in 2004 and subsequent monitoring suggested no significant decrease in luminosity from 2009 to 2012. In 2017, \citet{fischer17} suggested a decline in the NIR luminosity. The authors used SED modelling to predict the NIR flux (J, H and K$_{s}$ band), and found no detection at HOPS 383 in the NIR imaging. Based on the post-outburst SED, HOPS 383 is classified as a Class 0 protostar \citep{safron15}. 

\citet{galvan15} found no corresponding radio outburst between 1998 and 2014 after searching the \textit{VLA} archives for radio counterparts of HOPS\,383. In late 2017, \citet{grosso20} detected a hard X-ray source with \textit{Chandra}, which had not been observed in an earlier, January 2000, epoch. 
The newly found X-ray source is spatially coincident with the radio source (JVLA-NW) imaged by \citet{galvan15}.
Meanwhile, \citet{grosso20} monitored the same region at near IR wavelengths and identified an H$_{2}$ knot $\sim$ 15\arcsec away. The H$_{2}$ knot was cross-matched with a previously observed source, SMZ\,1-2B, obtaining a proper motion of 1\farcs.8 in the southeastern direction over 20 years. According to this proper motion, \citet{grosso20} estimated the dynamical timescale of the outflow shocked H$_{2}$ knot at ~180 $\pm$ 100 years.

\begin{figure}[htp]
	\centering
	{\includegraphics[trim={0.2cm 0.0cm 0.0cm 0.0cm},clip,width=0.90\columnwidth]{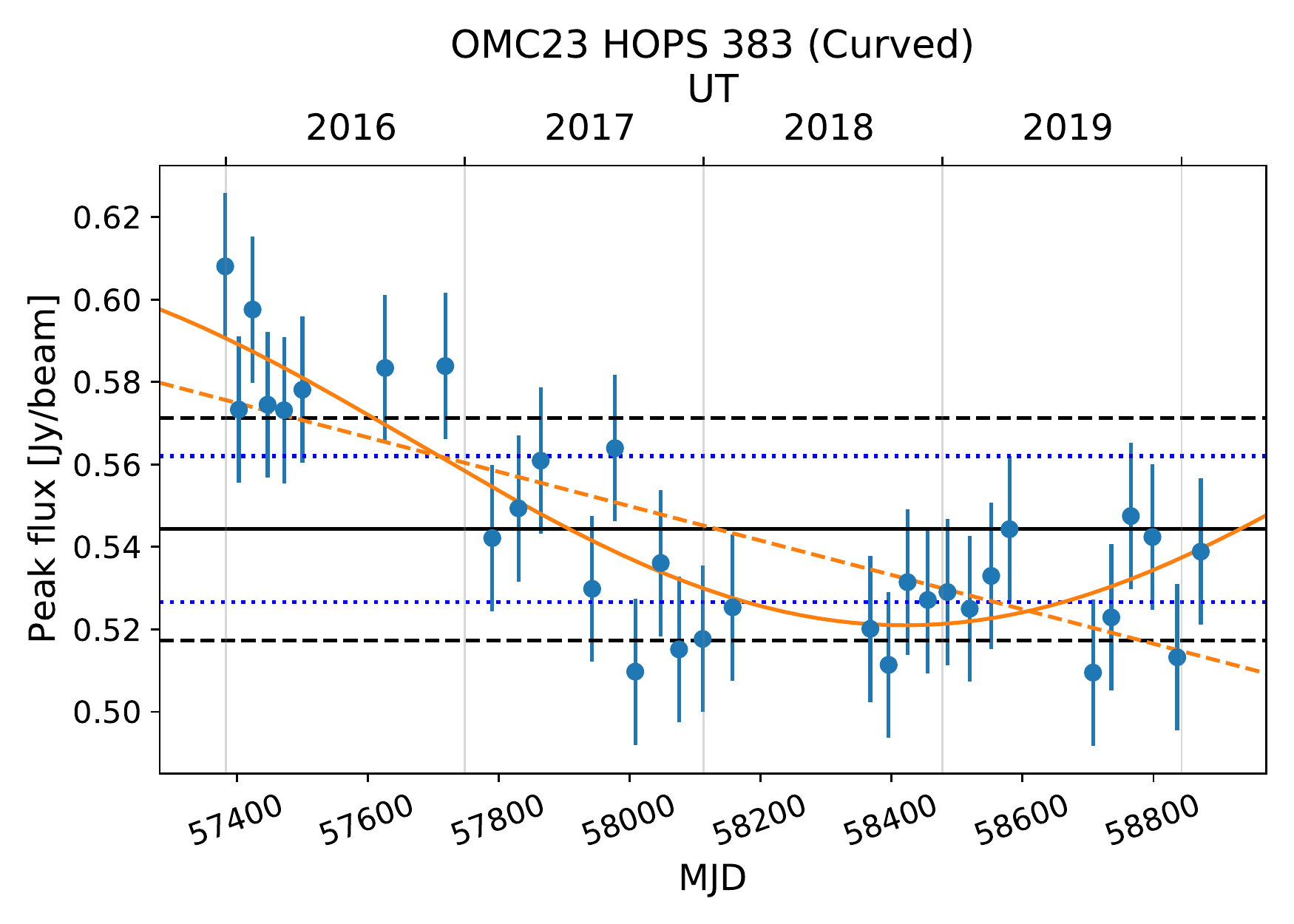}}
	\caption{Same as Figure \ref{fig:LC_IC348_MMS1} but for HOPS 383 in OMC 23. The orange solid line denotes the sinusoidal fitting provided by LSP while the orange dashed line denotes the linear fitting result. Both the linear and sinusoidal fits are robust, FAP $< 0.1$\%.} 
\label{fig:LC_OMC23_HOPS383}
\end{figure}

\newpage
\subsection{Perseus NGC\,1333, West\,40: Curved Group}

The submillimeter light curve for secular source Perseus NGC\,1333, West\,40 is presented in Figure B.12.
Also referred to as Per-emb-15 \citep{enoch09}, West\,40 has been observed from the infrared \citep{dunham15} through the radio, where it is unresolved by the VLA \citet{tobin16}. The source is classified by \citet{dunham15}  as Class\,0.

\begin{figure}[htp]
	\centering
	{\includegraphics[trim={0.2cm 0.0cm 0.0cm 0.0cm},clip,width=0.90\columnwidth]{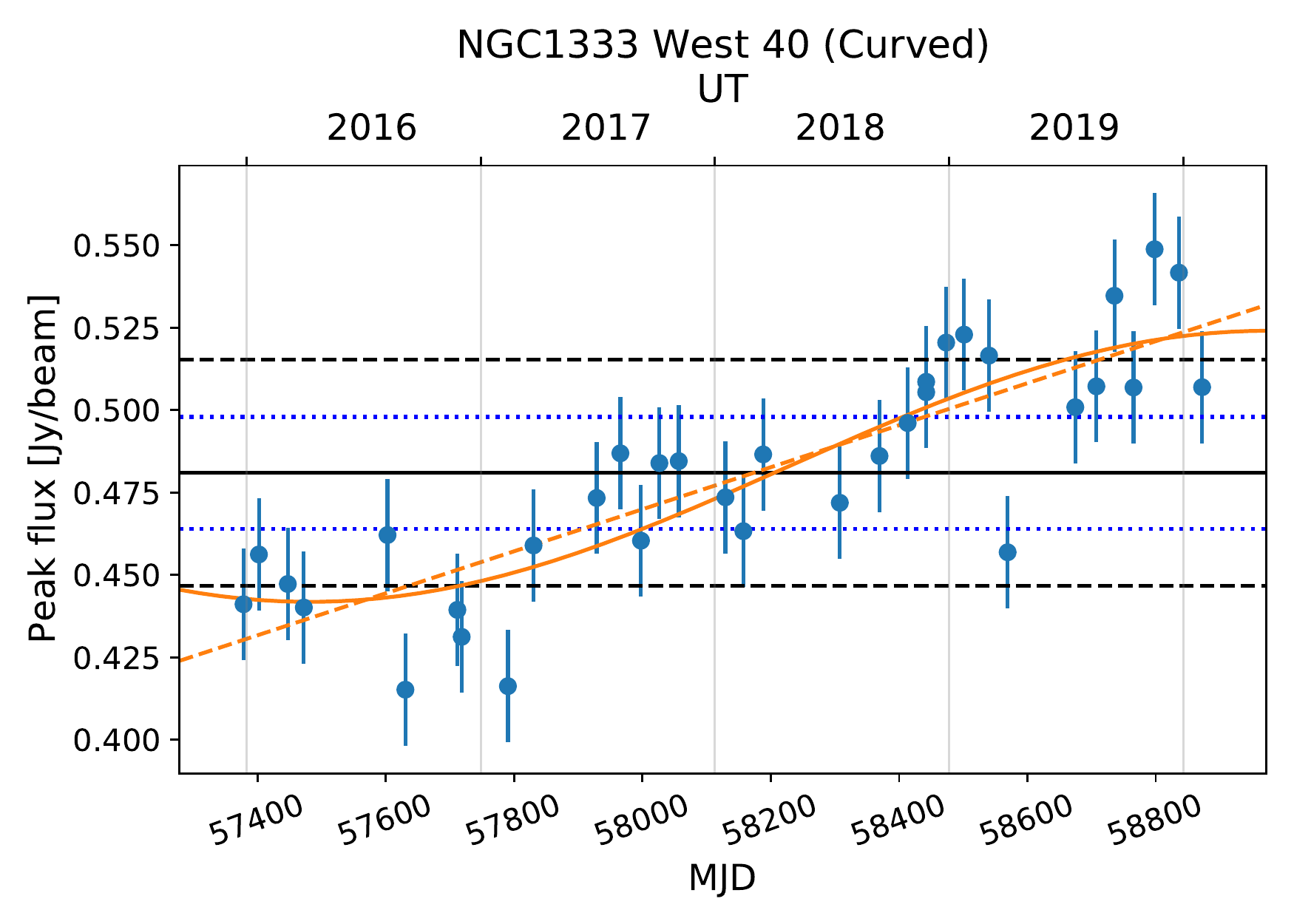}}
	\caption{Same as Figure \ref{fig:LC_IC348_MMS1} but for West 40 in NGC 1333. The orange solid line denotes the sinusoidal fitting provided by LSP while the orange dashed line denotes the linear fitting result. Both the linear and sinusoidal fits are robust, FAP $< 0.1$\%. } 
\label{fig:LC_NGC1333_West 40}
\end{figure}

\newpage
\subsection{Serpens South, IRAS\,18270-0153: Curved Group}

The submillimeter light curve for secular source Serpens South, IRAS\,18270-0153 is presented in Figure B.13.
The location of the submillimeter peak has not been observed by either the SMA or ALMA. The object is identified as a Class\,0 source by \citet{dunham15}. As noted in the footnote to Section \ref{sec:FUors}, previously \citet{johnstone18} mis-identified this source with the FU Ori candidate IRAS 18270-0153W \citep{connelley10} but closer inspection shows that it is beyond 15\arcsec\ from that infrared-bright source.

\begin{figure}[htp]
	\centering
	{\includegraphics[trim={0.2cm 0.0cm 0.0cm 0.0cm},clip,width=0.9\columnwidth]{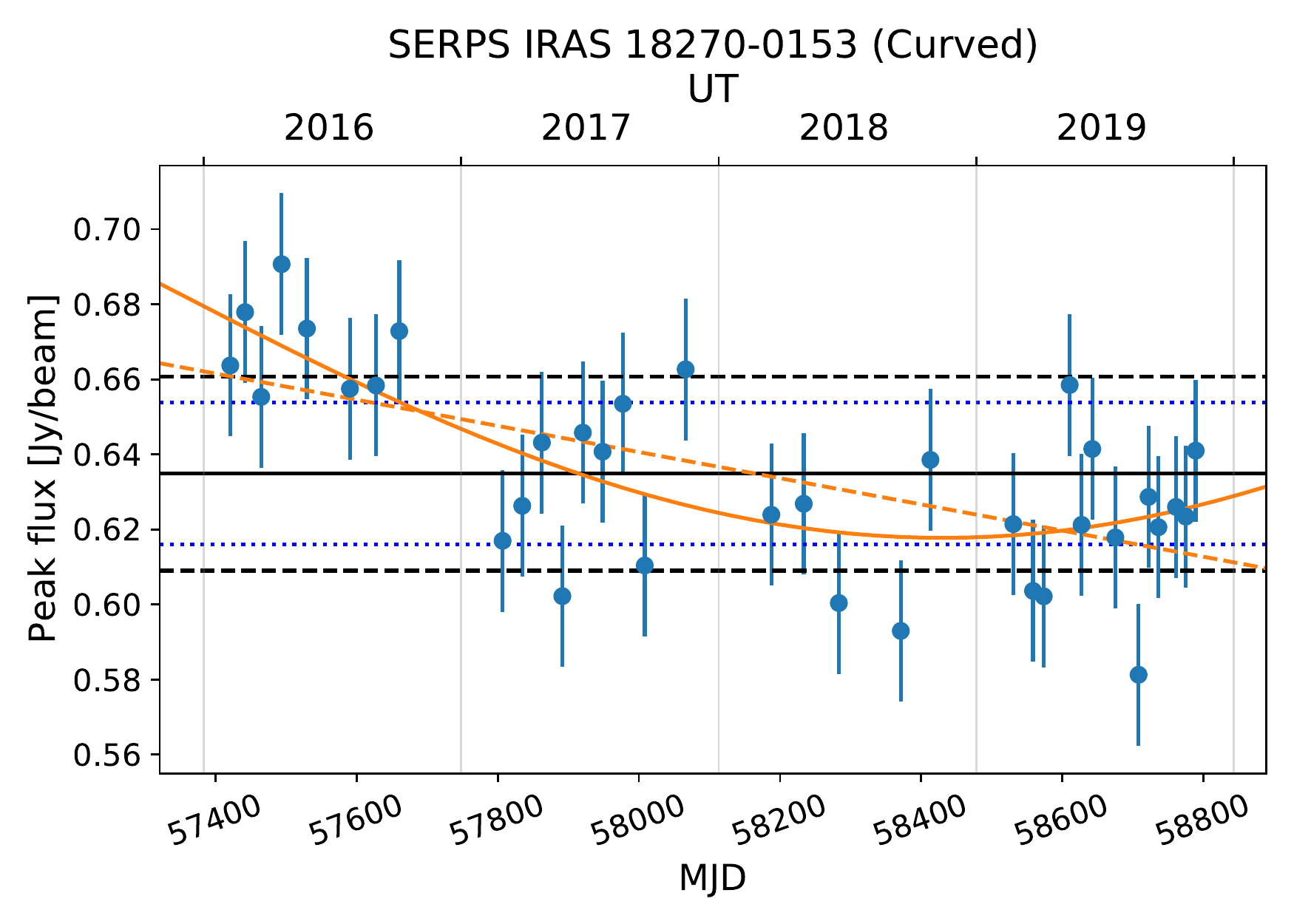}}
	\caption{Same as Figure \ref{fig:LC_IC348_MMS1} but for IRAS18270-0153 in Serpens South. The orange solid line denotes the sinusoidal fitting provided by LSP while the orange dashed line denotes the linear fitting result. Both the linear and sinusoidal fits are robust, FAP $< 0.1$\%.} 
\label{fig:LC_SERPS_IRAS18270-0153}
\end{figure}

\newpage
\subsection{Serpens Main, SMM\,1: Linear Group}

The submillimeter light curve for secular source Serpens Main, SMM\,1 is presented in Figure B.14.
SMM\,1 \citep{casali93} is also known as Serpens\,FIRS\,1 \citep{harvey84} and
Ser-emb-6 \citep{enoch09}. 
This source is the brightest Class\,0 in the Serpens Main region and one of the most extensively studied. Observations with the 12\,m  ALMA array reveal two
extremely bright resolved sources, SMM1-a and the relatively
fainter SMM1-b, surrounded by complex extended structure
associated with outflow cavities 
\citep{hull16,francis19}. \citet{hull16} find high-velocity $\sim80$ km\,s$^{-1}$ CO jets emanating from SMM1-a and -b, and interpret a C-shaped structure in the dust continuum around SMM1-a as walls of a cavity carved by precession of the jet. The same cavity is
also seen in free–free emission in VLA observations, which \citet{hull16} suggest to be caused by ionization of gas in shocks
at the cavity walls. Polarization measurements with ALMA suggest
that the jets are also playing a role in shaping the local
magnetic field \citep{hull17}. Lower velocity ($\sim$10–20 km\,s$^{-1}$) wide angle outflows are also seen in the CO 
emission around the high-velocity jets \citep{hull14,hull17}.
Studies of the chemistry of the outflows have identified SiO in the
high velocity jet and wide angle outflows, while HCN and H$_2$CO are
only seen in the slower wings, consistent with a lower C/O ratio
in the jet \citep{tychoniec19}. Mid-infrared Spitzer observations
also find jets in H$_2$ and various atomic emission lines (e.g., [Fe
II]), however, interpreting which source is driving each outflow
is complicated by the complexity of the outflows and the lower
Spitzer resolution \citep{dionatos14}.

\begin{figure}[htp]
	\centering
	{\includegraphics[trim={0.2cm 0.0cm 0.0cm 0.0cm},clip,width=0.9\columnwidth]{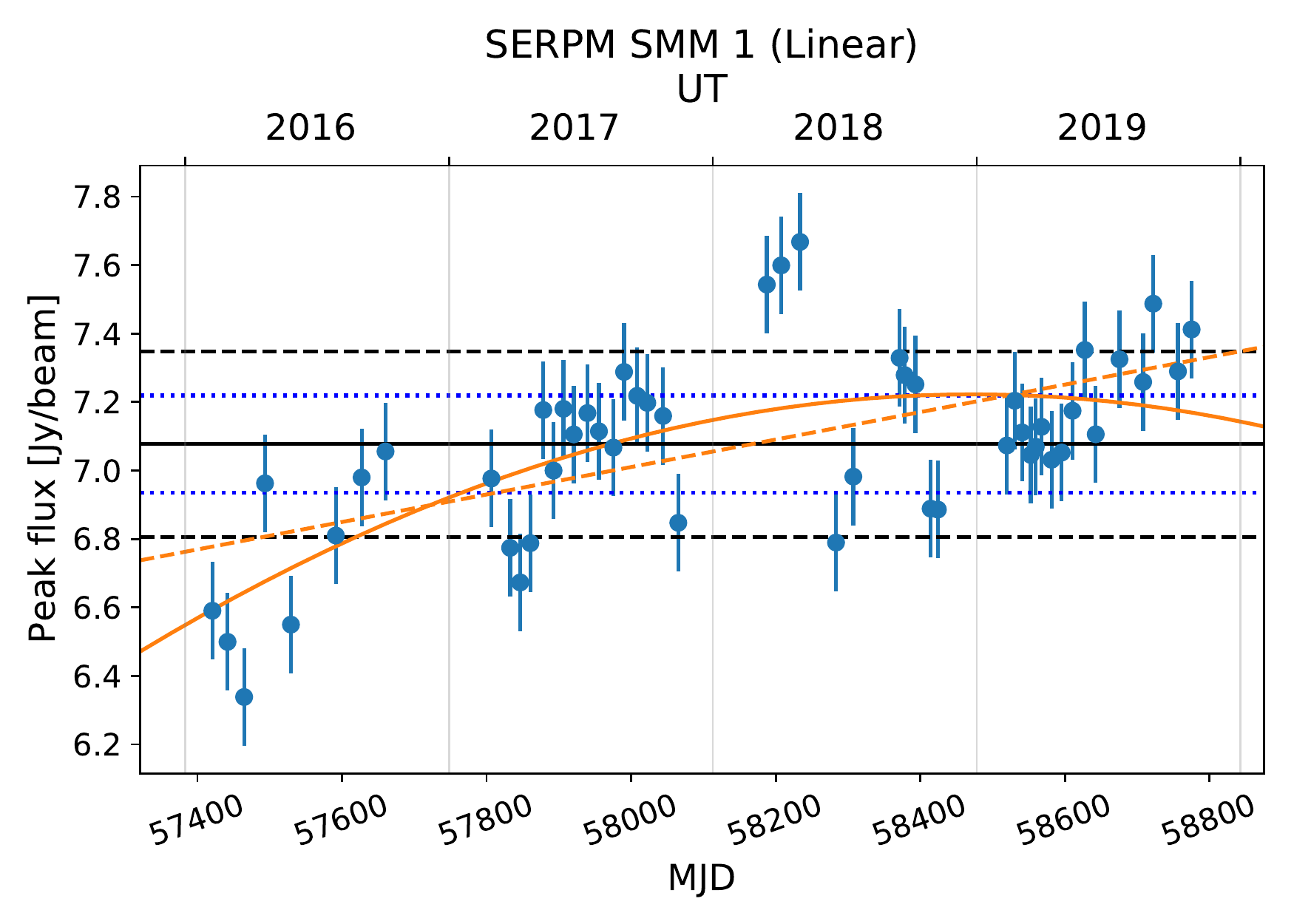}}
	\caption{Same as Figure \ref{fig:LC_IC348_MMS1} but for SMM 1 in Serpens Main. The orange solid line denotes the sinusoidal fitting provided by LSP while the orange dashed line denotes the linear fitting result. Both the linear and sinusoidal fits are robust, FAP $< 0.1$\%.} 
\label{fig:LC_SERPM_SMM1}
\end{figure}

\newpage
\subsection{Serpens Main, SH\,2-68\,N: Linear Group}

The submillimeter light curve for secular source Serpens Main, SH\,2-68\,N is presented in Figure B.15.
SH\,2-68\,N, also known as Ser-emb 8 \citep{enoch09} is an embedded protostar within an extended structure that also includes Ser-emb 8N. This Class 0+1 source lies about 2\arcsec\ north of the bright submillimeter peak \citep{francis19}. \citet{hull17} reported on ALMA observations of polarized dust emission. They found a weak and randomly oriented magnetic field, on scales of 100 -- 1000\, au, which did not exhibit the expected hourglass shape, likely due to the enhanced role of turbulence at these large scales. Sh\,2-68\,N has a slow bipolar molecular outflow observed in SiO 5 -- 4 \citep{hull14} powered by an extremely high-velocity jet. The outflow has a wide opening angle in $^{12}$CO 2 -- 1, surrounded by SO emission tracing the cavity walls  \citep[see also][]{aso19}. 

\begin{figure}[htp]
	\centering
	{\includegraphics[trim={0.2cm 0.0cm 0.0cm 0.0cm},clip,width=0.9\columnwidth]{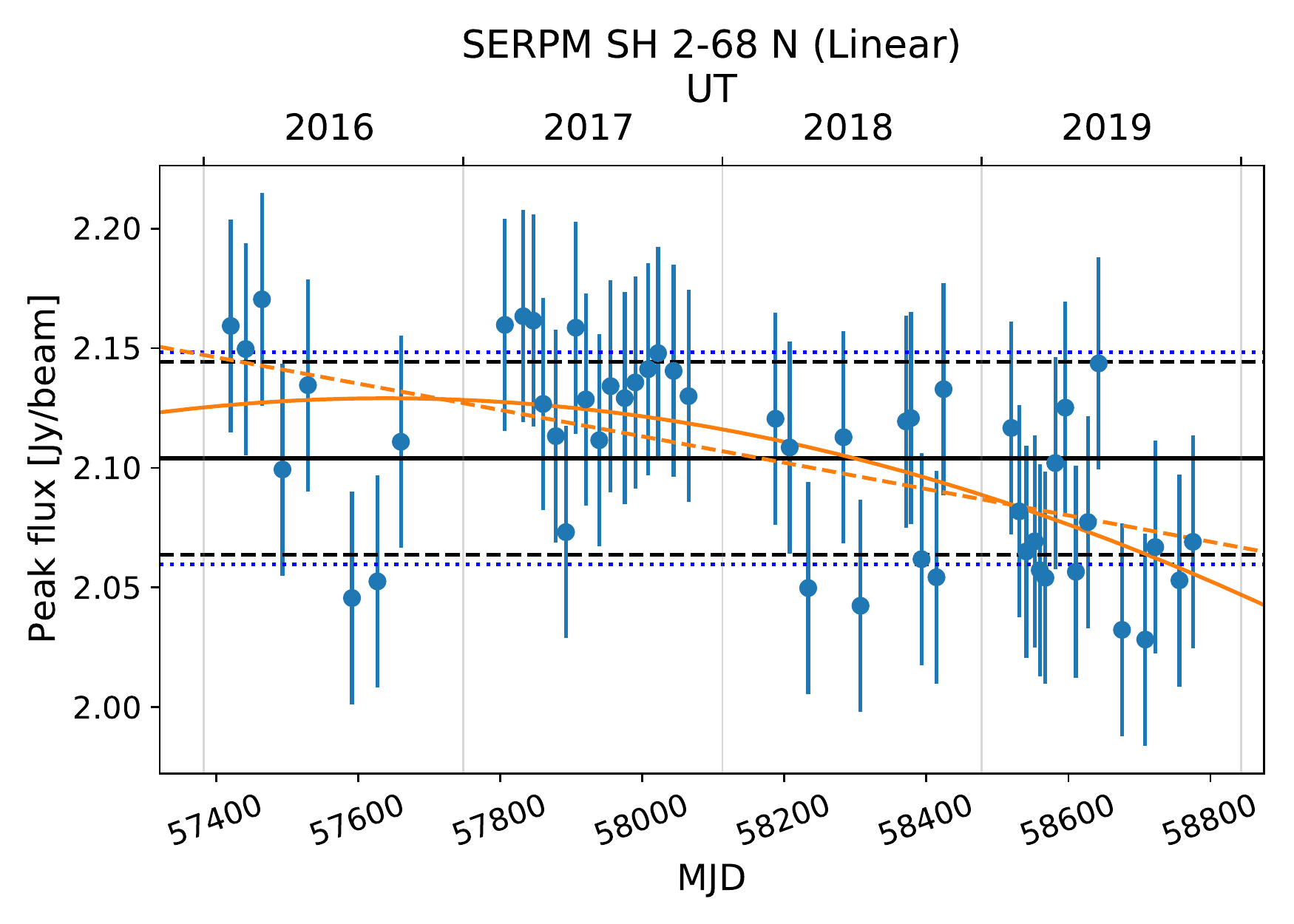}}
	\caption{Same as Figure \ref{fig:LC_IC348_MMS1} but for SH 2-68 N in Serpens Main. The orange solid line denotes the sinusoidal fitting provided by LSP while the orange dashed line denotes the linear fitting result. Both the linear and sinusoidal fits are robust, FAP $< 0.1$\%.} 
\label{fig:LC_SERPM_SH 2-68 N}
\end{figure}

\newpage
\subsection{Serpens South, CARMA\,7: Linear Group}

The submillimeter light curve for secular source Serpens South, CARMA\,7 is presented in Figure B.16.
The submillimeter peak lies in a crowded region, right in the middle of the Serpens South Cluster.
The brightest source in the JCMT field seen with ALMA is serp45 \citep{plunkett18}, and we identify the JCMT source with it.  This source is also known as CARMA\,7 \citep[3-mm;][]{plunkett15} and VLA\,12 \citep{kern16}. This source is a Class\,0 protostar with clear evidence for an episodic jet \citep{plunkett15x}.  The source was not identified as a protostar by \citet{dunham15}; however, Herschel observations reveal that the source is both luminous and buried within a cold envelope \citep[identified as SerpS-mms18 by][]{maury11}.

\begin{figure}[htp]
	\centering
	{\includegraphics[trim={0.2cm 0.0cm 0.0cm 0.0cm},clip,width=0.9\columnwidth]{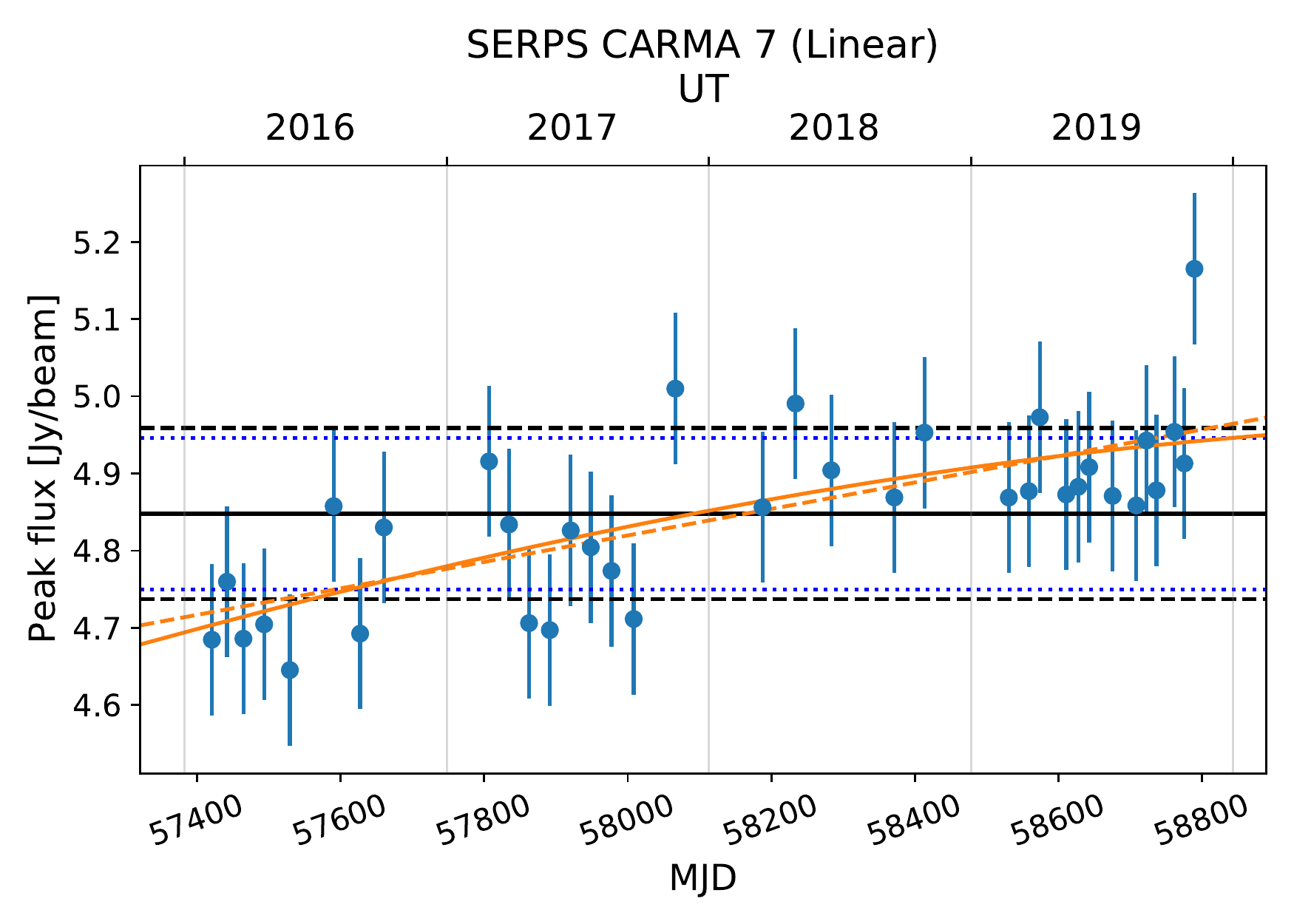}}
	\caption{Same as Figure \ref{fig:LC_IC348_MMS1} but for CARMA 7 in Serpens South. The orange solid line denotes the sinusoidal fitting provided by LSP while the orange dashed line denotes the linear fitting result. Both the linear and sinusoidal fits are robust, FAP $< 0.1$\%.}
\label{fig:LC_SERPS_CARMA7}
\end{figure}

\newpage
\subsection{Orion NGC\,2068, HOPS\,358: Linear Group}

The submillimeter light curve for secular source Orion NGC\,2068, HOPS\,358 is presented in Figure B.17.
The variable submillimeter source is associated with the protostar  HOPS\,358. \citet{stutz13} classified this source as a “PACS Bright Red sources” (PBRS) with a 70\,$\mu$m to 24\,$\mu$m flux ratio greater than 1.65. The authors suggest that PBRs sources are extreme Class 0 objects with a higher envelope density compared with typical Class 0 sources, and equivalently, higher mass infall rates. \citet{furlan16} also classified this source as a Class 0 protostar via SED fitting.  \citet{nagy20} investigated the outflow properties, finding a dynamical timescale of $10^4$\,yr and a mass-loss rate of $5 \times 10^{-6}$\,M$_\odot$\,yr$^{-1}$. The authors also showed an infall asymmetry. ALMA observations of the source and outflow are also presented by \citet{dutta20}.
\citet{tobin20} derived protostellar dust disk properties, finding a roughly 135\,au dust disk.


\begin{figure}[htp]
	\centering
	{\includegraphics[trim={0.2cm 0.0cm 0.0cm 0.0cm},clip,width=0.9\columnwidth]{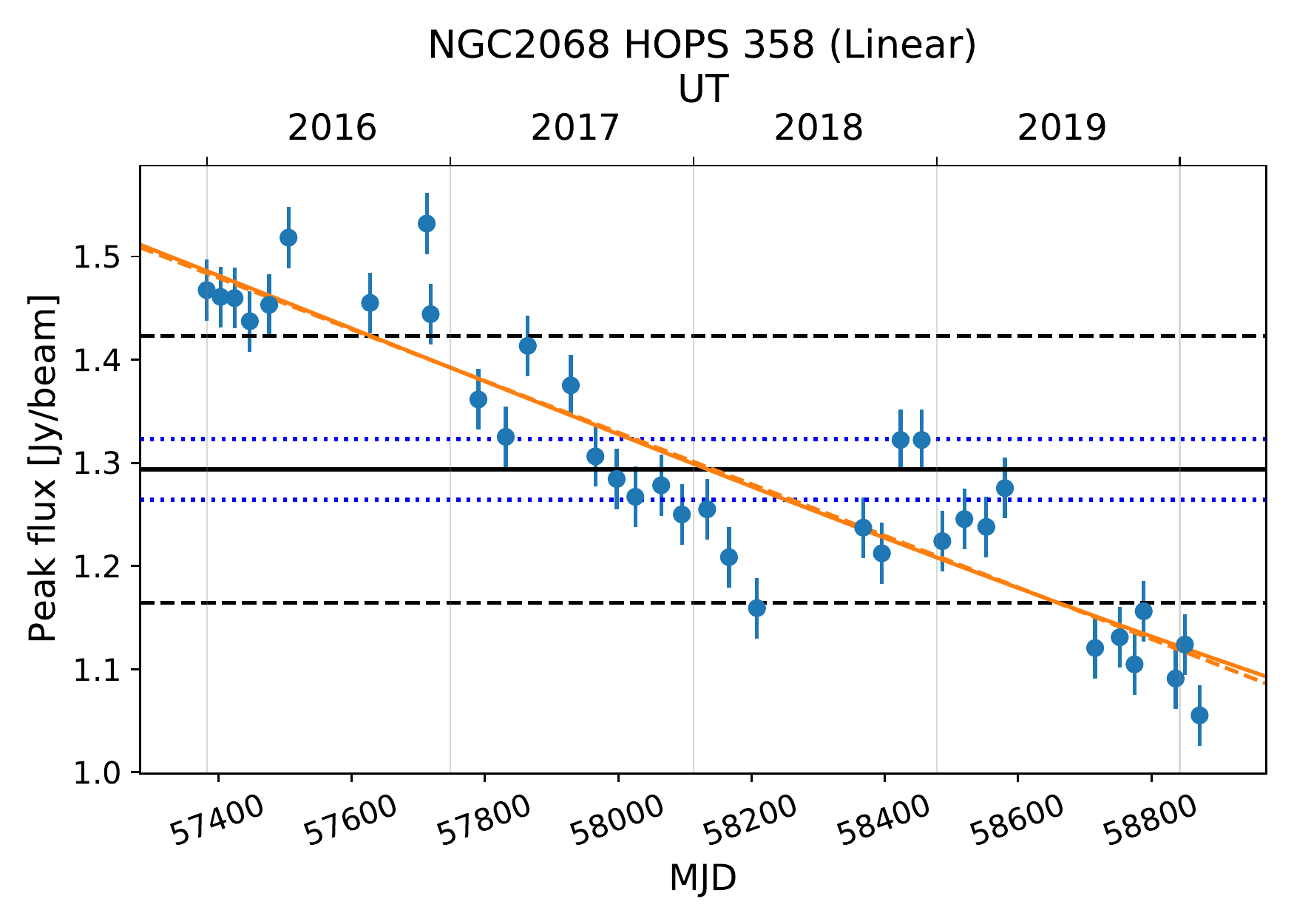}}
	\caption{Same as Figure \ref{fig:LC_IC348_MMS1} but for HOPS 358 in NGC 2068. The orange solid line denotes the sinusoidal fitting provided by LSP while the orange dashed line denotes the linear fitting result. Both the linear and sinusoidal fits are robust, FAP $< 0.1$\%.} 
\label{fig:LC_NGC2068_HOPS358}
\end{figure}

\newpage
\subsection{Perseus NGC\,1333, VLA\,3: Linear Group}

The submillimeter light curve for secular source Perseus NGC\,1333, VLA\,3 is presented in Figure B.18.
The submillimeter source was previously identified by \citet{enoch09} as Per-emb-44. It is a Class\,I protostar according to \citet{dunham15}. Although identified as a single source by \citet{tobin16} it separated into two peaks at cm wavelengths wavelengths \citep{tychoniec18}.  
The source was monitored by Spitzer at 3.6 and 4.5$\mu$m for $\sim$35 days as part of YSOVAR \citep{rebull15}), but was not found to be variable in the mid-infrared on that timescale.

\begin{figure}[htp]
	\centering
	{\includegraphics[trim={0.2cm 0.0cm 0.0cm 0.0cm},clip,width=0.9\columnwidth]{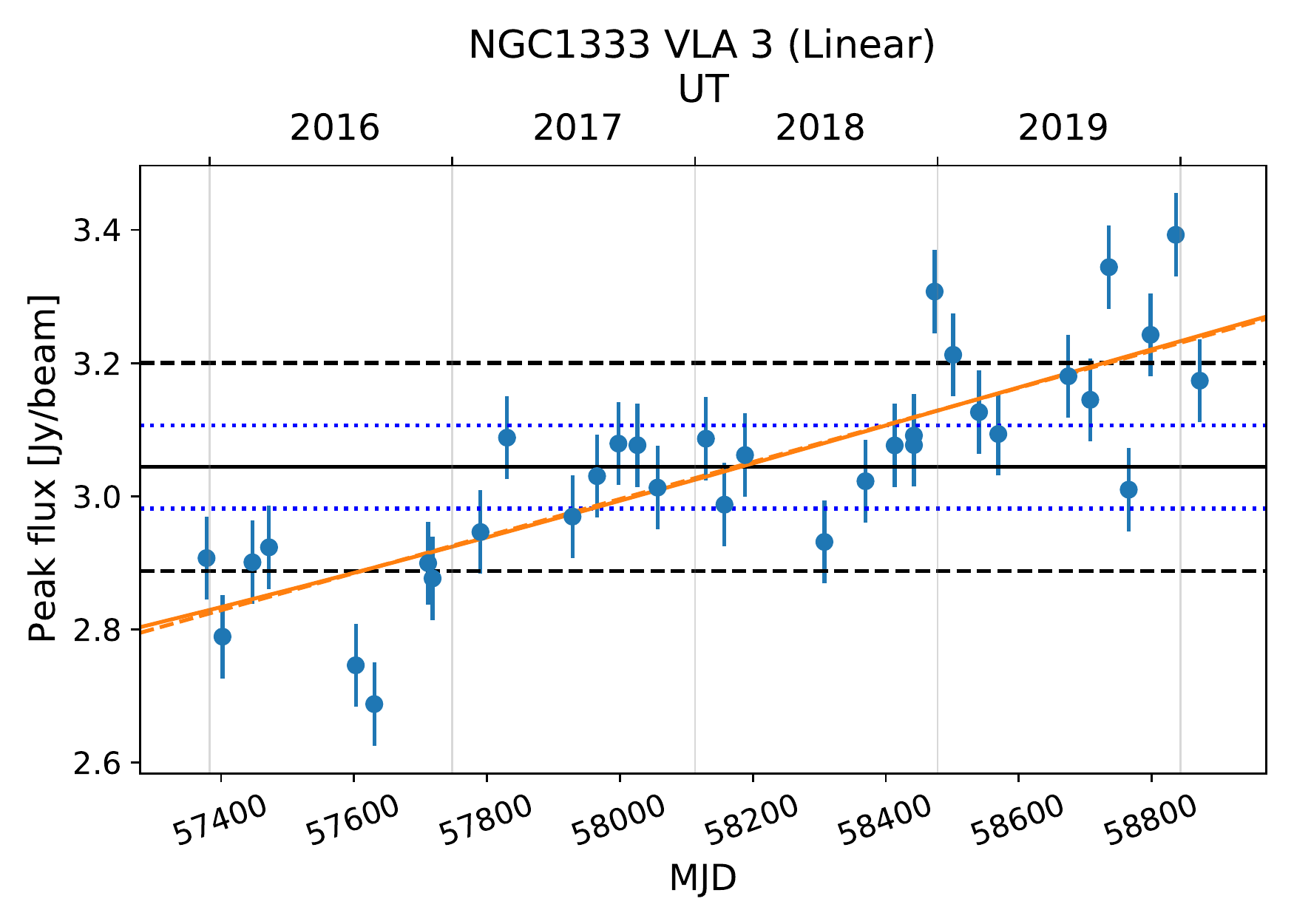}}
	\caption{Same as Figure \ref{fig:LC_IC348_MMS1} but for VLA 3 in NGC 1333. The orange solid line denotes the sinusoidal fitting provided by LSP while the orange dashed line denotes the linear fitting result. Both the linear and sinusoidal fits are robust, FAP $< 0.1$\%.} 
\label{fig:LC_NGC1333_VLA3}
\end{figure}


\end{document}